\documentclass[english,prd,nofootinbib,twocolumn,floatfix,showpacs]{revtex4}
\usepackage{hyperref}
\usepackage{amsfonts}
\usepackage{amsmath}
\usepackage{amssymb}
\usepackage{bm}
\usepackage{graphicx}
\usepackage{color}
\usepackage{epsfig}
\usepackage{remreset}
\usepackage{hyperref}
\usepackage{amsmath,amsthm}
\usepackage{mathtools}
\usepackage{mathrsfs}
\usepackage{color}
\usepackage{float}
\usepackage{tensor}
\usepackage{lipsum}
\usepackage[caption=false]{subfig}
\usepackage{hyperref}
{
 \definecolor{BLACK}{gray}{0}
 \definecolor{WHITE}{gray}{1}
 \definecolor{RED}{rgb}{1,0,0}
 \definecolor{GREEN}{rgb}{0,1,0}
\definecolor{dgreen}{rgb}{.1,.6,.1}
\definecolor{BLUE}{rgb}{0,0,1}
 \definecolor{CYAN}{cmyk}{1,0,0,0}
 \definecolor{MAGENTA}{cmyk}{0,1,0,0}
 \definecolor{YELLOW}{cmyk}{0,0,1,0}
 \definecolor{aw}{rgb}{0.2,0.5,0.75}
 }
 \hypersetup{
  colorlinks,%
  citecolor=blue,%
  filecolor=blue,%
  linkcolor=magenta,%
  urlcolor=blue
}
\definecolor{MyDarkRed}{rgb}{0.71,0.14,0.07}

\definecolor{YZBlue}{rgb}{0.0,0.0,1.0}

\definecolor{MyGreen}{rgb}{0.0,0.5,0.0}

\definecolor{amethyst}{rgb}{0.6, 0.4, 0.8}

\definecolor{grey}{gray}{0.5}
\newcommand{\init}[1]{{{#1}_{\mathbf i}}}
\newcommand{\rd}{{\mathrm d}}
\newcommand{\ddt}[1]{\frac{{\rm d}#1}{{\rm d}t}}
\newcommand{\CD}{\mathcal{D}}
\newcommand{\CJ}{\mathcal{J}}
\newcommand{\CR}{\mathcal{R}}
\newcommand\eprintarXiv[1]{\href{http://arXiv.org/abs/#1}{#1}}
\def\tLam{\tilde\Lambda}
\def\Ptil{\bar{P}}  \def\Pold{P}

\begin{document}
\title{Lagrangian theory of structure formation in relativistic cosmology. \\V. Irrotational fluids}
\author{Yong-Zhuang Li$^{1}$}\author{Pierre Mourier$^{2}$}
\author{Thomas Buchert$^{2}$\footnote{Corresponding author. buchert@ens--lyon.fr}}
\author{David L.~Wiltshire$^{1}$}
\affiliation{$^{1}$School of Physical \& Chemical Sciences, University of Canterbury, Private Bag 4800, Christchurch 8140, New Zealand}
\affiliation{$^{2}$Univ Lyon, Ens de Lyon, Univ Lyon1, CNRS, Centre de Recherche Astrophysique de Lyon UMR5574,
F--69007, Lyon, France}
\begin{abstract}
We extend the general relativistic Lagrangian perturbation theory,
recently developed for the formation of cosmic structures in a dust continuum,
to the case of model universes containing a single fluid with a single--valued
analytic equation of state. Using a coframe--based perturbation approach,
we investigate evolution equations for structure formation
in pressure--supported irrotational fluids that generate their rest--frame spacetime foliation.
We provide master equations to first order for the evolution of the trace and traceless parts of barotropic perturbations
that evolve in the perturbed space, where the latter describes the propagation of gravitational waves in the fluid. 
We illustrate the trace evolution for a linear equation of state and for a model equation of state describing isotropic velocity dispersion, and we discuss differences to the dust matter model, to the Newtonian case, and to standard perturbation approaches.
\end{abstract}
% four PACS numbers selected by PRD, but not included into the print version
%98.80.-k 	Cosmology (see also section 04 General relativity and gravitation; for origin and evolution of galaxies, see 98.62.Ai; for elementary particle and nuclear processes, see 95.30.Cq; for dark matter, see 95.35.+d; for dark energy, see 95.36.+x; for superclusters and large-scale structure of the Universe, see 98.65.Dx)
%98.65.Dx 	Superclusters; large-scale structure of the Universe (including voids, pancakes, great wall, etc.)
%95.35.+d 	Dark matter (stellar, interstellar, galactic, and cosmological) (see also 95.30.Cq Elementary particle processes; for brown dwarfs, see 97.20.Vs; for galactic halos, see 98.35.Gi or 98.62.Gq; for models of the early Universe, see 98.80.Cq)
%95.36.+x 	Dark energy (see also 98.80.-k Cosmology)
%98.80.Es 	Observational cosmology (including Hubble constant, distance scale, cosmological constant, early Universe, etc)
%98.80.Jk 	Mathematical and relativistic aspects of cosmology
%04.20.-q 	Classical general relativity (see also 02.40.-k Geometry, differential geometry, and topology)
%04.20.Cv 	Fundamental problems and general formalism
%04.25.Nx 	Post-Newtonian approximation; perturbation theory; related approximations
%04.30.-w 	Gravitational waves (see also 04.80.Nn Gravitational wave detectors and experiments)
\maketitle
\section{Introduction}
Relativistic cosmological perturbation theory is based on
evolving the Einstein equations with a global foliation of the spacetime
metric, via the $3+1$ formalism \cite{ADM, 3plus1bases}. In the standard approach
a spatially homogeneous and isotropic Friedmann--Lema\^{\i}tre--Robertson--Walker
(FLRW) geometry is assumed as the unperturbed global background spacetime, and Einstein's equations are then
solved to some order on this predefined background \cite{HideoMisao1984PTP}.
The standard approach is Eulerian in the sense that perturbations are represented
and propagate on this background that corresponds, in the Newtonian limit, to Eulerian perturbation theory. In this latter case, a perturbation method for the density and velocity fields is used to solve the
Euler--Poisson system of equations that governs the fluid evolution
\cite{Bernardeau2002PR}. Cosmological structure formation in the nonperturbative regime
is also generally modeled within the Newtonian framework.

An alternative approach to structure formation has also been developed,
principally in the Newtonian regime, which
is directly tied to fluid elements. It is consequently known as Lagrangian
perturbation theory \cite{Zeldovich1970a,BuchertGG1987,Buchert1989AA,bildhaueretal,buchert92,buchert93L,buchert94,buchert:varenna,
buchertehlers,Bouchet1995AA,BuchertJE1997,sharvari11,sharvari13,Eleonora2014JCAP,Cornelius2012JCAP,HAMK1999,Hideki2000,rampfbuchert}
to distinguish it from the Eulerian approach based on coordinates on an assumed global background.
The Lagrangian approach uses a single perturbation variable, the
fluid's deformation field.
This gives it the advantage of also applying in the nonlinear
regime, where Eulerian density perturbations are large. In recent years, Lagrangian
perturbation theory has been generalized to general relativistic cosmologies
with a dust continuum \cite{rza1,rza2,rza3,rza4}; see also \cite{Kasai1995PRD,Heinz1996PRD,Sabino1996David,salopek,ellistsagas,rampf:zeldovich,Cornelius2014PRD}.

In the Newtonian regime, an extension of Lagrangian perturbation
theory to fluids with dynamic pressure was considered first in terms
of isotropic pressure \cite{adlerbuchert}. The resulting
Lagrangian perturbation equations have been solved up to second order for a
polytropic fluid \cite{MMTT,TTMS}. For third order perturbative solutions in
Newtonian Lagrangian perturbation theory with pressure, see Ref.~\cite{TT1}.
Models with isotropic pressure can also be considered as phenomenological models for the generally anisotropic pressure originating from the velocity dispersion of dust particles \cite{M1,M2,M3},
by taking velocity moments of the collisionless Vlasov equation \cite{TBAD,RN50}.
For a sequence of modeling assumptions used in nonperturbative extensions of Lagrangian perturbation theory, see the
summary \cite{buchert06}.

In this paper we will extend relativistic Lagrangian perturbation theory for a dust matter model
\cite{rza1,rza2,rza3,rza4} to the
case of irrotational perfect fluids, and also to cases that are relevant for the modeling of multistream regimes where the dust approximation breaks down. This will provide a framework not only to deal with
a relativistic generalization of Newtonian Lagrangian perturbation theory with pressure
at late epochs, but also to the fully relativistic situation of the early Universe.

A primary motivation for such an investigation is to establish
a framework which is better suited to studies of the backreaction of
inhomogeneities in cosmology as compared to standard perturbation theory. In
particular, standard cosmological perturbation theory conventionally assumes
that average cosmic evolution is {\em exactly} described by a solution to
Einstein's equations with a prescribed energy--momentum tensor on a global
hypersurface irrespective of the scale of coarse--graining of the matter fields.
No fundamental physical principle demands such an outcome \cite{w14}.

The scalar averaging scheme introduced in \cite{b00,b01,TBDarkenergy,bmr18} is an example
of an approach to backreaction of inhomogeneities in cosmology, in which
the Einstein equations are assumed to hold on small scales, where they are well--tested, but not for the
average cosmic evolution on arbitrarily large spatial scales. This is a
consequence of the fact that a generic averaging operation
includes nonlocal fluctuation terms, and it should not be confused with
modified gravity approaches which change the Einstein-Hilbert action.
A variety of
phenomenological interpretations of the Buchert scheme are possible
\cite{bla06,labkc09,multiscale,w07a,w07b,w09,chaplygin,phasespace,inflation,Zimdahl2016},
since additional ingredients are required to relate statistical quantities
to physical observables determined from our own cosmological observations.

To date, no phenomenological approach to backreaction has fully utilized
the general scalar averaging framework for perfect fluids \cite{b01}.
In the timescape scenario \cite{w07a,w07b,w09}, solutions to the corresponding
system of averaged scalar equations
have been given with matter and radiation \cite{dnw13} extending
smoothly into the early radiation--dominated epoch in the early Universe.
However, in deriving these solutions it was assumed that backreaction is
insignificant before photon--electron decoupling, so that backreaction
involving pressure terms was neglected.

Neglecting backreaction in the primordial plasma may seem to be a reasonable
approximation for the evolution of the background universe to leading order,
given that it is extremely close to being spatially homogeneous and
isotropic at early times. However, backreaction can
nonetheless make a significant difference when considering the growth of
perturbations. In particular, even if the difference from the Friedmann
equation is of order $10^{-5}$ as a fraction of energy density
at decoupling, this is nonetheless of the same order as the density
perturbations.
A recent study of Cosmic Microwave Background (CMB) anisotropies
in the timescape model found that neglecting such small differences in initial conditions at last scattering leads to systematic uncertainties of $8$--$13$\% for particular cosmological parameters
at the present epoch \cite{nw15}.
This remark applies to the conservative assumption that the background universe
does not already contain backreaction arising from earlier epochs that could be compatible with
large--scale homogeneity and isotropy \cite{inflation}.

For these reasons, we desire a new approach to cosmological perturbation theory
which is intrinsic to the fluid and not anchored to an embedding space.
Relativistic Lagrangian perturbation theory represents
a promising avenue, as it is intimately tied to physical particles. To proceed
to a fully realistic theory will require important steps beyond those which
we investigate in this paper. Such steps will include:
\begin{itemize}
\item An extension from one fluid to the many fluids pertinent to the early
Universe, which requires considering a tilt between various fluid flow vectors and the normal to the spatial hypersurfaces;\footnote{Note that in the standard approach, the same FLRW frame is used for different matter components. (Even in this idealized case there are important differences to be respected for the different matter components \cite{Wessel}.)}
\item Identifying relevant physical scales and volume partitioning the model universe into
regions whose average evolves by averaged dynamical equations, rather than by global Friedmann equations;
\item Aiming at a background--free description. While perturbations
are still formulated in the present paper as deviations from a fixed background cosmology, a general volume partitioning
can be implemented without referring to a background \cite{buchertcarfora:curvature,multiscale}.
\end{itemize}
As a first step towards these goals, in the present paper we will firstly
consider relativistic Lagrangian perturbation theory for the same system that
was considered in Ref.\ \cite{b01},
namely a single component perfect fluid with barotropic equation of state.
We will also include an explicit cosmological constant term.

This paper is organized as follows.
In Section \ref{SecII} we employ a $3+1$ formalism \cite{ADM, 3plus1bases} with Lagrangian spatial coordinates, presenting the general framework and foliation structure for a general irrotational matter model. We then restrict our attention to a barotropic fluid and discuss in detail the fluid variables and their equation of state. In this context, in Section \ref{SecIII} we introduce Cartan's coframe formalism, proceeding with the relativistic Lagrangian perturbation approach. We develop the first--order Lagrangian scheme and derive master equations for the trace and trace--free parts of the perturbation field. In Section \ref{SecIV} we apply the first--order Lagrangian scheme to particular matter models, allowing us to explicitly derive solutions for the trace part, and we illustrate and discuss the results. Particular solutions for the gravitoelectric traceless part are studied in Appendix \ref{solutiontraceless}. We summarize our main results in Section \ref{SecV}.
\vspace{-5pt}
\section{Spacetime foliation structure and $\textbf{3}\textbf{+}\textbf{1}$ Einstein equations}
\label{SecII}
\vspace{-5pt}
In this paper we will consider a model universe containing a single irrotational fluid, so that a foliation of spacetime
into flow--orthogonal hypersurfaces can be introduced.
\vspace{-5pt}
\subsection{Decomposition of Einstein's equations for flow--orthogonal hypersurfaces}
The irrotationality assumption on the fluid amounts to the existence of two scalar functions,
$N$ and $t$, such that the $1-$form dual to the normalized $4-$velocity vector $u^\mu$ of the fluid can be written
as:\footnote{%
In the convention we use here, greek letters $\mu, \nu, \cdots$ are spacetime indices running from $0$ to $3$, while lowercase latin letters $i, j, \cdots$ are spatial indices running from $1$ to $3$. We use units in which $c=1$, if not
otherwise stated.}
\begin{equation}
u_{\mu}=- N\,\partial_{\mu}t \qquad;\qquad N := \left(-\partial^{\mu}t \,
\partial_{\mu}t \right)^{-1/2} \; .
\label{4velocityfluid}
\end{equation}
The level sets of $t$ then define flow--orthogonal hypersurfaces, labeled $\Sigma_t$,
which foliate spacetime, with unit normal $u^\mu$, $u^{\mu}u_{\mu}=-1$.
We will now follow the $3+1$ formalism \cite{ADM,3plus1bases} and define our time coordinate as coinciding with this function $t$. In this case, $N(t,x^i)$ is the {\em lapse function.}

In addition, we choose the spatial coordinates
to be spatial Lagrangian (or comoving) coordinates, denoted $X^i$, that
are assumed to be constant along each flow line. In the set of coordinates
$(X^\mu) = (t,X^i)$, the components of the fluid $4-$velocity vector and its dual
are then respectively:
\begin{equation}
u^{\mu}=\frac{1}{N}(1, 0, 0, 0)\quad;\quad u_{\mu}=(-N, 0, 0, 0)\,,
\label{unitvector}
\end{equation}
while the line element can be written as
\begin{equation}
\rd s^2= g_{\mu\nu}\, \rd X^{\mu} \rd X^{\nu} = -N^{2}\rd t^2+g_{ij}\, \rd X^{i} \rd X^{j} \; .
\label{metric1}
\end{equation}
Here, $g_{ij}$ corresponds both to the spatial coefficients
of the $4-$metric $g_{\mu \nu}$ and to the components of the
$3-$metric that it induces
on the hypersurfaces $\Sigma_t$.
Introducing the projector onto $\Sigma_{t}$, $h_{\mu\nu}=g_{\mu\nu}+u_{\mu}u_{\nu}$, this $3-$metric is indeed
\begin{equation}
h_{ij}:=g_{\mu\nu}\tensor{h}{^\mu_i}\tensor{h}{^\nu_j} = g_{i j} \; .
\end{equation}
The spatial metric and the lapse function $N$ together encode
the inhomogeneities. (We will later use the more elementary coframe coefficients instead of the $3-$metric coefficients.)
We use $\tensor{\mathcal{R}}{^i_j}$ to denote the Ricci tensor coefficients of this spatial metric, with $\mathcal{R}$ the corresponding Ricci scalar.

Without loss of generality, the energy--momentum tensor of the fluid is given by
\begin{equation}
 T_{\mu\nu}=(\epsilon+p)u_{\mu}u_{\nu}+pg_{\mu\nu}+\pi_{\mu\nu}+q_{\mu}u_{\nu}+q_{\nu}u_{\mu}\;,\label{en-mom}
\end{equation}
where $\pi_{\mu\nu}$ is an anisotropic pressure, with
$\pi_{[\mu \nu]} = 0$, $u^\mu \pi_{\mu \nu} = 0$ and $\tensor{\pi}{^\mu_\mu}=0$,
and $q_{\mu}$ the heat flux, with $q_{\mu}u^{\mu}=0$.

Introducing the expansion tensor (as minus the extrinsic curvature) of the hypersurfaces,
\begin{equation}
\Theta_{ij} :=\nabla_{\nu}n_{\mu}\tensor{h}{^\mu_i}\tensor{h}{^\nu_j}=\frac{1}{2N}\partial_{t}g_{ij}\;,
\end{equation}
Einstein's equations with a cosmological constant may be cast into a set of constraint
and evolution equations. The constraint equations are the energy and momentum constraints:\footnote{%
The symbol $\parallel$ denotes the covariant derivative with respect to the $3-$metric $h_{ij}$.
When applied to scalars it reduces to a partial derivative, denoted $\mid$, with respect to the Lagrangian coordinates, $X^i$.}
\begin{equation}
\begin{aligned}
\mathcal{R}+\Theta^{2}-\tensor{\Theta}{^i_j}\tensor{\Theta}{^j_i}&=16\pi G \, \epsilon \,+\, 2\, \Lambda\; ; \\
\tensor{\Theta}{^i_j_\parallel_i}-\Theta_{\mid j}&=-8 \pi G q_j \; .
\label{constraint1}
\end{aligned}
\end{equation}
The propagation equations are given by
\begin{equation}
\begin{aligned}
\tensor{\Theta}{^i_j}= &\, \frac{1}{2N}g^{ik}\partial_{t}g_{kj} \; ; \\
N^{-1}\partial_{t}\tensor{\Theta}{^i_j}=&-\Theta \tensor{\Theta}{^i_j} -\tensor{\mathcal{R}}{^i_j} + \tensor{\mathcal A}{^i_j} \\
& + 4\pi G \left[(\epsilon-p) \, \tensor{\delta}{^i_j}+2\tensor{\pi}{^i_j} \right] \, + \, \Lambda \, \tensor{\delta}{^i_j} \; ,
\label{evolution1}
\end{aligned}
\end{equation}
where $a_{\mu}:=u^{\nu}\boldsymbol{\nabla}_{\nu}u_{\mu} = N^{-1} N_{||\mu}$ is the covariant acceleration of the fluid
(with $\boldsymbol{\nabla}$ denoting the $4-$covariant derivative), and
$\tensor{\mathcal A}{^i_j} := a^{i}_{\ \parallel j}+a^{i}a_{j} = N^{-1}N^{\parallel i}_{\;\;\; \parallel j}$.
Combining the trace of the second equation with the energy constraint yields the Raychaudhuri equation:
\begin{equation}
N^{-1}\partial_{t}\Theta=-\frac{1}{3}\Theta^{2}-2\sigma^{2}-4\pi G (\epsilon+3p)+\mathcal{A} \, + \, \Lambda \; ,
\label{traceevolution11}
\end{equation}
where $\mathcal{A}:= \tensor{\mathcal A}{^i_i} =\boldsymbol{\nabla}_{\mu}a^{\mu}= N^{-1}N^{\parallel i}_{\;\;\; \parallel i}$.

With the spacetime described by the given metric, the energy--momentum conservation laws are expressed as follows:
\begin{align}
& \!\!\! \partial_{t}\epsilon+N\Theta(\epsilon+p) =-N \left(\tensor{q}{^\mu_{||\mu}} + 2 q^\mu a_\mu
		+ \tensor{\sigma}{_\mu_\nu}\tensor{\pi}{^\mu^\nu} \right)\, ;
\label{conservationlaw1}\\
& \!\!\! (\epsilon + p) \, a_{\mu}+ p_{||\mu} = - \left(\tensor{\pi}{_\mu_\nu^|^|^\nu} +a^{\nu}\tensor{\pi}{_\mu_\nu} \right) \nonumber \\
	& \qquad - \left( \frac{4}{3} \Theta \, q_\mu + q^\nu \sigma_{\mu \nu} + u^\nu \boldsymbol{\nabla}_\nu q_\mu - q^\nu a_\nu \, u_\mu \right) \, .
\label{conservationlaw2}
\end{align}
In what follows, we will specialize to the case of isotropic pressure, $\pi_{\mu\nu}=0$, and vanishing heat flux,
$q_\mu=0$. Note that with these assumptions we do still allow for some nonperfect fluids, since $p$ is not necessarily
the local thermodynamic equilibrium pressure \cite{Ellis2012}.
Such a restriction is required here since both extra terms in general create vorticity,
which cannot be covered by the class of flow--orthogonal foliations considered in this work.

Let us illustrate this by considering more closely the irrotationality condition for a fluid with negligible heat flux, $q_\mu = 0$, to see how this condition constrains the equation of state and the anisotropic pressure.
The vanishing of the vorticity $2-$form implies vanishing of the antisymmetrized projected gradient of the acceleration,
$a_{[\nu|| \mu]} = 0$, since $a_\mu = (\ln N)_{||\mu}$ from \eqref{4velocityfluid}, being a consequence of the existence of the fluid--orthogonal foliation.
From this, one obtains through \eqref{conservationlaw2} the following constraint on the energy--momentum components:
\begin{align}
 \epsilon_{||[\mu} \,\, p_{||\nu]} + (\epsilon + p)_{||[\mu} \, \tensor{h}{^\rho_\nu_]} \nabla_\sigma \tensor{\pi}{^\sigma_\rho} \hspace{15ex} \nonumber \\
		- (\epsilon + p) \, \tensor{h}{^\rho_[_\mu} \tensor{h}{^\sigma_\nu_]} \nabla_\rho \nabla_\tau \tensor{\pi}{^\tau_\sigma} = 0 \; .
\end{align}
Since $\nabla_\mu \tensor{\pi}{^\mu_\nu} = 0$ would imply the vanishing of the right hand sides of \eqref{conservationlaw1}--\eqref{conservationlaw2}, an anisotropic pressure that does contribute to the dynamics will satisfy $\nabla_\mu \tensor{\pi}{^\mu_\nu} \neq 0$ and thus will not fulfill the above condition in general, producing a vortical flow.
Conversely, a barotropic fluid flow with $\pi_{\mu \nu} = 0$ and an effective equation of state of the form $p = \beta(\epsilon)$, automatically satisfies the above constraint. Moreover, for such a fluid, \eqref{conservationlaw2} allows one to write the acceleration as a flow--orthogonal projected gradient, and it will indeed obey the relativistic equivalent of the Kelvin--Helmholtz theorem,
so that irrotationality will be preserved along the flow lines \cite{Ehlers1961,Ellis2012}. 

\subsection{Barotropic perfect fluid spacetimes}

For the remainder of this paper we will only consider fluids with $q_\mu=0$ and $\pi_{\mu\nu}=0$. 
The energy--momentum tensor (\ref{en-mom}) then reduces to perfect fluid form:
\begin{equation}
T_{\mu\nu}=(\epsilon+p)u_{\mu}u_{\nu}+pg_{\mu\nu} \; ,
\label{energymomentum1}
\end{equation}
while its conservation equations \eqref{conservationlaw1}--\eqref{conservationlaw2} become, respectively
\begin{gather}
\label{conservationlaw1_PF}
\partial_t \epsilon + N \Theta (\epsilon + p) = 0 \; ; \\
\label{conservationlaw2_PF}
a_\mu = - \frac{p_{||\mu}}{\epsilon + p} \; .
\end{gather}
As a further restriction we will assume that the fluid flow is {\em barotropic}, \textit{i.e.,} we assume a local relation of the form $p= \beta(\epsilon)$ to effectively hold throughout the entire fluid,\footnote{%
\label{spatialdependence}%
Considering the local dynamical solution for these variables, there is always a freedom of integration constant that depends on the Lagrangian coordinates, \textit{i.e.}, on the particular fluid element. We assume here that the same relation holds for all fluid elements. Only this assumption makes the dynamical relation an apparent equation of state that is valid throughout the fluid flow. All related variables then also depend on this assumption, which is a restriction imposed on initial data.%
}
that we will henceforth call \emph{the equation of state} or EoS. As noted earlier, such a relation will ensure that the flow remains irrotational. For such a fluid, setting some reference constant energy and rest mass density values $\epsilon_1, \varrho_1$, we may use the EoS to define a formal rest mass density $\varrho(\epsilon)$ and a related specific enthalpy $h(\epsilon)$ -- as an injection energy per fluid element and unit formal rest mass \cite{Israel1976} -- respectively, by
\begin{gather}
 \label{def_varrho}
 \varrho := F(\epsilon) := \varrho_1 \exp \int_{\epsilon_1}^\epsilon \frac{{\mathrm d}x}{x + \beta(x)} \; ; \\
 \label{def_h}
 h (\epsilon) := \frac{\epsilon + \beta(\epsilon)}{F(\epsilon)} = \frac{\epsilon + p}{\varrho} \; .
\end{gather}
The energy--momentum conservation equations \eqref{conservationlaw1} and \eqref{conservationlaw2} then, respectively, provide a conservation law for $\varrho$,
\begin{equation}
 \partial_{t}\varrho+N\Theta\varrho=0 \; ,
 \label{restmass_conservation}
\end{equation}
and a relation between the specific enthalpy (\ref{def_h}) and the lapse,
\begin{equation}
 \frac{N_{||\mu}}{N} = a_\mu = - \frac{h_{||\mu}}{h} \; : \; (N h)_{|i} = 0 \; .
\end{equation}
By an appropriate choice of the hypersurface--labeling function $t$, the lapse can thus be rescaled so that
 \cite{b01,Ellis2012}
\begin{equation}
N = \frac{1}{h} = \frac{F(\epsilon)}{\epsilon + \beta(\epsilon)} \; .
\label{enthalpy1}
\end{equation}
If we assume that the fluid remains in thermodynamic equilibrium locally, and if it has a nonvanishing rest mass density, then this density will follow the same evolution law \eqref{restmass_conservation} as $\varrho = F(\epsilon)$, by rest mass conservation. This formal $\varrho$ and the actual rest mass density will then coincide up to a possible different spatial dependence (\textit{cf.,} footnote \ref{spatialdependence}). These two quantities may be made equal by a suitable choice of initial conditions for the rest mass density or local thermodynamic equilibrium assumptions.\footnote{%
Let us take the local state of the fluid to belong to a thermodynamic Gibbs space admitting the equation of state $u(s,v)$, where $s$ is the specific entropy, $v$ is the specific volume and $u = \epsilon v$ is the specific internal energy. If we now assume that $p$ is the local thermodynamic equilibrium pressure of the fluid, it can then be expressed as $p(s,v) = - \partial u / \partial v$. Provided that a specific equation of state does not render the above relations degenerate, then these relations may be inverted to provide $v(\epsilon,p)$. Within a barotropic flow satisfying $p = \beta(\epsilon)$, the actual rest mass density $v^{-1}$ thus only depends on the energy density $\epsilon$, which fully determines its initial conditions. From the conservation equations of both quantities, $\partial_t \epsilon / (\epsilon + \beta(\epsilon)) = -N \Theta = \partial_t (v^{-1}) / v^{-1}$, this dependency must be $v^{-1} = F(\epsilon)$, for $\Theta$ not identically vanishing, up to a constant prefactor which can be absorbed in the choice of $\varrho_1$. Hence, in this case, $F(\epsilon)$ is indeed the rest mass density of the fluid with no further loss of generality. Also note that under the same assumptions, $s$ is also a function of $\epsilon$, preserved along the flow lines as the flow is adiabatic: $\partial_t s = 0 = (\rd s / \rd \epsilon) \, \partial_t \epsilon$, while $\partial_t \epsilon$ is not identically vanishing. The flow is thus {\em isentropic}, $s$ being a constant $s_1$ that depends neither on time nor on the fluid element. The barotropic relation then corresponds to the equation $p(\epsilon,s)$ deduced from the thermodynamic equation of state, and taken at constant $s$, $\beta(\epsilon) = p(\epsilon, s=s_1)$ (see \cite{Israel1976,Israel1979,Ehlers1961,Friedrich1998,Stephani2003,Ellis2012}).\label{footnoteTD}}
This would then ensure the validity of the interpretation of
$\varrho$ and $h$ as the physical rest mass density (or particle number density) and specific enthalpy of the fluid, respectively. We will not, however, make such assumptions in the following Section \ref{SecIII}, to keep its level of generality. This will allow us to consider the case of a zero rest mass fluid (for which $F(\epsilon) \neq 0$ and $h(\epsilon)$ are still well--defined), as well as that of a nonzero rest mass density with less constrained initial conditions. It will also allow us to consider the variable $p$ as an effective pressure term --- \textit{e.g.}, modeling velocity dispersion --- instead of the local thermodynamic equilibrium pressure. For the general treatment including these cases it will suffice to formally define $\varrho$ and $h$ from Equations \eqref{def_varrho}--\eqref{def_h} using the single barotropic assumption $p = \beta(\epsilon)$.
We follow the notation of Ref.~\cite{b01} here.

\section{Lagrangian perturbation scheme}
\label{SecIII}

In this section we will introduce the coframe formalism to describe spacetime, which is a set of four deformation $1-$form fields dual to a generally noncoordinate basis of vectors at every point of the manifold \cite{Marc1999, Ellis1967,Zakharov2006}.
A general relativistic version of a coframe--based perturbative approach for an irrotational dust continuum
has been proposed in Ref.~\cite{Kasai1995PRD}, developed further in Ref.~\cite{Sabino1996David} and in final form,
featuring only the coframes as the single perturbation variable in Ref.~\cite{rza1}.

\subsection{Coframe formulation}

Following \cite{rza2,rza3,rza4}, we construct a set of three spatial coframes $\boldsymbol{\eta}^{a}$
such that the spatial metric can be rewritten in the form
\begin{equation}
\tensor[^(^3^)]{\mathbf{g}}{ }=G_{ab}\,\boldsymbol{\eta}^{a}\otimes\boldsymbol{\eta}^{b} \quad :
		\quad g_{ij}=G_{ab}\tensor{\eta}{^{a}_i}\tensor{\eta}{^{b}_j} \; .
\end{equation}
Here $G_{ab}(\bm{X})$ is the Gram matrix that encodes all the initial spatial metric perturbations,
$G_{ab}(\bm{X}) := \tensor{\delta}{_a^i} \tensor{\delta}{_b^j} G_{i j}(\bm X)$, with the initial metric coefficients, $G_{i j}(\bm X) := g_{ij}(\init t, \bm{X})$.
On the other hand we can also include the temporal component into the matrix and rewrite it as
\begin{equation}
\tensor{\tilde{G}}{_{\alpha}_{\beta}}=\left(
\begin{array}{cc}
-1 & 0 \\
0 & G_{ab} \\
\end{array}\right) \; .
\end{equation}
With this we introduce a full set of four spacetime coframes $\boldsymbol{\eta}^{\alpha}$ to describe the $4-$metric $\tensor[^(^4^)]{\mathbf{g}}{}$:
\begin{equation}
\tensor[^(^4^)]{\mathbf{g}}{} = \tensor{\tilde{G}}{_{\alpha}_{\beta}} \, \boldsymbol{\eta}^{\alpha}\otimes\boldsymbol{\eta}^{\beta} \; ,
\end{equation}
by defining the coframe components as
\begin{equation}
\tensor{\eta}{^0_\mu}=(-N, 0, 0, 0) \quad ; \quad \tensor{\eta}{^a_\mu}=(0, \tensor{\eta}{^a_i})  \; .
\label{component11}
\end{equation}
We now define the transformation between coordinate and noncoordinate bases as:
$\mathcal{J}=\sqrt{-g}/\sqrt{-\tilde{G}}=\sqrt{-g}/\sqrt{G}$ (the signature adopted here being $(-1, 1, 1, 1)$, and using the notation $g := \det({}^{(4)} \mathbf{g})$, $\tilde{G} := \det(\tilde{G}_{\alpha \beta})$ and $G := \det(G_{ab})$).
This corresponds to $\mathcal{J}=-\det(\tensor{\eta}{^{\alpha}_\mu})$,~or,
\begin{equation}
\begin{aligned}
\frac{1}{4!}\epsilon_{\alpha\beta\gamma\delta}\,\boldsymbol{\eta}^{\alpha}&\wedge\boldsymbol{\eta}^{\beta}\wedge\boldsymbol{\eta}^{\gamma}\wedge\boldsymbol{\eta}^{\delta}=\\
&-\frac{1}{4!}\mathcal{J}\epsilon_{\mu\nu\rho\sigma} \, \rd X^{\mu}\wedge \rd X^{\nu} \wedge \rd X^{\rho}\wedge \rd X^{\sigma} \; .
\end{aligned}
\end{equation}
From Eq.~(\ref{component11}), in terms of the spatial
components of the coframes, $\mathcal{J}$ becomes
\begin{equation}
\mathcal{J}=\frac{1}{3!}N\epsilon_{abc}\epsilon^{ijk}\tensor{\eta}{^a_i}\tensor{\eta}{^b_j}\tensor{\eta}{^c_k}
	 {} = N \det(\tensor{\eta}{^{a}_i}) \; ,
\label{newtransformation2}
\end{equation}
while correspondingly, the dual vector basis can be described by the four frames $\bm{e}_{\alpha}=\tensor{e}{_{\alpha}^\mu} \, \partial/\partial X^{\mu}$:
\begin{equation}
\begin{gathered}
\tensor{e}{_{\alpha}^\mu} \, \tensor{\eta}{^{\alpha}_\nu}=\tensor{\delta}{^\mu_\nu} \;\; ; \;\; \tensor{e}{_\alpha^\mu} \, \tensor{\eta}{^\beta_\mu} = \tensor{\delta}{_\alpha^\beta} \; ; \\
\tensor{e}{_{\alpha}^\mu}=-\frac{1}{6\mathcal{J}}\,\epsilon_{\alpha\beta\gamma\delta}\,\epsilon^{\mu\nu\varrho\sigma}\tensor{\eta}{^{\beta}_\nu}\tensor{\eta}{^{\gamma}_\varrho}\tensor{\eta}{^{\delta}_\sigma} \; ;\\
\tensor{e}{_a^i}=\frac{1}{2\mathcal{J}}N\epsilon_{abc}\epsilon^{ijk}\tensor{\eta}{^b_j}\tensor{\eta}{^c_k} \; ;\\
\tensor{e}{_0^\mu}=\frac{1}{N}(-1, 0, 0, 0) \; ; \; \tensor{e}{_a^\mu}=(0, \tensor{e}{_a^i}) \; .
\end{gathered}
\end{equation}
With this choice, the evolution equations for $\mathcal{J}$ and the expansion tensor coefficients $\tensor{\Theta}{^i_j}$ read:
\begin{equation}
\begin{aligned}
\partial_{t}\mathcal{J}&=\frac{\partial_{t}N}{N}\mathcal{J}+\mathcal{J}N\Theta \; ;\\
\tensor{\Theta}{^i_j}&=\frac{1}{2\mathcal{J}}\epsilon_{abc}\epsilon^{ikl}\left(\partial_{t}\tensor{\eta}{^a_j}\right)\tensor{\eta}{^b_k}\tensor{\eta}{^c_l} \; ;\\
\frac{\partial_{t}\tensor{\Theta}{^i_j}}{N}&=-\Theta \tensor{\Theta}{^i_j}+\frac{1}{2\mathcal{J}}\epsilon_{abc}\epsilon^{ikl}\partial_{t}\Big(\frac{1}{N}\partial_{t}\tensor{\eta}{^a_j}\Big)\tensor{\eta}{^b_k}\tensor{\eta}{^c_l}\\
&\quad+\frac{1}{N\mathcal{J}}\epsilon_{abc}\epsilon^{ikl}\left(\partial_{t}\tensor{\eta}{^a_j}\right)\left(\partial_{t}\tensor{\eta}{^b_k}\right)\tensor{\eta}{^c_l} \;.
\label{newevolutionexpan2}
\end{aligned}
\end{equation}
From the constraint and evolution equations \eqref{constraint1}--\eqref{traceevolution11}, together with the definition of $\mathcal{J}$ and Eqs.~(\ref{newevolutionexpan2}),
the Lagrange--Einstein system of an irrotational barotropic fluid model is cast into the following form:
\begin{gather}
\label{LES2_symmetry}
G_{ab} \, \partial_{t}\tensor{\eta}{^a_[_i}\tensor{\eta}{^b_j_]}=0 \; ; \\
\frac{1}{2\mathcal{J}} \epsilon_{abc}\epsilon^{ikl}
	\, \partial_{t}\!\left(\frac{1}{N}\left(\partial_{t}\tensor{\eta}{^a_j}\right)\tensor{\eta}{^b_k}\tensor{\eta}{^c_l} \right)
	=\tensor{\mathcal{A}}{^i_j} - \tensor{\mathcal{R}}{^i_j} \nonumber \\
\label{LES2_evol}
	 \hspace{20ex} + \left[ 4 \pi G (\epsilon - p) + \Lambda \right] \tensor{\delta}{^i_j} \; ; \\
\label{LES2_hamilton}
\epsilon_{abc}\epsilon^{ijk}\left(\partial_{t}\tensor{\eta}{^a_i}\right)\left(\partial_{t}\tensor{\eta}{^b_j}\right)\tensor{\eta}{^c_k}
	=(16\pi G \epsilon \, + \, 2 \, \Lambda -\mathcal{R})N\mathcal{J} \; ; \\
\label{LES2_momentum}
\left[\tfrac{1}{\CJ}\epsilon_{abc}\epsilon^{ikl}\left(\partial_{t}\tensor{\eta}{^a_j}\right)\tensor{\eta}{^b_k}\tensor{\eta}{^c_l}\right]_{\parallel i}
=\left[\tfrac{1}{\CJ}\epsilon_{abc}\epsilon^{ikl}\left(\partial_{t}\tensor{\eta}{^a_i}\right)\tensor{\eta}{^b_k}\tensor{\eta}{^c_l}\right]_{\mid j};\\ p = \beta(\epsilon) \; .
	\label{LES2_EoS}
\end{gather}
Equations \eqref{LES2_symmetry}--\eqref{LES2_momentum} are not closed unless an EoS, here \eqref{LES2_EoS}, is specified. Recall that the lapse appearing above can be replaced by its expression in terms of $\epsilon$, $N = (\epsilon + \beta(\epsilon))^{-1}\, F(\epsilon)$.
The evolution equation \eqref{LES2_evol} may be split into a trace part, which we then combine
with the energy constraint \eqref{LES2_hamilton} to obtain the Raychaudhuri equation, and a traceless part, yielding respectively:
\begin{gather}
\label{LES2_raychaudhuri}
\frac{1}{2\mathcal{J}}\epsilon_{abc}\epsilon^{ikl}\partial_{t}\!
	\left(\frac{1}{N}\partial_{t}\tensor{\eta}{^a_i}\right)\tensor{\eta}{^b_k}\tensor{\eta}{^c_l}
	=\mathcal{A}-4\pi G (\epsilon+3p) \, + \, \Lambda \; ; \\
\frac{1}{2\mathcal{J}}\left[\epsilon_{abc}\epsilon^{ikl}
	\partial_{t}\left(\frac{1}{N}\left(\partial_{t}\tensor{\eta}{^a_j}\right)\tensor{\eta}{^b_k}\tensor{\eta}{^c_l}\right)\right. 
	\hspace{17ex}
	\nonumber \\
	-\left.\frac{1}{3}\epsilon_{abc}\epsilon^{mkl}
	\partial_{t}\left(\frac{1}{N}\left(\partial_{t}\tensor{\eta}{^a_m}\right)\tensor{\eta}{^b_k}\tensor{\eta}{^c_l}\right)\tensor{\delta}{^i_j}\right] = \tensor{\xi}{^i_j}-\tensor{\tau}{^i_j} \; ,
\label{LES2_tracefree}	
\end{gather}
where $\tensor{\tau}{^i_j} := \tensor{\mathcal{R}}{^i_j}-\frac{1}{3}\mathcal{R}\tensor{\delta}{^i_j}$
are the coefficients of the traceless part of the spatial Ricci tensor, and
$\tensor{\xi}{^i_j} :=\tensor{\mathcal{A}}{^i_j}-\frac{1}{3}\mathcal{A}\,\tensor{\delta}{^i_j}$.

The Lagrange--Einstein system, Eqs.~(\ref{LES2_symmetry})--(\ref{LES2_EoS}), is closed and provides the components $\tensor{\eta}{^a_i}$ of coframes,
from which one can calculate the evolution of the perturbations. The system comprises $14$ equations, where $9$ equations describe the evolution
for the coefficient functions of $3$ spatial Cartan coframe fields, and the remaining $5$ equations originate from the $4$ constraints and the EoS defining the properties of the fluid.

\subsection{Perturbation ansatz}
\subsubsection{Background}
We will choose
a spatially flat, homogeneous and isotropic model universe as the background spacetime, with the same barotropic EoS, and including a possible constant curvature term into the first--order perturbations, (\textit{cf.,} e.g., \cite{rza3}). Accordingly, the spatial metric coefficients of the background will be $a^2(t)\delta_{ij}$, $a(t)$ being the background scale factor. We prescribe a homogeneous lapse $N_H(t)$ for this homogeneous and isotropic background, by setting its relation to the background energy density $\epsilon_{H}$,
formal rest mass density $\varrho_{H} := F(\epsilon_H)$ and pressure $p_{H} = \beta(\epsilon_H)$ as being the same relations as those for the inhomogeneous quantities,
\begin{equation}
 N_H = \frac{\varrho_H}{\epsilon_H + p_H} = \frac{F(\epsilon_H)}{\epsilon_H + \beta(\epsilon_H)} \; .
 \label{background_lapse_def}
\end{equation}
We may then write the background line element as
\begin{equation}
\rd s_H^{2}=-N_H^{2}(t)\rd t^{2}+a^{2}(t)\,\tensor{\delta}{_i_j} \rd X^{i}\rd X^{j}\; .
\label{metricb1}
\end{equation}
Note that the evolution of the background lapse function
$N_H(t)$ will be given by its definition \eqref{background_lapse_def} and the EoS, making it time--dependent for $p_{H} \neq 0$.
One should keep in mind that our choice of time coordinate $t$ will consequently not coincide in general with the usual `cosmic time' coordinate for the background, and will evolve at a different rate.
The usual cosmic time $\tilde t$ would rather be defined by $\rd \tilde{t}=N_H (t)\rd t$, so that the background line element (\ref{metricb1})
would take the usual Friedmannian form for homogeneous and isotropic model universes:\footnote{%
The notation $a\!\left[\tilde t\right]$ signifies that the scale factor still takes the same values, $a[\tilde t] := a(t)$, but has a different functional dependence on the alternative time coordinate.
}
\begin{equation}
\rd s_H^{2}=-\rd \tilde{t}^{2}+ a^{2}\!\left[\tilde t\right] \tensor{\delta}{_i_j}\,\rd X^{i}\rd X^{j} \; .
\label{metricb2}
\end{equation}
With this time variable, the standard Friedmann equations would indeed be recovered:
\begin{align}
& 3 \, \frac{\partial_{\tilde{t}}^{2}a}{a} = {} -4\pi G(\epsilon_{H}+3p_{H}) \, + \, \Lambda \; ; \nonumber \\
& 3 \left(\frac{\partial_{\tilde{t}} a}{a}\right)^{2} = {} 8\pi G \epsilon_{H} \, + \, \Lambda \; ; \nonumber \\
& \partial_{\tilde t} \epsilon_H + {} 3 \, \frac{\partial_{\tilde t} a}{a} (\epsilon_H + p_H) = 0 \; .
\end{align}
However, for consistency with the lapsed foliation used for the full inhomogeneous spacetime, in what follows we include the homogeneous lapse $N_H$ into the background and use the coordinate $t$. In terms of this variable, the acceleration and Friedmann equations are respectively:
\begin{align}
&\frac{3}{N_H^2}\frac{\partial_{t}^{2}a}{a}= -4\pi G(\epsilon_{H}+3p_{H}) \, + \, \Lambda + 3 \frac{\partial_{t}a}{a}\frac{\partial_{t}N_H}{N_H^3} \; ; \nonumber \\
&\frac{3}{N_H^{2}}\left(\frac{\partial_{t} a}{a}\right)^{2} \! = \, 8 \pi G \epsilon_{H} \, + \, \Lambda \; ,
\label{backgroundeqn1}
\end{align}
while the energy--momentum conservation equation is formally unchanged:
\begin{equation}
 \partial_t \epsilon_H + 3 \, \frac{\partial_t a}{a} (\epsilon_H + p_H) = 0 \; .
\end{equation}

\subsubsection{Coframes decomposition}
It is important to express the full set of equations in terms of a single perturbation variable,
the coframes, so that the Lagrangian perturbation approach is well--defined. Although this is not made fully explicit in the Lagrange--Einstein system \eqref{LES2_symmetry}--\eqref{LES2_EoS}, it is implicitly the case as the Ricci tensor and covariant derivatives are functionals of the metric, and hence of the coframes, and $\epsilon$, $p$, $N$ and $\tensor{\mathcal{A}}{^i_j}$ can be expressed in terms of the coframes and initial energy density data. The latter relations are obtained \textit{via} the conservation equation \eqref{restmass_conservation} for $\varrho = F(\epsilon)$ and the evolution equation for
$J := \mathcal{J}/N = \det(\tensor{\eta}{^a_i})$ from the first equation in \eqref{newevolutionexpan2}:
\begin{equation}
 N \Theta = - \frac{\partial_t F(\epsilon)}{F(\epsilon)} = \frac{\partial_t J}{J} \;\; ;\;\; \epsilon = F^{-1} \left(\frac{F(\init \epsilon)}{J} \right) \; ,
 \label{epsilon_of_J}
\end{equation}
where for any quantity $A$,  $A_{\mathbf i}$ denotes the quantity at initial time $t_{\mathbf i}$. Here $J_{\mathbf i}=1$ as a result of the choice of initial conditions for the coframes.
The barotropic EoS and choice of $N$ then allow us to determine $p$, $N$ and $\tensor{\mathcal{A}}{^i_j} = N^{-1}\tensor{N}{^|^|^i_|_|_j}$, and to express these fields as functions of $J = \det(\tensor{\eta}{^a_i})$.

We then follow the previous papers \cite{rza1}--\cite{rza4} and decompose the coframes into a FLRW coframe set and deviations thereof,
\begin{equation}
\bm{\eta}^a =\tensor{\eta}{^a_i}\rd X^{i} = a(t) \left( \tensor{\delta}{^a_i} + \tensor{P}{^a_i} \right) {\rm d}X^i \; .
 \label{coframe}
\end{equation}
At this nonperturbative level, the metric coefficients are then related to the deformation field by
\begin{equation}
g_{ij}=a^{2}(t)\left(G_{ij} +2 P_{(ij)} + G_{ab} \tensor{P}{^a_i} \tensor{P}{^b_j} \right)\;,
\label{FLRWfirstorder1}
\end{equation}
where we have defined
\begin{equation}
\tensor{P}{^i_j} := \tensor{\delta}{_a^i}\tensor{P}{^a_j}\; ; \; P := \tensor{P}{^k_k}=\tensor{\delta}{_a^k}\tensor{P}{^a_k} \; ;
	\; P_{ij} := G_{ai}\tensor{P}{^a_j} \; .
\label{definitiontrace1}
\end{equation}
Recall that the Gram matrix coefficients $G_{ab}$ have been defined to encode the initial metric inhomogeneities, so that the coefficients $\tensor{P}{^a_i}$ can be set to zero initially. Also recall that this coframe split is made with respect to a FLRW background with a nontrivial lapse included, and that the functional dependence of $a$, or of the deformation field, on the time coordinate $t$ will be affected accordingly.

We then expand the deformation fields $\tensor{P}{^a_i}$ into a perturbative sum,
so that the coframes are given by:
\begin{equation}
\boldsymbol{\eta}^{a}=a(t)\left(\tensor{\delta}{^a_i}+\sum^{\infty}_{m=1}\tensor{P}{^a_i^(^m^)}\right)\rd X^{i},
\end{equation}
where the m{\it th}--order deformation field coefficients $\tensor{P}{^a_i^(^m^)}$ are of order $\varepsilon^m$ for some bookkeeping parameter $\varepsilon \ll 1$. In this paper we will only focus on first--order deformations.

\subsubsection{Initial conditions}
We will follow the steps of Refs.~\cite{rza3,rza4} to prescribe the initial data.
The deformation field and its time--derivatives are given at some initial time $\init t$ by:
\begin{equation}
\begin{aligned}
\tensor{P}{^a_i}(\init{t})&=0 \; ; \\
\left(\partial_{t}\tensor{P}{^a_i} \right)(\init t) &=: \tensor{U}{^a_i} \; ; \\
\left(\partial_{t}^{2}\tensor{P}{^a_i} \right)(\init t) &=: \tensor{W}{^a_i}-2\init{H}\tensor{U}{^a_i} \; ,
\label{initialdeformation1}
\end{aligned}
\end{equation}
where $H :=\partial_{t}a/a$ is the Hubble function. Hereafter, we will normalize the scale factor as $\init{a}=1$. 
The six $1-$form fields $\boldsymbol{U}^{a}=\tensor{U}{^a_i}\rd X^{i}$ and $\boldsymbol{W}^{a}=\tensor{W}{^a_i}\rd X^{i}$
are $1-$form generalizations of the initial Newtonian peculiar--velocity and peculiar--acceleration gradient fields, respectively.

The Lagrange--Einstein system with its split into trace and traceless parts according to \eqref{LES2_symmetry}--\eqref{LES2_tracefree} then translates into constraints on the initial data:
\begin{align}
\label{init_cond_symmetry}
& U_{[i j]} = 0 \quad ; \quad W_{[i j]} = 0 \; ; \\
& W - U \left( \frac{\partial_t N}{N} \right)_{\!\mathbf i} = 3 \init H \left[ \left(\frac{\partial_t N}{N} \right)_{\!\mathbf i} - \left( \frac{\partial_t N_H}{N_H} \right)_{\!\mathbf i \;} \right] \nonumber \\
	& \qquad {} + \Lambda \left( \init N^2 - \init{N_H^2} \right) + \init{N}^2 \, \init{\mathcal{A}} \nonumber \\
\label{init_cond_raychaudhuri}
	& \qquad {} - 4 \pi G \left[ (\init \epsilon + 3 \init p) \init N^2 - (\init{\epsilon_H} + 3 \, \init{p_H}) \init{N^2_H} \right] \; ; \\
& \tensor[^{\rm{tl}}]{W}{^a_j} \tensor{\delta}{_a^i} + \left(\init H - \left(\frac{\partial_t N}{N} \right)_{\!\mathbf i} \right) \tensor[^{\rm{tl}}]{U}{^a_j} \tensor{\delta}{_a^i} \nonumber \\
& {\;} + U \,\tensor[^{\rm{tl}}]{U}{^a_j} \tensor{\delta}{_a^i} - \left(\tensor{U}{^a_k} \tensor{\delta}{_a^i} \tensor{U}{^b_j} \tensor{\delta}{_b^k} - \frac{1}{3} \tensor{U}{^a_l} \tensor{\delta}{_a^k}\tensor{U}{^b_k} \tensor{\delta}{_b^l} \tensor{\delta}{^i_j} \right) \nonumber \\
\label{init_cond_traceless}
& {\qquad} = N^2_{\mathbf i} \big(\tensor{\xi}{^i_j}(\init t) - \tensor{\tau}{^i_j}(\init t) \big) \; ; 
\end{align}
\begin{align}
& U^{2}-\tensor{U}{^a_i}\tensor{\delta}{_a^j}\tensor{U}{^b_j}\tensor{\delta}{_b^i} +4\init{H}U \nonumber \\
\label{init_cond_hamilton}
	& = 16 \pi G \! \left(\init \epsilon \init N^2 \! - \init{\epsilon_H} \init{N^2_H} \right) + 2 \Lambda \! \left(\init N^2 \! - \init{N_H^2} \right) - \init{\mathcal{R}} \init N^2 \; ; \\
\label{init_cond_momentum}
& \left(\init{N}^{-1} \tensor{U}{^a_j}\tensor{\delta}{_a^i}\right)_{\parallel i} =
	\left(\init{N}^{-1} U\right)_{\mid j}+2\init{H} \left(\init{N}^{-1}\right)_{\mid j} \; ; \\
\label{init_cond_EoS}
	& \init p = \beta(\init \epsilon) \quad ; \quad \init{p_H} = \beta(\init{\epsilon_H}) \; .
\end{align}
The abbreviations $U:=\tensor{U}{^a_k}\tensor{\delta}{_a^k}$, $W:=\tensor{W}{^a_k}\tensor{\delta}{_a^k}$, and $\tensor[^{\rm{tl}}]{W}{^a_i} := \tensor{W}{^a_i} - (1/3) W \tensor{\delta}{^a_i}$, $\tensor[^{\rm{tl}}]{U}{^a_i} := \tensor{U}{^a_i} - (1/3) U \tensor{\delta}{^a_i}$, are used for the trace and traceless parts, respectively.

\subsection{First--order Lagrange--Einstein system}
\label{LES1}

We now expand the above Lagrange--Einstein system and its initial conditions to first order\footnote{Note that initial data can be assumed, without loss of generality, to be first order.} in the only dynamical variable in this Lagrangian perturbation approach, namely the deformation field $\tensor{P}{^a_i}$. In what follows we omit the index ${}^{(1)}$ for the first--order deformation field and the associated initial conditions $U_{i j}$, $W_{i j}$, but keep the index for the other variables, as functionals of $\tensor{P}{^a_i}$.
We first need to express these functionals explicitly at first order.

\subsubsection{Dependent variables at first order}
\label{generalEOS}

In order to express the first--order Ricci tensor and scalar curvature in terms of the coframes, we expand the initial metric coefficients to first order as $G_{i j}(\bm X) = \delta_{i j} + G^{(1)}_{i j}(\bm X)$ since they reduce to $\delta_{i j}$ at the unperturbed zero--order level. Introducing the first--order quantities
$G^{(1) i j} := \tensor{\delta}{^i^k} \tensor{\delta}{^j^l} G^{(1)}_{kl}$, $P^{i j} := \tensor{\delta}{^i^k} \tensor{\delta}{^j^l} P_{k l}$ for the inverse metric, we can then substitute the metric and its inverse, truncated to first order,
\begin{align}
g_{ij}&=a^{2}\left(\delta_{ij}+G^{(1)}_{ij}+2P_{(ij)}\right) \; ;
\label{metricfirstorder1} \\
g^{ij}&=a^{-2}\left(\delta^{ij}-G^{(1)ij}-2P^{(ij)}\right) \; ,
\label{metricfirstorder2}
\end{align}
into the definitions of the spatial Christoffel symbols and of the spatial Ricci tensor. We thereby obtain:
\begin{align}
\tensor{\Gamma}{^k_i_j^(^1^)}&=\frac{1}{2}\delta^{kl}\left(G^{(1)}_{il\mid j}+G^{(1)}_{lj\mid i}-G^{(1)}_{ij\mid l}\right)\\ \nonumber
	&+\delta^{kl}\left(P_{(il)\mid j}+P_{(lj)\mid i}-P_{(ij)\mid l}\right);\\
\label{initialriccitens1}
\mathcal{R}^{(1)}_{ij}&= \mathscr{R}_{i j} + {} \tensor{P}{^k_[_j_\mid_k_]_\mid_i}+\tensor{P}{_[_j^k_\mid_k_]_\mid_i}
	+ \tensor{P}{_(_i_k_)_\mid_j^\mid^k}-\tensor{P}{_(_i_j_)^\mid^k_\mid_k};\\
\label{initialriccitrace1}
\mathcal{R}^{(1)}&= a^{-2}\mathscr{R} +2a^{-2}\left(\tensor{P}{^k^i_{\mid i}_{\mid k}}
	-\tensor{P}{^{\mid k}_{\mid k}}\right),
\end{align}
where $\mathscr{R}_{i j} := G^{(1) \mid k}_{i[k \mid j]} + G^{k (1)}_{\,\, [j \mid k] \mid i}$,
and $\mathscr{R} := \tensor{\delta}{^i^j} \tensor{\mathscr{R}}{_i_j} = 2 \, \tensor{G}{^l_[_k_{\mid l}_]^{\mid}^k}^{(1)}$
are the initial conditions for the curvature tensor coefficients and their trace, respectively.

An important difference from the dust case is that here the spatial Ricci scalar will in general not be constrained to evolve as $\mathscr{R}(\bm X) \,a(t)^{-2}$ at first order, due to the contributions from the lapse in the momentum constraints. As will be shown below, these contributions give rise to a nonzero evolution for the (initially vanishing) second term $(\tensor{P}{^k^i_{\mid i}_{\mid k}} -\tensor{P}{^{\mid k}_{\mid k}})$, or equivalently a nonconserved scalar curvature, $\partial_t \CR^{(1)} + 2 H \CR^{(1)} = a^{-2} \partial_t (a^2 \CR^{(1)}) \ne 0$, in contrast to the dust case.

Using the barotropic EoS and the corresponding solution \eqref{epsilon_of_J} to the energy conservation equation \eqref{conservationlaw1_PF}, we can also
expand $\epsilon$, $p$, $N$ and $\tensor{\mathcal{A}}{^i_j}$ in terms of the first--order deformation field.
We write $\init{\epsilon} := \init{\epsilon_H} (1 + \delta \init \epsilon)$ at first order, and expand $J^{-1} = a^{-3} \det(\tensor{\delta}{^a_i} + \tensor{P}{^a_i})^{-1}$ at the same order as $a^{-3} (1-P)$. The solution \eqref{epsilon_of_J} for $\epsilon$ as a function of $J$ can then be expanded to first order in the perturbation as
\begin{align}
 \epsilon &= F^{-1} \left( \frac{F(\init{\epsilon_H}) + F'(\init{\epsilon_H}) \, \init{\epsilon_H} \, \delta \init \epsilon - F(\init{\epsilon_H}) \, P}{a^3} \right) \nonumber \\
 & = F^{-1}\left(\frac{F(\init{\epsilon_H})}{a^3}\right) \nonumber \\
	& {\!\!} + \left[\frac{1}{a^3} \init{\epsilon_H} \, F'(\init{\epsilon_H}) \delta \init\epsilon - P \frac{F(\init{\epsilon_H})}{a^3} \right] \left(F^{-1} \right)' \!\left(\frac{F(\init{\epsilon_H})}{a^3}\right) \, .
\label{generalEOSepsilon_lin_1}
\end{align}
The energy--momentum conservation equation \eqref{epsilon_of_J} still holds for background quantities, giving
\begin{equation}
 F(\epsilon_H) = \frac{F(\init{\epsilon_H})}{a^3} \; .
\end{equation}
This can be substituted into \eqref{generalEOSepsilon_lin_1} to give
\begin{equation}
 \epsilon = \epsilon_H \left[ 1 + \frac{F(\epsilon_H)}{\epsilon_H \, F'(\epsilon_H)}
		\left( \frac{\init{\epsilon_H} \, F'(\init{\epsilon_H})}{F(\init{\epsilon_H})} \, \delta\init\epsilon - P \right) \right] \, .
\label{generalEOSepsilon_lin_2}
\end{equation}
The further use of the definition of $F$, Eq.~\eqref{def_varrho}, allows us to simplify the above to
\begin{equation}
 \epsilon = \epsilon_H \left[ 1 - \left(1+ \frac{p_H}{\epsilon_H} \right)\Ptil
\right] \, ,
\label{generalEOSepsilon_lin}
\end{equation}
which we have written for convenience in terms of a shifted deformation trace,
\begin{equation}
\Ptil:= P \, -\, \init{\alpha_H} \, \delta\init\epsilon \; ,
\label{Ptil}
\end{equation}
where $\init{\alpha_H} := \left(\init{\epsilon_H}+\beta(\init{\epsilon_H})\right)^{-1}\init{\epsilon_H}$ is a constant, and $\delta\init\epsilon$ is the initial energy perturbation.

The pressure can in turn be expanded to first order as $p = \beta(\epsilon)$, yielding
\begin{equation}
p = p_H - \beta'(\epsilon_H) \, \big(\epsilon_H + p_H \big) \Ptil \, .
\end{equation}
Note that the factor
$\beta'(\epsilon_H)$ corresponds to the (generally time--dependent) dimensionless ratio of the background speed of sound to speed of light squared, $\beta'(\epsilon_H) =: c_S^2(t)/c^2$, if $p_H$ is the thermodynamic equilibrium pressure for the background fluid.

We then expand the lapse $N = (\epsilon + p)^{-1} F(\epsilon)$ as
\begin{equation}
 N = N_H \left[ 1 + \beta'(\epsilon_H) \,\Ptil \right]
\end{equation}
at first order in the deformation field.
At this order, one will then have (with $\partial_t P = \partial_t \Ptil$):
\begin{equation}
\begin{aligned}
\frac{\partial_{t}N}{N}={}&\frac{\partial_{t}N_{H}}{N_{H}} +
\beta'\left(\epsilon_{H}\right)\partial_{t}\Ptil\\ & -3H
\left(\epsilon_{H}+\beta\left(\epsilon_{H}\right)\right)\beta''\left(\epsilon_{H}\right) \Ptil\; ,
\label{firstorderN1}
\end{aligned}
\end{equation}
with
\begin{equation}
\frac{\partial_{t}N_{H}}{N_{H}}=3H \beta'\left(\epsilon_{H}\right) \; .
\label{firstorderevolNH}
\end{equation}
This also allows one to obtain the first--order expression for $\tensor{\mathcal{A}}{^i_j} = N^{-1} \tensor{N}{^{||i}_{||j}}$:
\begin{equation}
\tensor{\mathcal{A}}{^i_j^{(1)}}= a^{-2} \beta'(\epsilon_H) \, \tensor{\delta}{^i^k}{\Ptil}_{|j|k} \; .
\label{4divergence11}
\end{equation}

\subsubsection{First--order system}

Using the above expansions, the Lagrange--Einstein system \eqref{LES2_symmetry}--\eqref{LES2_momentum} can be rewritten at first order in the deformation field as follows:
\begin{gather}
\partial_t \Pold_{[ij]} = 0 \; ; \label{LE1}\\
\begin{aligned}
\partial_t^2 \tensor{{\Pold}}{^i_j} &+ 3 H \big[1-\beta'(\epsilon_H)\big] \, \partial_t \tensor{{\Pold}}{^i_j} \\
	&{} + H \big[1- \beta'(\epsilon_H) - \mathcal{V}(t)\big] \partial_t \Ptil \, \tensor{\delta}{^i_j} \\
	 &{} = N_H^2 \tensor{\mathcal{A}}{^i_j^{(1)}} - N_H^2 \left(\tensor{\mathcal{R}}{^i_j^{(1)}} - \frac{\mathcal{V}(t)}{4} \mathcal{R}^{(1)} \, \tensor{\delta}{^i_j} \right) \; ;
\label{LES3_evol}
\end{aligned} \\
\partial_{t} \left(\tensor{{\Pold}}{^i_{j|i}}-\Ptil_{|j} \right) =
- 2 H \beta'\left(\epsilon_{H}\right) {\Ptil} _{|j} \; ,
\label{LES3_momentum}\\
H\,\partial_{t}\Ptil +
4 \pi G \left[\epsilon_H+p_H -(2 \epsilon_H + \tLam) \,\beta'(\epsilon_H) \right] N_H^2 \Ptil \nonumber\\
	= - \frac{1}{4}N_{H}^{2}\tensor{\mathcal{R}}{^(^1^)} \; ,
\label{HamiltonEOS}
\end{gather}
with $\partial_t \Pold=\partial_t\Ptil$, and where $\tensor{\mathcal{A}}{^i_j^{(1)}}$, $\tensor{\mathcal{R}}{^i_j^{(1)}} = a^{-2} \tensor{\delta}{^i^k} \mathcal{R}^{(1)}_{kj}$ and $\mathcal{R}^{(1)}$ are expressed as functions of $\tensor{{\Pold}}{^a_i}$
according to the formulas given above,
$\tilde\Lambda:=\Lambda/(4\pi G)$, and we introduce the abbreviation
\begin{align}
&\mathcal{V}(t) := \Bigl[\epsilon_H + p_H - \bigl(2\epsilon_H + \tilde\Lambda\bigr) \beta'(\epsilon_H)\Bigr]^{-1}\nonumber\\
&\times \ \left\{\epsilon_H + p_H- \bigl(3 \epsilon_H - p_H + 2\tilde\Lambda\bigr)\beta'(\epsilon_H)\right.\nonumber\\
&
\left. + \bigl(2\epsilon_H + \tilde\Lambda \bigr) \bigl(\epsilon_H + p_H\bigr) \beta''(\epsilon_H)\right\} \; .
\label{Vt}
\end{align}
Equation \eqref{LES3_evol} has been obtained from the first--order expansion of the extrinsic curvature evolution equation
\eqref{LES2_evol} by combining it with the first--order energy constraint \eqref{HamiltonEOS}.
The EoS \eqref{LES2_EoS} has already been used to expand $\epsilon$, $p$ and $N$ in terms of the first--order deformation field.

\subsection{First--order master equations}
\label{mastereqs}

Following the approach of Ref.~\cite{rza4} the above system can be reexpressed by decomposing the deformation fields into trace, trace--free symmetric and antisymmetric parts:
\begin{equation}
\tensor{P}{^i_j}=\frac{1}{3}P\tensor{\delta}{^i_j}+\tensor{\Pi}{^i_j}+\tensor{\mathfrak{P}}{^i_j}\;,
\label{newrepresentation1}
\end{equation}
where $\tensor{\Pi}{_i_j}=P_{(ij)}-\frac{1}{3}P\delta_{ij}$ and $\tensor{\mathfrak{P}}{_i_j}=P_{[ij]}$.

We will now derive the governing equations for these parts, named {\it master equations}. For the trace part
we use the new variable $\Ptil$ from Eq.~\eqref{Ptil}.
Accordingly, (\ref{LE1})--(\ref{LES3_evol}) become:
\begin{gather}
\label{antisymmetricpart1}
\partial_{t}\mathfrak{P}_{ij}=0 \; : \quad \mathfrak{P}_{ij} = \mathfrak{P}_{ij}(\init t) = 0 \; ; \\
\partial^2_t \Ptil
+3 H \big[2 - 2 \beta'(\epsilon_H) - \mathcal{V}(t)\big] \partial_t \Ptil
\hspace{18ex} \nonumber \\
\label{traceevol1}
	\hspace{12ex} {} = N_H^2 \mathcal{A}^{(1)} - N_H^2 \left(1 - \frac{3}{4} \mathcal{V}(t) \right) \mathcal{R}^{(1)} \;; \\
\label{tracefreeEOS2}
\partial_{t}^{2}\tensor{\Pi}{^i_j}+3H[1-\beta '(\epsilon_H)]\partial_{t}\tensor{\Pi}{^i_j} =
	N_H^{2}\left(\tensor{\xi}{^i_j^(^1^)}-\tensor{\tau}{^i_j^(^1^)}\right) \; ; \\
\label{momentumconstraintEOS}
\!\! \partial_{t} \left(\tensor{\Pi}{^i_{j|i}}- \frac{2}{3} \Ptil_{|j}
\right) = - 2 H \beta'\left(\epsilon_{H}\right) \Ptil_{|j} \; .
\end{gather}
Once again the first--order quantities $\mathcal{A}^{(1)}$, $\tensor{\xi}{^i_j^(^1^)}$, $\mathcal{R}^{(1)}$ and $\tensor{\tau}{^i_j^(^1^)}$ are used as shorthand notations but are meant to be expressed in terms of the deformation field. These expressions are obtained from the results above, Eqs.~\eqref{initialriccitens1}, \eqref{initialriccitrace1}, \eqref{4divergence11}, as follows:\footnote{%
The expression given for $\tensor{\tau}{^i_j^(^1^)}$ makes use of the momentum constraints \eqref{momentumconstraintEOS}, which imply, through their spatial derivative, $\partial_t \tensor{\Pi}{_k_[_i_|_j_]^|^k} = 0$, and thus $\tensor{\Pi}{_k_[_i_|_j_]^|^k} = \tensor{\Pi}{_k_[_i_|_j_]^|^k}(\init t) = 0$.
Also note that since $\Pold$ and $\Ptil$ differ by an initial spatial function, we can express (\ref{4divergencetrace})--(\ref{curvaturetracefree}) in terms of either variable. Here we have adopted the most compact possibility, noting that the initial value of $\Ptil$ is nonzero, whereas (\ref{curvaturetrace2}) and (\ref{curvaturetracefree}) involve the initial curvature which is independent of the initial perturbation field.\label{footnote_tau_ij}}
\begin{gather}
\label{4divergencetrace}
a^2 \tensor{\mathcal{A}}{^(^1^)} = \beta'(\epsilon_H) \, \Delta_0 \Ptil
\; ; \\
\label{4divergencetracefree}
a^2 \tensor{\xi}{^i_j^(^1^)} = \beta'(\epsilon_H) \left(
\Ptil^{|i}_{\ |j} - \frac{\tensor{\delta}{^i_j}}{3} \Delta_0 \Ptil
\right)\! ; 
\end{gather}
\begin{gather}
\label{curvaturetrace2}
 a^2 \mathcal{R}^{(1)} = \mathscr{R} + 2 \left( \tensor{\Pi}{^k^i_|_k_|_i} - \frac{2}{3} \tensor{P}{^|^k_|_k} \right) \; ; \\
 a^2 \tensor{\tau}{^i_j^(^1^)} = \tensor{\mathscr{T}}{^i_j} + 2 \, \tensor{\Pi}{^i_k_|_j^|^k} - \tensor{\Pi}{^i_j^|^k_|_k} \hspace{20ex} \nonumber \\
\label{curvaturetracefree}
 \hspace{8ex} {} - \frac{1}{3} \left(2 \, \tensor{\Pi}{^k_l_|_k^|^l} \, \tensor{\delta}{^i_j} + \tensor{P}{^|^i_|_j} - \frac{1}{3} \Delta_0 P \, \tensor{\delta}{^i_j} \right) \; ,
\end{gather}
with $\tensor{\mathscr{T}}{^i_j} := \tensor{\mathscr{R}}{^i_j} - \frac{1}{3} \mathscr{R} \tensor{\delta}{^i_j} = \tensor{\tau}{^i_j^(^1^)} (\init t)$, and with $\Delta_0$ the coordinate Laplacian operator in the Lagrangian coordinates $\{X^i \}$, $\Delta_0 := \tensor{\delta}{^i^j} \partial_i \partial_j$.

\subsubsection{Master equation for the trace}

Contracting the momentum constraints \eqref{momentumconstraintEOS} with a spatial derivative ${}_{|j}$ yields the
first--order evolution equation for the nontrivial part of the scalar curvature:
\begin{align}
\partial_t \left(\tensor{{\Pold}}{^k^i_|_k_|_i} - \tensor{{\Pold}}{^|^k_|_k} \right) &
= \partial_t \left(\tensor{\Pi}{^k^i_|_k_|_i} - \frac{2}{3} \tensor{{\Ptil}}{^|^k_|_k} \right) \nonumber
\\ & = - 2 H \beta'(\epsilon_H) \, \Delta_0 \Ptil \; .
\end{align}
From the respective expressions \eqref{initialriccitrace1}, \eqref{4divergencetrace} for $\mathcal{R}^{(1)}$ and $\mathcal{A}^{(1)}$, this amounts to the following evolution for $\mathcal{R}^{(1)}$:
\begin{align}
\label{evofirstRicci1}
\partial_{t}\tensor{\mathcal{R}}{^(^1^)}+2 H \tensor{\mathcal{R}}{^(^1^)}& {} =
-4 H a^{-2} \beta'(\epsilon_H) \Delta_0 \Ptil
 \nonumber \\
& {} = -4 H \tensor{\mathcal{A}}{^(^1^)} \; ,
\end{align}
which unlike the case of dust does remain coupled to the dynamics of the inhomogeneous perturbation.

Combining this evolution equation with the linearized energy constraint \eqref{HamiltonEOS} and its time--derivative one then obtains the {\it master equation for the evolution of the trace} (\ref{Ptil}) of the first--order deformation field:\footnote{%
This equation can also be derived by combining the energy constraint \eqref{HamiltonEOS} with the trace \eqref{traceevol1} of the evolution equation to eliminate $\mathcal{R}^{(1)}$, or equivalently by directly expanding the Raychaudhuri equation \eqref{LES2_raychaudhuri} to first--order. In both cases, the master equation for the trace would then be recovered after replacing the first--order acceleration divergence $\mathcal{A}^{(1)}$ by its explicit expansion \eqref{4divergencetrace}.%
}
\begin{align}
\label{mastereqtrace}
 & \partial_{t}^{2}\Ptil+2H (1-3\beta' \!\left(\epsilon_H\right)) \, \partial_{t}\Ptil
- \mathcal{W}(t) N_H^{2} \Ptil
\nonumber \\
& \qquad = a^{-2}N_H^{2}\,\beta'\left(\epsilon_{H}\right) \Delta_{0} \Ptil ,
\end{align}
where $p_H = \beta(\epsilon_H)$ and $N_H = F(\epsilon_H)/(\epsilon_H+p_H)$
still, and
\begin{align}
\mathcal{W}(t) & := 4\pi G\big[\epsilon_H + p_H - (2 \epsilon_H + \tLam) \beta'(\epsilon_H) \big]
\big[4-3 \mathcal{V}(t) \big] \nonumber \\
& {} = 4 \pi G \left[ \epsilon_H+p_H + \big(\epsilon_H-3p_H + 2 \tLam \big)\beta '(\epsilon_H) \right]
\nonumber \\ & \qquad - 12\pi G \, \bigl(2\epsilon_H + \tLam \bigr)
\big(\epsilon_H+p_H \big) \beta ''(\epsilon_H) \; .\label{Wdef}
\end{align}
To avoid potential confusion, since the time coordinate $t$ used in this paper has a different rate as compared to the conventional cosmic time, it will sometimes be convenient for further applications to use the (time--coordinate--independent) background scale factor $a$ as the time variable instead. With this change of parametrization, the energy constraint~(\ref{HamiltonEOS}) and the master equation for the trace~(\ref{mastereqtrace}) may be rewritten as follows:
\begin{align}
&a\frac{\partial \Ptil}{\partial a}+{\alpha_{0}}\Ptil=-\frac{N_H^2}{4 \,H^2}\mathcal{R}^{(1)}\; ;
\label{firstorderhamilton-a}\\
&\frac{\partial^{2}\Ptil}{\partial a^{2}}+\frac{\alpha_{1}}{a}\frac{\partial \Ptil}{\partial a}-\frac{\alpha_{2}}{a^{2}}\Ptil=\frac{\alpha_{3}}{a^{4}}\Delta_{0}\Ptil,
\label{master-trace-a}
\end{align}
respectively, with time--dependent coefficients,
\begin{equation}
\begin{aligned}
\alpha_{0}&:=4\pi G \frac{N_H^2}{H^2} \left[\epsilon_{H}+p_{H}-\left(2\epsilon_{H}+\tilde{\Lambda}\right)\,\beta'\left(\epsilon_{H}\right)\right]\; ;\\
\alpha_{1}&:=\alpha_{0}+4\pi G \frac{N_H^2}{H^2}\left[\tilde{\Lambda}-2p_{H}\right]\; ;\\
\alpha_{2}&:=N_H^2 \mathcal{W}(t)/H^2\quad ; \quad \alpha_{3}:=N_H^2 \beta'\left(\epsilon_{H}\right)/H^2\; ,
\end{aligned}\label{poly-coeffs}
\end{equation}
where we recall that from the background Friedmann equation we have $H^2/N_H^2 = 4\pi G\, (2\epsilon_H + \tilde\Lambda)/3$.

From Eq.~(\ref{master-trace-a})
we can introduce a time--dependent background Jeans wave number $k_J(\epsilon_H)$ by\footnote{We include the factor $c$ explicitly so that the dimensional
content of this relation is clear. The right hand side of (\ref{Wdef}) must be divided by $c^2$ if units $c\ne1$ are restored.}
\begin{equation}
\label{jeans}
 k_J(\epsilon_H):= \frac{1}{c}\sqrt{\frac{\alpha_{2}}{\alpha_{3}}}=
\frac{1}{c}\sqrt{\frac{\mathcal{W}(t)}{\beta'(\epsilon_H)}}\; ,
\end{equation}
provided that the term in the square root is positive. Pressure should be positive for sound waves to resist gravitational collapse, and the existence of the Jeans length is intimately related to the energy conditions satisfied by the matter field. 

A remark is in order here.
In general, one would expect the evolution of the inhomogeneous deformation to be affected by the local, inhomogeneous speed of sound and density, so that a nonperturbative Lagrangian realization would rather feature a local Jeans wave number $k_J(\epsilon)$ \cite{buchert06}. The dynamics in the presence of a significant density contrast will thus only be partially captured by the above first--order equation, where $\epsilon$ has been expanded in $\tensor{P}{^a_i}$ and, accordingly, only zero--order background factors such as $k_J(\epsilon_H)$ survive in front of the first--order $\Ptil$.

As in the dust case, the advantages of the Lagrangian approach are only fully realized \textit{via} nonlinear extrapolation, e.g., by computing the energy density as a full nonlinear functional from the first--order deformation. This is part of the Relativistic Zel'dovich Approximation scheme, as defined for dust fluids in \cite{rza1}. As in the dust case and in contrast to standard Eulerian linear perturbation schemes, applying this procedure to compute the energy density out of the solution to first--order equations such as \eqref{mastereqtrace}, will already capture part of the nonlinear features. This is due to the nonlinear extrapolation and to the use of Lagrangian spatial coordinates which follow the fluid propagation in an exact manner. Further nonlinear effects of inhomogeneous pressure will, however, still be missed due to the absence of local Jeans length contributions in the equation used for $\Ptil$, compared to what should appear in the nonperturbative evolution equation. 

We will not go beyond this procedure in the present paper. Let us nonetheless suggest here a possible direction for improvement. It would require properly defining the local Jeans length in the relativistic context as a functional of the deformation. This quantity would then replace the background Jeans length in the trace master equation. The corresponding nonlinear master equation could then be solved in an iterative manner, by computing at each step the local Jeans length \textit{via} functional extrapolation out of the previous estimate for the deformation field. Note that each step would also involve a search for the traceless part of the deformation, as all of its components would be required for the extrapolation.

The evolution equation (\ref{mastereqtrace}) may be rewritten in an alternative form \textit{via} a time--dependent rescaling of the variable $\Ptil \mapsto \Ptil/N_H(t)$. Using the variation rate \eqref{firstorderevolNH} of the background lapse one finds the more transparent form:
\begin{align}
 \partial_t^2 \left( \frac{\Ptil}{N_H} \right) &{}+ 2 H \partial_t \left( \frac{\Ptil}{N_H} \right) - 4 \pi G (\epsilon_H + p_H) N_H^2 \left( \frac{\Ptil}{N_H} \right) \nonumber \\
 &{}= a^{-2} N_H^2 \, \beta'(\epsilon_H) \, \Delta_0 \!\left( \frac{\Ptil}{N_H} \right) \; . \label{altjeans}
\end{align}
\paragraph*{$\bullet$ Dust limit:}
Setting $p_H = \beta(\epsilon_H) = 0$, we find 
$\mathcal{W}(t) = 4 \pi G \epsilon_H = 4 \pi G \varrho_H = 4 \pi G \init{\varrho_H} a^{-3}$ and $N_H(t) = (\epsilon_H + p_H)^{-1} \varrho_H = 1$, and consequently 
both $t$--variable forms of the trace master equation, Eqs.~\eqref{mastereqtrace} and \eqref{altjeans}, reduce to the dust deformation trace evolution equation of \cite{rza1}--\cite{rza4}. The trace master equation becomes:
\begin{equation}
 \partial_t^2 P + 2 H \partial_t P - 4 \pi G \init{\varrho_H} a^{-3} P = - 4 \pi G \init{\varrho_H} a^{-3} \delta\init\epsilon \; .
 \label{dustlimittrace}
\end{equation}
With $N_H = 1$ the time variable used is the standard FLRW time coordinate $\tilde t =t$, so that the above
time--derivatives coincide with those used in \cite{rza1}--\cite{rza4} (denoted there by overdots).
Finally, as evaluating Eq.~\eqref{dustlimittrace} at the initial time gives $W = - 4 \pi G \init{\varrho_H} \delta\init\epsilon$, its right hand side can always be rewritten as $W a^{-3}$, and the dust--case master equation for the trace
(e.g., Eq.~(41) of \cite{rza4}) is thus recovered.

\medskip

\paragraph*{$\bullet$ Newtonian limit:}
The Newtonian limit is obtained by the joint application of the \textit{Minkowski Restriction} (MR) for the deformation field, as introduced for dust in \cite{rza1,rza2}, and of the $c \rightarrow \infty$ limit together with the assumption of a nonrelativistic pressure.

The latter two assumptions imply that the pressure is no longer
a source of the gravitational field, as the energy density is then $\epsilon \simeq \varrho c^2 \gg p$ (where the constant 
$c$ has been temporarily restored), so that all source terms reduce to the contribution of $\varrho$. Note that $\varrho$ can be considered as equal to the actual rest mass density in this limit. A further consequence of this is that the lapse becomes trivial, $N = \varrho c^2 / (\epsilon + p) \simeq 1$, consistent with its spatial variation rate, $N^{-1} N_{|i} = - (\epsilon + p)^{-1} \,p_{|i} \simeq -(\varrho c^2)^{-1} \,p_{|i} \rightarrow 0$ when $c \rightarrow \infty$, for any pressure spatial gradient. It is also the case for the (already homogeneous, but generally time--dependent) background lapse that $N_H \simeq 1$. Consequently, the fluid--orthogonal hypersurface time label $t$ now coincides with the fluid's proper time $\tau$ (since $1 \simeq N=\partial_t \tau$) as well as with the standard background cosmic (proper) time $\tilde t$. All these notions thus consistently define a time reference that can be used as the Newtonian absolute time. We will denote the corresponding Lagrangian time--derivative operator by an overdot.

With $N=1$ the line element \eqref{metric1} reduces to the one used in \cite{rza1,rza2} for irrotational dust, and one can thus directly use the corresponding definition of the MR in this context.\footnote{%
Note that the Minkowski Restriction introduced for the dust case is in principle independent of a possible $c \rightarrow \infty$ limit and can still otherwise be applied in a Minkowskian regime, as the name suggests. In the present case, when $c$ is still finite, this procedure would need to be extended to the presence of pressure and consequently of an inhomogeneous lapse. We believe, however, that such an extension to this case would require a modification of the perturbation framework used so far in this paper, through the use of a spacetime foliation better adapted to this purpose, and we will consequently not attempt to provide such a generalization here.
}
This restriction amounts to assuming that the initial metric is Euclidean and that the spatial coframes are exact in the three--dimensional hypersurfaces,  \textit{i.e.}, that there exist spatial coordinates $x^a = f^a(X^i,t)$ such that $G_{ab} = \delta_{ab}$ and
\begin{equation}
\tensor{\eta}{^a_i}=a(t) \left(\tensor{\delta}{^a_i}+\tensor{P}{^a_i}\right) = \tensor{f}{^a_|_i} \; .
\label{MRdef}
\end{equation}
In any $t=const$ hypersurface, the spatial line element then reads $\rd s^2 = \delta_{a b} \,\rd x^a \rd x^b$. The coordinates $x^a$ thus define Cartesian--type Eulerian coordinates in which the metric coefficients are manifestly Euclidean at each time, and they can be used to define a Newtonian spatial reference frame. Through its second equality, Eq.~\eqref{MRdef} also implies that the deformation 1--forms $P^a$ are also exact and accordingly define a deformation vector $\mathbf{P}$, with components $P^a$,
\begin{equation}
\mathbf{x}=a(t)\big[\mathbf{X}+\mathbf{P} \left(\mathbf{X}, t\right)\big] \quad , \quad \tensor{P}{^a_i}=: \tensor{P}{^a_|_i}\; .
\end{equation}
With these two assumptions the master equation \eqref{altjeans} on the trace $P = \tensor{\delta}{_a^i} \tensor{P}{^a_i}$ becomes an equation on the Lagrangian divergence $\boldsymbol{\nabla}_0 \cdot \mathbf{P} := \tensor{\delta}{_a^i} \tensor{P}{^a_|_i}$ of $\mathbf{P}$:
\begin{align}
\boldsymbol{\nabla}_0\cdot \ddot{\mathbf{P}}+2H \,\boldsymbol{\nabla}_0\cdot\dot{\mathbf{P}}-4 \pi G \varrho_{H}\big(\boldsymbol{\nabla}_0\cdot \mathbf{P}-\delta\init\varrho \big) \qquad \nonumber\\
 = a^{-2}\, \frac{\mathrm{d}p_H}{\mathrm{d}\varrho_H} \, \Delta_{0}\big(\boldsymbol{\nabla}_0\cdot\mathbf{P}-\delta\init\varrho \big) \; ,
\label{NLtauequation}
\end{align}
with $\varrho_H = a^{-3} \init{\varrho_H}$ still, and $\init\varrho =: \init{\varrho_H} (1+\delta\init\varrho)$. Note that, although the pressure itself no longer contributes as a source of gravitation, its spatial gradient still produces an acceleration (as obviously expected in a Newtonian framework), which is why it still affects the dynamics of $\boldsymbol{\nabla}_0 \cdot \mathbf{P}$ above through the sound speed squared factor $\mathrm{d}p_H / \mathrm{d}\varrho_H$ in front of its Laplacian.

The above Eq. \eqref{NLtauequation} matches\footnote{%
Eq.~\eqref{NLtauequation} features additional contributions from the initial density perturbations $\delta\init\varrho$ as compared to the original Newtonian result obtained in \cite{adlerbuchert}. These perturbations were actually neglected there, by assuming $\init\varrho = \init{\varrho_H}$, as is also assumed in Zel'dovich's original work for the dust case \cite{Zeldovich1970a}. However, as is demonstrated in Appendix A of \cite{adlerbuchert}, such an assumption can be made without loss of generality in the Newtonian context within the first--order perturbation scheme in the deformation vector, through a suitable change of Lagrangian coordinates, making both approaches equivalent.
}
the corresponding equation for the deformation vector obtained in the Newtonian Lagrangian framework, Eq.~(24b) in \cite{adlerbuchert} or Eq.~(45) in \cite{RN50}
written for the longitudinal part of the deformation vector. By definition, this part obeys the same evolution equation as the Lagrangian divergence of the vector, as can be seen in the unnumbered equations involving that divergence before Eq.~(24a) in \cite{adlerbuchert} . Note that in this reference, the Laplacian term features a local sound speed squared (related to the local Jeans length) $\mathrm{d}p / \mathrm{d}\varrho$, but it is already noted there that it should actually be replaced by the background value for consistency with the first--order expansion, and it is indeed replaced by the corresponding background expression in \cite{RN50}.

\subsubsection{Master equation for the traceless part}

The first--order evolution of the traceless symmetric part $\tensor{\Pi}{^i_j}$ is given by Eq.~\eqref{tracefreeEOS2}, with $\tensor{\xi}{^i_j^{(1)}}$ and $\tensor{\tau}{^i_j^{(1)}}$ replaced by their expressions \eqref{4divergencetracefree} and \eqref{curvaturetracefree}, respectively. Eliminating the initial traceless curvature $\tensor{\mathscr{T}}{^i_j}$ by evaluation of the evolution equation at the time corresponding to the initial condition \eqref{initialconditiontl}, then first yields the following evolution equation for the traceless symmetric part:
\begin{align}
 & \partial_t^2 \tensor{\Pi}{^i_j} + 3 H \big[ 1 - \beta'(\epsilon_H)\big] \partial_t \tensor{\Pi}{^i_j} \nonumber \\
 & \quad {} + \frac{N_H^2}{a^{2}} \left( 2\, \tensor{\Pi}{^i_k_|_j^|^k} - \tensor{\Pi}{^i_j_|_k^|^k} - \frac{2}{3}\, \tensor{\Pi}{^k_l_|_k^|^l} \,\tensor{\delta}{^i_j} \right) \nonumber \\
 	& \; {} = \frac{N_H^2}{3 a^{2}} \Big( \big[1 + 3 \beta'(\epsilon_H)\big] \, \tensor{\CD}{^i_j} \Ptil
-\big[1 + 3 \beta'(\init{\epsilon_H})\big] \, \tensor{\CD}{^i_j} \init\Ptil \Big)
\nonumber \\
 	& \quad {}
	+ \frac{N_H^2}{a^2 N^{2}_{H_\mathbf{i}}} \, \Big( \tensor[^{\rm{tl}}]{W}{^i_j} + \init H \, \big[1- 3 \beta'(\init{\epsilon_H})\big] \tensor[^{\rm{tl}}]{U}{^i_j} \Big) \; .
 \label{firstordertraceless}
\end{align}
Here $\init\Ptil=-\init{\alpha_H}\,\delta \init \epsilon$ due to the vanishing of the initial spatial perturbation field, and we have introduced
the coordinate traceless spatial Hessian operator $\tensor{\CD}{^i_j} := \delta^{i k} \partial_k \partial_j - (1/3) \tensor{\delta}{^i_j} \Delta_0$.

This equation still explicitly features the trace, but it can be fully expressed in terms of $\tensor{\Pi}{^i_j}$ by making use of the momentum constraints \eqref{momentumconstraintEOS}. This can be achieved by rewriting \eqref{momentumconstraintEOS} as
\begin{equation}
 \frac{1}{N_H} \,\partial_t \tensor{\Pi}{^i_j_|_i} = \frac{2}{3} \,\partial_t \!\left( \frac{\Ptil_{|j}}{N_H} \right) \; .
 \label{momentumconstraintEOS2}
\end{equation}
A time--integration and spatial differentiation of this equation allows one to express $\tensor{\mathcal D}{^i_j}\Ptil$ as
\begin{equation}
\label{substitutionoftrace}
 \frac{\tensor{\mathcal D}{^i_j}\Ptil}{N_H} = \frac{\tensor{\mathcal D}{^i_j}\init\Ptil}{\init{N_H}} + \frac{1}{2} \int_{\init t}^{t}{\frac{\partial_t \left(3\, \tensor{\Pi}{^k_j_|_k^|^i} - \tensor{\Pi}{^k_l_|_k^|^l} \,\tensor{\delta}{^i_j} \right)}{N_H} \, {\rm d}t'} \, .
\end{equation}
The pair of equations $\lbrace$\eqref{firstordertraceless},\eqref{substitutionoftrace}$\rbrace$ together comprise the
{\it master equation for the traceless part.}
When $p_H=0$, one simply has $N_H(t)=1$ and $\beta'(\epsilon_H) =0$ so that this master equation
reduces to the corresponding one in the dust case, Eq.~(43) in \cite{rza4}.

\subsubsection{Master equations for free and scattered gravitational waves}

Following the approach developed in \cite{rza3,rza4}, we can gain more insight into the evolution of $\tensor{\Pi}{^i_j}$ by splitting the full master equation for the traceless variable into gravitoelectric and gravitomagnetic parts.

To this end, we first define a corresponding split of the initial conditions for the traceless variables:
\begin{gather}
\label{EHsplitinitial}
 \tensor[^{\mathrm{tl}}]{U}{^i_j} = \tensor[^{\mathrm{tl},E}]{U}{^i_j} + \tensor[^{\mathrm{tl},H}]{U}{^i_j} \; ; \; \tensor[^{\mathrm{tl}}]{W}{^i_j} = \tensor[^{\mathrm{tl},E}]{W}{^i_j} + \tensor[^{\mathrm{tl},H}]{W}{^i_j} \; ; \\
 \label{magneticconstraints}
 \tensor[^{\mathrm{tl},H}]{U}{^i_j_|_i} = 0 \; ; \; \tensor[^{\mathrm{tl},H}]{W}{^i_j_|_i} = 0 \; ; \\
 \label{electricconstraintU}
 2 \,\Delta_0 \tensor[^{\mathrm{tl},E}]{U}{^i_j} + \tensor[^{\mathrm{tl},E}]{U}{^k_l_|_k^|^l} \,\tensor{\delta}{^i_j} - 3 \,\tensor[^{\mathrm{tl},E}]{U}{^i_k_|_j^|^k} = 0 \; ; \\
 \label{electricconstraintW}
 2 \,\Delta_0 \tensor[^{\mathrm{tl},E}]{W}{^i_j} + \tensor[^{\mathrm{tl},E}]{W}{^k_l_|_k^|^l} \,\tensor{\delta}{^i_j} - 3 \,\tensor[^{\mathrm{tl},E}]{W}{^i_k_|_j^|^k} = 0 \; .
\end{gather}
These conditions can be jointly required because of the following geometric identity (taking its first two time--derivatives and evaluating them at the initial time):
\begin{equation}
 \big( 2 \, \Delta_0 \tensor{\Pi}{^i_j} + \tensor{\Pi}{^k_l_|_k^|^l} \,\tensor{\delta}{^i_j} - 3 \, \tensor{\Pi}{^i_k_|_j^|^k} \big)_{|i} = 0 \; .
\end{equation}
This in turn is due to $\tensor{\Pi}{^k_[_i_|_j_]_|_k} = 0$, which is a consequence of the momentum constraints (see footnote \ref{footnote_tau_ij}).

We can then define the gravitoelectric and gravitomagnetic traceless parts, respectively, $\tensor[^E]{\Pi}{^i_j}$ and $\tensor[^H]{\Pi}{^i_j}$, from their vanishing initial values and their respective initial first time--derivatives $\tensor[^{\mathrm{tl},E}]{U}{^i_j}$ and $\tensor[^{\mathrm{tl},H}]{U}{^i_j}$, requiring them to obey the following evolution equations:
\begin{align}
 & \partial_t^2 \tensor[^H]{\Pi}{^i_j} + 3 H \big[ 1 - \beta'(\epsilon_H)\big] \partial_t \!\tensor[^H]{\Pi}{^i_j} - \frac{N_H^2}{a^{2}} \Delta_0 \!\tensor[^H]{\Pi}{^i_j} \nonumber \\
 & \; {} = \frac{N_H^2}{a^2 N^{2}_{H_\mathbf{i}}} \, \Big( \tensor[^{\rm{tl},H}]{W}{^i_j} \!+ \init H \, \big[1- 3 \beta'(\init{\epsilon_H})\big] \tensor[^{\rm{tl},H}]{U}{^i_j} \Big) \; ;
 \label{firstordertracelessH} \\
 & \partial_t^2 \tensor[^E]{\Pi}{^i_j} + 3 H \big[ 1 - \beta'(\epsilon_H)\big] \partial_t \!\tensor[^E]{\Pi}{^i_j} + \frac{N_H^2}{3 a^{2}} \Delta_0 \!\tensor[^E]{\Pi}{^i_j} \nonumber \\
 & \; {} = \frac{N_H^2}{3 a^{2}} \Big( \big[1 + 3 \beta'(\epsilon_H)\big] \, \tensor{\CD}{^i_j} \Ptil
-\big[1 + 3 \beta'(\init{\epsilon_H})\big] \, \tensor{\CD}{^i_j} \init\Ptil \Big)
\nonumber \\
 & \quad {}
+ \frac{N_H^2}{a^2 N^{2}_{H_\mathbf{i}}} \, \Big( \tensor[^{\rm{tl},E}]{W}{^i_j} \!+ \init H \, \big[1- 3 \beta'(\init{\epsilon_H})\big] \tensor[^{\rm{tl},E}]{U}{^i_j} \Big) \; .
 \label{firstordertracelessE}
\end{align}
Equation \eqref{firstordertracelessH} is the {\it master equation for free gravitational waves}, while
Equation \eqref{firstordertracelessE}, after elimination of the coupling to the trace, is the {\it master equation for
the gravitational wave part that is scattered at the fluid source.} We will discuss the coupling to the trace of this latter equation in more detail below.

The above evolution equations ensure that we indeed get a decomposition of the traceless deformation field obeying \eqref{firstordertraceless} at all times:
\begin{equation}
 \tensor{\Pi}{^i_j} = \tensor[^E]{\Pi}{^i_j} + \tensor[^H]{\Pi}{^i_j} \; .
\end{equation}
They will also propagate the initial constraints \eqref{EHsplitinitial}--\eqref{electricconstraintW} that define the split of $\tensor[^{\rm{tl}}]{U}{^i_j}$ and $\tensor[^{\rm{tl}}]{W}{^i_j}$. This will ensure the preservation at all times of the divergence--free nature of free gravitational waves as well as the geometric identity on their scattered part, similar to the dust case (cf. \cite{rza3,rza4}):
\begin{gather}
 \tensor[^H]{\Pi}{^i_j_|_i} = 0 \; ; \\
 2 \,\Delta_0 \!\tensor[^E]{\Pi}{^i_j} + \tensor[^E]{\Pi}{^k_l_|_k^|^l} \, \tensor{\delta}{^i_j} - 3 \,\tensor[^E]{\Pi}{^i_k_|_j^|^k} = 0 \; .
\label{laplacianEPi}
\end{gather}
The (also propagating) momentum constraints \eqref{momentumconstraintEOS2} split as follows:
\begin{equation}
 \tensor[^H]{\Pi}{^i_j_|_i} = 0 \quad ; \quad \frac{1}{N_H} \,\partial_t \!\tensor[^E]{\Pi}{^i_j_|_i} = \frac{2}{3} \,\partial_t \!\left( \frac{\Ptil_{|j}}{N_H} \right) \quad .
\end{equation}
Observe that $\tensor[^H]{\Pi}{^i_j}$ decouples from the trace in both the momentum constraints and the evolution equation, while $\tensor[^E]{\Pi}{^i_j}$ remains coupled to the trace in both cases.

Alternatively, using a time integral of the momentum constraints,
\begin{equation}
\tensor[^E]{\Pi}{^i_j_|_i} = \frac{2}{3} \int_{\init t}^t \! N_H \, \partial_t \! \left( \frac{\Ptil_{|j}}{N_H} \right) {\rm d}t' \; ,
\end{equation}
the geometric constraint \eqref{laplacianEPi} on $\tensor[^E]{\Pi}{^i_j}$ can be expressed as follows:
\begin{equation}
 \Delta_0 \tensor[^E]{\Pi}{^i_j} = \tensor{\mathcal{D}}{^i_j} \left( \int_{\init t}^t \! N_H \, \partial_t \! \left( \frac{\Ptil}{N_H} \right) {\rm d}t' \right) \; .
 \label{laplacianEPi2}
\end{equation}
This is to be compared to the dust--case relation, Eq.~(51) in \cite{rza4}, to which it reduces when $p_H = 0$ and accordingly $N_H(t) = 1$: $\Delta_0 \tensor[^E]{\Pi}{^i_j} = \tensor{\mathcal{D}}{^i_j} (\Ptil - \init \Ptil) = \tensor{\mathcal{D}}{^i_j} P$.
Hence, in the presence of pressure, in contrast to the dust case, the gravitoelectric traceless part and the trace, although still tightly coupled, will in general have different time behaviors.

With the antisymmetric part vanishing at all times, the evolution equations for the trace and for the gravitoelectromagnetic split of the traceless symmetric part, coupled through the momentum constraints, characterize the behavior of the first--order Lagrangian deformation field for this general barotropic single fluid.
These evolution equations have yet to be complemented by the set of initial constraints \eqref{init_cond_symmetry}--\eqref{init_cond_EoS}, to which we turn now.

\subsection{First--order initial conditions}
\label{SubsecInitCond}
The constraints on the initial conditions for the deformation field, the density and the spatial curvature are expressed at the first--order level as follows:
\begin{align}
\label{initialconditionsymm}
& U_{[i j]} = 0 \quad ; \quad W_{[i j]} = 0 \; ; \\
 & W-6\init{H}\,\beta'(\init{\epsilon_{H}}) \, U = {} \nonumber \\
 & \qquad - \init{N_H^2} \,\init{\alpha_H} \, \Big[ \mathcal{W}(\init t) \, \delta\init\epsilon + \beta'(\init{\epsilon_H}) \, \Delta_0 (\delta\init\epsilon) \Big] \;
\label{initialconditiontrace} ; \\
& \tensor[^{\rm{tl}}]{W}{^i_j} + \init H \,\big[1- 3 \,\beta'(\init{\epsilon_H}) \big] \tensor[^{\rm{tl}}]{U}{^i_j} = - \init{N_H^2} \, \tensor{\mathscr{T}}{^i_j} \nonumber \\
\label{initialconditiontl}
& \qquad {} - \init{N_H^2} \,\init{\alpha_H} \,\beta'(\init{\epsilon_H}) \left[ \tensor{\left(\delta\init\epsilon\right)}{^|^i_|_j} - \frac{1}{3} \Delta_0 \! \left(\delta\init\epsilon\right) \tensor{\delta}{^i_j} \right] ; \\
& \init{H}U = - \frac{1}{4} \mathscr{R} \init{N_H^2}+ 4 \pi G \,
\init{N_H^2} \, \init{\alpha_H} \, \delta \init \epsilon \, \times \nonumber \\
\label{initialconditionenergy}
& \hbox to67pt{\hfil} \Big[\init{\epsilon_H} + \init{p_H} - (2 \init{\epsilon_H} + \tLam) \, \beta'(\init{\epsilon_H}) \Big]
\; ; \\
\label{initialconditionmomentum}
& \tensor{U}{^i_j_|_i} - U_{\mid j} = 2\init{H} \, \init{\alpha_H} \, \beta'(\init{\epsilon_H}) \left(\delta\init\epsilon\right)_{|j} \; ;\\
& \init p = \init{p_H} + \init{\epsilon_H} \, \beta'(\init{\epsilon_H}) \,\delta\init\epsilon
\quad ; \quad \init{p_H} = \beta(\init{\epsilon_H}) \; .
\label{initialconditionEoS}
\end{align}
This set of initial conditions can also be obtained by evaluating the linearized Lagrange--Einstein system at the initial time.
It can be complemented by the requirements \eqref{EHsplitinitial}--\eqref{electricconstraintW} which define the initial split into gravitoelectric and gravitomagnetic parts of the traceless deformation field.

Note that the above set keeps more variables coupled than the corresponding ones in \cite{rza4}. This is to be expected, since in the dust case
a vanishing pressure and a constant lapse allowed for the elimination of $\epsilon$ and $\Lambda$ between the first two constraints,
leaving only a relation among $U$, $W$ and $\mathscr{R}$. Here, we also have contributions from $p$, $\Lambda$ (due to the lapse factor in the $\Lambda$ term) and the nonvanishing $\init{\mathcal{A}}^{(1)}$. Accordingly, the dependence on the initial energy density $\init\epsilon$ and its spatial derivatives can no longer be explicitly removed in general. However, as in the dust case, the scalar constraints \eqref{initialconditiontrace} and \eqref{initialconditionenergy}, together with the initial EoS \eqref{initialconditionEoS}, show that only two independent first--order initial conditions need to be given for the scalar variables $U$, $W$, $\mathscr{R}$, $\init\epsilon$, and $\init p$. One could for instance only specify $U$ and $W$ as can be done in the dust case, fully determining the other scalar initial conditions. In contrast to the dust case, however, determining $\init\epsilon$ in this situation would involve solving for the Laplacian differential equation \eqref{initialconditiontrace}.

\section{Application to specific equations of state}
\label{SecIV}

Concrete results can be obtained by looking at special cases of the barotropic EoS. In this section, we will first consider the family of linear relations between the pressure and the energy density. We then proceed to a special nonlinear polytropic EoS that allows one to model the isotropic part of a velocity dispersion field up to late epochs of nonlinear structure formation.

\subsection{Case of a linear Equation of State: $p=w\epsilon$}
\label{seclinearEoS}

In the previous section we have derived the evolution equations for the first--order deformation field, sourced by a general barotropic fluid. In this section we will consider as an example the simplest barotropic EoS, $p= \beta(\epsilon)=w\epsilon$ with $w$ a constant parameter obeying the dominant energy condition, $-1\le w\le1$. In addition to the radiation fluid, with $w = 1/3$, other interesting cases include a ``stiff fluid'' corresponding to a free scalar field source, with $w=1$, and a ``curvature'' or ``string gas'' equation of state, with $w=-1/3$.
For this class of linear EoS we can readily apply the procedure suggested in \cite{rza3,rza4} to find the relativistic Lagrangian first--order solutions.

The formal rest mass density $F(\epsilon)$ and the lapse are found to be as follows:
\begin{equation}
 F(\epsilon) = \varrho_1 \left( \frac{\epsilon}{\epsilon_1} \right)^{\!1/(1+w)} ; \; N = \frac{\varrho_1}{\epsilon_1 (1+w)} \left( \frac{\epsilon}{\epsilon_1} \right)^{\!-w/(1+w)}\label{lapsew},
\end{equation}
if $w\neq -1$. (The case $w=-1$ for a ``vacuum energy equation of state'' can be treated separately by the explicit cosmological term.)

The solution \eqref{epsilon_of_J} of the energy conservation law then yields the energy density, and the lapse as deduced from (\ref{lapsew}), as the following respective functionals of the coframes, with $J = \det(\tensor{\eta}{^a_i})$:
\begin{equation}
\epsilon =\init{\epsilon} \,J^{-(1+w)} \; ; \; N = \init{N} \, J^{w} \; .
\label{reducejacobian1}
\end{equation}
Similar equations hold for the background spacetime,
\begin{equation}
\epsilon_H = \init{\epsilon_H} \, a^{-3(1+w)} \; ; \; N_H = \init{N_H} \, a^{3w} \; ; \; \frac{\partial_t N_H}{N_H} = 3 w H \; .
\end{equation}
Given the linear barotropic EoS, the pressure and background pressure are immediately deduced from the expression of the corresponding energy densities, and will share their functional dependencies.

\subsubsection{First--order equations}
With the linear EoS $\beta(\epsilon) = w \epsilon$, $\beta'(\epsilon_H)$ reduces to the constant value $w$, $\beta''(\epsilon_H)$ vanishes at all times, and $\init{\alpha_H}=(1+w)^{-1}$. Consistent with a first--order evaluation of the exact formulas above, the first--order expressions \eqref{generalEOSepsilon_lin}--\eqref{firstorderN1} for $\Ptil$, $\epsilon$, $p$, $F(\epsilon)$, $N$ (and its rate of evolution) thus simplify to
\begin{gather}
\Ptil = P - (1+w)^{-1} \delta\init\epsilon \; ; \nonumber\\
\epsilon = \epsilon_H \Big[1 - (1+w)\Ptil \Big] \quad ; \quad
p = p_H - w (1+w) \,\epsilon_H \,\Ptil \; ; \nonumber\\
F(\epsilon) = F(\epsilon_H) \big[1 - \Ptil \big] \; ; \nonumber\\
N = N_H \big[1 + w \Ptil \big] \quad ; \quad
\frac{\partial_t N}{N} = 3 w H + w \,\partial_t \Ptil \; .
\end{gather}
Eq.~(\ref{Vt}) reduces to
\begin{equation}
\mathcal{V}(t) = \frac{\epsilon_H (1-w)^2 - 2 w \tLam}{\epsilon_H (1-w) - w \tLam} \; ,
%\frac{4 \pi G \epsilon_H (1-w)^2 - 2 w \Lambda}{4 \pi G \epsilon_H (1-w) - w \Lambda} \; ,
\end{equation}
and the first--order Lagrange--Einstein system \eqref{HamiltonEOS}, \eqref{antisymmetricpart1}--\eqref{momentumconstraintEOS} becomes:\footnote{It is worth noting in the case when $\Lambda = 0$, $\mathcal{V}(t)$ simplifies further and reduces to the constant $1 - w$, so that \eqref{LEStrace1linearEoS} becomes
$$
\partial_t^2 \Ptil + 3 H (1-w) \partial_t \Ptil = \init{N_H^2}\, a^{6w} \! \left[ \mathcal{A}^{(1)} - \frac{1+3w}{4} \mathcal{R}^{(1)} \right]\; .
$$}
\begin{gather}
\partial_{t}\mathfrak{P}_{ij}=0 \; : \quad \mathfrak{P}_{ij} = \mathfrak{P}_{ij}(\init t) = 0 \; ; \nonumber\\
\partial_t^2 \Ptil + 3 H \,
\frac{\epsilon_H (1-w)^2 + 2 w^2 \tLam}{\epsilon_H (1-w) - w \tLam}
%\frac{4 \pi G \epsilon_H (1-w)^2 + 2 w^2 \Lambda}{4 \pi G \epsilon_H (1-w) - w \Lambda}
\, \partial_t \Ptil \hspace{15ex} \nonumber \\
\label{LEStrace1linearEoS}
	{} = \init{N_H^2} \;\!a^{6w} \;\!\!\!\left[
\mathcal{A}^{(1)} - \frac{\epsilon_H (1-w)(1+3w) + 2 w \tLam}{4 \epsilon_H (1-w) - 4 w \tLam}
%\frac{4 \pi G \epsilon_H (1-w)(1+3w) + 2 w \Lambda}{16 \pi G \epsilon_H (1-w) - 4 w \Lambda}
\mathcal{R}^{(1)} \right] ; 
\end{gather}
\begin{gather}
\partial_t^2 \tensor{\Pi}{^i_j} + 3 H (1-w) \partial_t \tensor{\Pi}{^i_j} = \init{N_H^2}\, a^{6w} \left( \tensor{\xi}{^i_j^{(1)}} - \tensor{\tau}{^i_j^{(1)}} \right) \; ; \\
H\,\partial_t \Ptil + 4\pi G \big[ \epsilon_H (1-w) - w \tLam \big] \init{N_H^2}\, a^{6w} \, \Ptil\nonumber\\
%\times \hspace{13ex} \nonumber \\ \hspace{9ex} \big[P - (1+w)^{-1} \delta\init\epsilon \big]
\label{firstorderhamilton2}
 = - \frac{1}{4} \init{N_H^2}\, a^{6w} \;\!\mathcal{R}^{(1)} \; ; \\
\partial_t \left( \tensor{\Pi}{^i_j_|_i} - \frac{2}{3}\Ptil_{|j} \right) = - 2 w H
\Ptil \; .
\label{firstordermomentum2}
%(P - (1+w)^{-1} \delta\init\epsilon) \; .
\end{gather}
The acceleration gradient and its trace and traceless parts are expressed in terms of the deformation field at first order according to Eqs.~\eqref{4divergence11}, \eqref{4divergencetrace}, and 
\eqref{4divergencetracefree}, yielding
\begin{gather}
\tensor{\mathcal{A}}{^i_j^{(1)}} = a^{-2} w \, \tensor{\Ptil
%\big(P - (1+w)^{-1} \delta\init\epsilon \big)
}{^|^i_|_j} \; ; \\
\mathcal{A}^{(1)} = a^{-2} w \, \Delta_0 \Ptil
%\big(P - (1+w)^{-1} \delta\init\epsilon \big)
\; ; \\
\tensor{\xi}{^i_j^{(1)}} = a^{-2} w \, \tensor{\mathcal{D}}{^i_j} \Ptil
%\big(P - (1+w)^{-1} \delta\init\epsilon \big)
\; ,
\end{gather}
while the first--order expressions \eqref{initialriccitrace1},\eqref{curvaturetrace2}, and \eqref{curvaturetracefree} of the Ricci tensor and its trace/traceless split are formally unchanged.
Since for the chosen EoS, $\mathcal{W}(t)$ yields
\begin{align}
\mathcal{W}(t) &{}= 4 \pi G \left[\epsilon_H (1-w)(1+3w) + 2 w \tLam\right] \hspace{15ex} \nonumber \\
&{}= 4 \pi G\left[ \init{\epsilon_H} \, a^{-3(1+w)} (1-w)(1+3w) + 2 w \tLam\right],
\end{align}
the master equation \eqref{mastereqtrace} for the trace of the perturbation now reads:
\begin{align}
& \partial_t^2 \Ptil + 2H (1-3w) \partial_t \Ptil\nonumber\\
%- \big[P - (1+w)^{-1} \delta\init\epsilon \big] \times \nonumber \\
%& \quad {} \big[ 4 \pi G \init{N_H^2} \,\init{\epsilon_H} (1-w)(1+3w) \,a^{3(w-1)} + 2 w \Lambda \init{N_H^2} \,a^{6w} \big] \nonumber \\
&\; -4 \pi G \init{N_H^2} \big[ \init{\epsilon_H} (1-w)(1+3w) \,a^{3(w-1)} + 2 w \tLam \,a^{6w} \big]\Ptil \nonumber \\
\label{firstordertrace2}
& \qquad {} = w \init{N_H^2} \,a^{6w-2} \Delta_0 \Ptil
%\big[P - (1+w)^{-1} \delta\init\epsilon \big]
\; .
\end{align}
In turn, the master equation \eqref{firstordertraceless} for the traceless symmetric part of the deformation field becomes
\begin{align}
& \partial_t^2 \tensor{\Pi}{^i_j} + 3 H (1-w) \partial_t \tensor{\Pi}{^i_j} \nonumber \\
& \
+ \init{N_H^2} \, a^{6w-2} \left\{ 2 \tensor{\Pi}{^i_k_|_j^|^k} - \tensor{\Pi}{^i_j_|_k^|^k} - \frac{2}{3} \tensor{\Pi}{^k_l_|_k^|^l} \tensor{\delta}{^i_j}\right.
%\right) \nonumber \\ & {} \quad = a^{6w-2} \bigg[ \frac{1}{3} (1+3w) \init{N_H^2} \,
\nonumber\\ &\hspace{68pt}\left.- \frac{1}{3}\big(1+3w\big)\tensor{\mathcal{D}}{^i_j} \left(\Ptil - \init{\Ptil}\right) \right\} \nonumber\\
&{} \quad=a^{6w-2} \left[
\tensor[^{\rm{tl}}]{W}{^i_j} + (1-3w) \init H \tensor[^{\rm{tl}}]{U}{^i_j} \right] \; ,
\end{align}
with, from the momentum constraints \eqref{firstordermomentum2},
\begin{equation}
 a^{-3w}\,\tensor{\mathcal D}{^i_j}\Ptil = \tensor{\mathcal D}{^i_j}\init\Ptil + \int_{\init t}^{t}{\frac{\partial_t \left(3\, \tensor{\Pi}{^k_j_|_k^|^i} - \tensor{\Pi}{^k_l_|_k^|^l} \,\tensor{\delta}{^i_j} \right)}{2 \,a^{3w}} \, {\rm d}t'} \, .
\end{equation}
We can finally rewrite the set of initial conditions \eqref{initialconditionsymm}--\eqref{initialconditionEoS} for the linear EoS:
\begin{align}
& U_{[ij]} = 0 \quad ; \quad W_{[ij]} = 0 \; ; \\
& W- 6 w \init H U = - \frac{\init{N_H^2}}{1+w} \Big( w\, \Delta_0(\delta\init\epsilon) \nonumber \\
	&\qquad{} + 4 \pi G \big[ \init{\epsilon_H} (1-w)(1+3w) + 2 w \tLam \big] \delta\init\epsilon \Big) \; ; \\
& \tensor[^{\rm{tl}}]{W}{^i_j}\! + (1-3w) \init H \tensor[^{\rm{tl}}]{U}{^i_j} \;\!\! = - \init{N_H^2}\! \left[ \tensor{\mathscr{T}}{^i_j} \!\!+ \frac{w}{1+w} \tensor{\mathcal{D}}{^i_j} (\delta\init\epsilon) \right]  ; \\
& \init H \,U = - \frac{1}{4} \mathscr{R} \init{N_H^2} + \frac{4 \pi G\init{N_H^2}}{1+w} \,\big[ \init{\epsilon_H} (1-w) - w \tLam \big] \delta\init\epsilon \; ; \\
& \tensor{U}{^i_j_|_i} - U_{|j} = 2 \,\frac{w}{1+w} \init H \,(\delta\init\epsilon)_{|j} \; ; \\
& \init p = \init{p_H} + w \,\init{\epsilon_H} \,\delta\init\epsilon \quad ; \quad \init{p_H} = w \,\init{\epsilon_H} \; .
\end{align}

\subsubsection{Solutions for the trace of the deformation field}
\label{sectracesolutionw}

Similarly to \cite{rza1,rza2}, we will now further investigate the behavior of the trace $P$ of the first--order deformation. For simplicity, we will restrict attention to the case of a vanishing cosmological constant, $\Lambda = 0$, as may be reasonably assumed during the radiation--dominated era. In this case Eqs.~\eqref{firstorderhamilton-a}--\eqref{poly-coeffs} reduce to
\begin{equation}
\begin{split}
a\frac{\partial \Ptil}{\partial a}+\frac{3}{2}(1-w)\Ptil=\frac{-3}{32 \pi G \init{\epsilon_H}} a^{3(1+w)}\mathcal{R}^{(1)} \; ;
\label{firstorderhamilton3}
\end{split}
\end{equation}
\begin{align}
 \frac{\partial^{2} \Ptil}{\partial a^{2}}+\frac{\alpha_1}{a}\frac{\partial \Ptil}{\partial a}-\frac{\alpha_{2}}{a^{2}}\Ptil=\init{\alpha_3} \,a^{3w-1}\Delta_{0}\Ptil \; ,
\label{firstordertrace3}
\end{align}
with the constant parameters
\begin{equation}
\begin{gathered}
\alpha_{1}=\frac{3(1-3w)}{2}\quad ; \quad \alpha_{2}=\frac{3(1-w)(3w+1)}{2}\; ;\\
 \init{\alpha_3}=\frac{3w}{8\pi G \init{\epsilon_{H}}} \; .
\end{gathered}
\end{equation}
If $w>0$ (implying $\init{\alpha_3} > 0$), as we will assume in the following, then Eq.~\eqref{firstordertrace3} is a second--order hyperbolic partial differential equation (PDE).\footnote{%
It would be an elliptic PDE for $w<0$ (\textit{i.e.}, $\init{\alpha_3} <0$), while for the parabolic case $w=0$ (and consequently $\init{\alpha_3} = 0$) it reduces, as expected, to the evolution equation for the dust case, with decoupled time and space variables.}
This equation is formally analogous to the standard Eulerian propagation equations for a linearized density contrast
\cite{Christos2008RC,Ellis2012,Patrick2009Ox} once those are reexpressed in terms of the variable $a$.\footnote{%
Note that in terms of the conventional cosmic time  $\tilde t$ introduced in \eqref{metricb2}, Eq.~\eqref{firstordertrace3} reduces to $\partial^2_{\tilde t}\Ptil+(2-3w)a^{-1}\partial_{\tilde t}a\partial_{\tilde t}\Ptil-4\pi G\left[(1-w)(1+3w)\epsilon_H+2w\tLam\right]\Ptil=wa^{-2}\Delta_0\Ptil$.
This is formally equivalent to the linearized Eulerian equation (3.2.17) of Ref.~\cite{Christos2008RC}
in that the coefficients agree, but both the dependent and (spatial) independent variables differ. 
}
In the Eulerian case, assuming global flat--space spatial coordinates, one can find the analytical general solution using a Fourier transformation. A discussion of the differences between the Eulerian and Lagrangian approaches has been given in \cite{rza4}. (See also the related discussion in \cite{Eleonora2014JCAP}.) Ref.~\cite{rza4} also elucidated a procedure for finding general--relativistic Lagrangian first--order solutions for the deformation field in the dust case. We show here that this procedure can be readily extended to the presence of pressure and apply it to the determination of a Lagrangian solution for the trace part.\footnote{%
A complementary picture of an equivalent procedure is shown in Appendix~\ref{appendixFouriersec} and applied to the search for a particular solution for the traceless part.
}

First, we can use the formal identity of Eq.~(\ref{firstordertrace3}), written in Lagrangian coordinates on the nontrivial spacetime manifold, with an equation written in Euclidean space. We can thus work within this flat space with its effective `Eulerian' Cartesian spatial coordinates $x^i$ and solve Equation (\ref{firstordertrace3}) with $\Delta_0 \mapsto \delta^{i j} \,\partial_{x^i} \partial_{x^j}$ for the unknown $\Ptil(a,\mathbf{x})$. On this space we can then apply an inverse Fourier transformation
\begin{equation}
\Ptil (a,\mathbf{x})=\iiint\Ptil_{k}(a,\mathbf{k})\,e^{-i \mathbf{k}\cdot \mathbf{x}}\,\rd^{3}\mathbf{k} \; ,
\label{fourier1}
\end{equation}
and thus get a second--order linear ordinary differential equation:
\begin{equation}
 \frac{d^{2} \Ptil_{k}}{d a^{2}}+\frac{\alpha_1}{a}\frac{d \Ptil_{k}}{da}-\left(\alpha_{2}a^{-2}-\init{\alpha_3}\,k^{2} a^{3w-1}\right)\Ptil_{k}=0 \; ,
\label{firstordertrace4}
\end{equation}
where we have used $\mathbf{k}\cdot \mathbf{x} := \delta_{i j} k^i x^j$ and $k := \left(\delta_{i j} k^i k^j \right)^{1/2}$.

In this case the background Jeans wave number \eqref{jeans} satisfies
\begin{equation}
\begin{aligned}
k_{J}(\epsilon_H)^{2} &{}= \frac{\alpha_{2}}{\init{\alpha_3}}a^{-3(1+w)} \\
&{}= 4\pi G \init{\epsilon_H} \, \frac{(1-w) (3w+1)}{w \, a^{3(1+w)}} \; ,
\end{aligned}
\label{Jeans1}
\end{equation}
where we recall that $0<w\le1$ is assumed.
The behavior of the solution to Eq.~(\ref{firstordertrace3}) will then depend on the relative values of $k$ and $a \, k_{J}(\epsilon_H)$.

One can first proceed by investigating the extreme cases, as is commonly done in the Eulerian analyses. When $k\ll a \, k_{J}(\epsilon_H)$, Eq.~\eqref{firstordertrace4} may be solved as
\begin{equation}
\Ptil_{k}=a^{1+3w} C_{k,1}+a^{\frac{3}{2}(w-1)} C_{k,2} \; , \label{jcase1}
\end{equation}
where $C_{k,1(2)}$ are two functions of $\mathbf{k}$ encoding the initial conditions.
This corresponds, as expected, to the unstable regime since the term with coefficient $C_{k,1}$ is a growing mode.

In the opposite situation when $k\gg a \, k_{J}(\epsilon_H)$, the solution reads
\begin{gather}
\Ptil_{k}=a^{\frac{9w-1}{4}} \Big[J_{\hat{\nu}}\left(B \,a^{\frac{1+3w}{2}} k\right)C_{k,1}+Y_{\hat{\nu}}\left(B \,a^{\frac{1+3w}{2}} k\right)C_{k,2} \Big] \, ; \nonumber \\
B:=\frac{2 \,\sqrt{\init{\alpha_3}}}{1+3w} \quad ; \quad \hat{\nu}:=\frac{9w-1}{2+6w} \quad ,
\label{case2}
\end{gather}
with different $\mathbf{k}$--dependent coefficients $C_{k,1(2)}$, and where $J_{\nu}(x)$ and $Y_{\nu}(x)$ denote the Bessel functions of the first and second kind, respectively. This corresponds to a `stable' regime of acoustic oscillations, although their amplitude will grow over time (as $a^{(3w-1)/2}$ for large $a$) for an unusual EoS with $w > 1/3$. The latter remark includes the ``stiff fluid'' EoS $w=1$, for which the above solution is exact at all times, since it corresponds to $k_J(\epsilon_H) = 0$.

From the expression \eqref{Jeans1} of $k_J(\epsilon_H)$, the noncomoving Jeans wave number $a \, k_J(\epsilon_H)$ decreases over time, so that even an initially unstable solution will eventually enter the stable regime. Such a solution will cross the threshold $k \simeq a \, k_J(\epsilon_H)$ and it may be useful to be able to describe this transition period as well.

As in the Newtonian case in the Eulerian approach, with different coefficients (see, \textit{e.g.}, \cite{GailisFrankel06}), the Bessel functions actually allow for an explicit solution of Eq.~\eqref{firstordertrace4} for any mode at all times. The general solution is the same as (\ref{case2}) up to a change of the order of the Bessel functions:
\begin{gather}
\Ptil_{k}=a^{\frac{9 w-1}{4}} \Big[J_{\nu}\left(B \, a^{\frac{1+3w}{2}} k\right)C_{k,1}+Y_{\nu}\left(B \,a^{\frac{1+3w}{2}}k\right)C_{k,2} \Big]\, ; \nonumber \\
B=\frac{2 \,\sqrt{\init{\alpha_3}}}{1+3w} \quad ; \quad \nu=\frac{5+3w}{2+6w} \; .
\label{fullsolution1}
\end{gather}
The integration constants $C_{k,1(2)}$ are derived from the initial conditions on $\Ptil$ and its time--derivative, $\init{\Ptil}(\mathbf{X})$ and $U(\mathbf{X})$. To this end, one formally replaces these quantities by functions of the `Eulerian' coordinates $x^i$ on the Euclidean space, with the same functional dependence, $\init{\Ptil}(\mathbf{x})$ and $U(\mathbf{x})$. One is then working on flat--space, and the respective Fourier transforms $\Ptil_k(a=\init a =1,\mathbf{k})$ and $(\partial_t \Ptil_k)(a=1,\mathbf{k}) = \init H (\partial_a \Ptil_k)(a=1,\mathbf{k})$ can be computed, from which $C_{k,1(2)}(\mathbf{k})$ are deduced. Knowing these, $\Ptil(a,\mathbf{k})$ is expressed as the full solution given by Eq.~\eqref{fullsolution1} and its inverse Fourier transform \eqref{fourier1} gives $\Ptil(a,\mathbf{x})$ in Euclidean space.

Finally, one can formally replace the Eulerian spatial coordinates by the Lagrangian ones in $\Ptil(a,\mathbf{x})$ while preserving the functional form. The resulting Lagrangian function $\Ptil(a,\mathbf{X})$ then gives a solution to the evolution equation \eqref{firstordertrace3} in the nonconstant curvature spatial sections, thanks to the algebraic identity of this equation with its Euclidean space counterpart.
It is now a Lagrangian solution, however, and must be interpreted as such: the coordinates $X^i$ are comoving with the inhomogeneous fluid flow. They are local coordinates on the perturbed manifold; thus the solution $P(a,\mathbf{X})$ describes perturbations as they evolve in the perturbed space. This perturbed space is in general not
isometric to Euclidean space.
Note that the Fourier modes $\Ptil(a,\mathbf{k})$ are only an intermediate resolution step as they only correspond to modes in the ancillary Euclidean space. As the inversion of the solution \eqref{fullsolution1} does not allow for an explicit general analytic expression, it requires the specification of the initial conditions and will usually involve numerical integration with the given $C_{k,1(2)}(\mathbf{k})$ to realize this solution procedure.

\subsection{Case of a polytropic Equation of State: $p=\kappa \varrho^{\gamma}$}
\label{secpolytrope}

As a second class of models we will now turn to the nonlinear case of polytropic equations of state.

\subsubsection{Equation of state and resolution procedure}
We consider the polytropic EoS, $p=\kappa \varrho^{\gamma}$, $\varrho = F(\epsilon)$, where $\kappa$ is the polytropic constant, and $\gamma > 1$ the polytropic exponent. For such flows the pressure and the energy density obey the relation \cite{Ellis2012,Luciano2013}
\begin{equation}
\begin{aligned}
\epsilon=\beta^{-1}(p)=\frac{1}{\gamma-1}\,p+A \,p^{1/\gamma} =\frac{1}{\gamma-1}\,\kappa\,\varrho^{\gamma}+A\,\kappa^{1/\gamma} \varrho \; ,
\label{polytropiceos}
\end{aligned}
\end{equation}
where $A$ is a constant parameter. We will assume in this section that the formal $\varrho = F(\epsilon)$ actually coincides with the rest mass density of the fluid, \textit{e.g.,} \textit{via} suitable initial conditions. For $A=0$, we again obtain the (nondust) linear case $p=w\epsilon$ with $w:=\gamma-1 > 0$. In the following, we will instead consider the case $A\,\kappa^{1/\gamma}=1$ (in particular $A > 0$), corresponding to an EoS of the type of a nonrelativistic adiabatic ideal gas, the energy density being the sum of the rest mass density and an internal energy density equal to $p/(\gamma-1)$.

As a relevant example, we will focus on the case $\gamma=5 /3$, which is an exact solution for a locally isotropic distribution with velocity dispersion, derived from the relativistic kinetic theory of collisionless matter \cite{Ehlers1969}. (See also \cite{TEllis} and references therein.) This EoS also coincides with the corresponding exact solution in Newtonian cosmology derived from kinetic theory \cite{TBAD,RN50}. In these latter papers it is also shown that this particular EoS arises in the inhomogeneous case by closing the hierarchy of kinetic equations through truncation of the third and higher reduced moments. In the inhomogeneous case this law is, however, phenomenological, since there is a nonvanishing anisotropic part. Neglecting this part strictly results in shear--free motion confirming the exactness of the law in the homogeneous case.

The conservation law (\ref{restmass_conservation}), combined with $p=\kappa \varrho^{\gamma}$, gives for the evolution of $p$:
\begin{equation}
\partial_{t}p+\gamma N\Theta p=0 \;\; ;\;\; \gamma=\frac{5}{3}\;\; .
\label{pressure-ploy-conservation}
\end{equation}
The same relation holds within the background spacetime, so that $p_{H}\,a^{5}=\init{p_{H}}\,\init{a}^{5}$.
The assumption of the background sources following the same EoS also gives, for $\gamma = 5/3$:
\begin{eqnarray}
%\begin{gathered}
\epsilon_{H}= \beta^{-1}(p_H) =\frac{3}{2}p_{H}+A\, p_{H}^{3/5} \; ; \nonumber \\ \label{poly-EOS-detail}
\beta'(\epsilon_{H})=\frac{2}{3}\, \frac{5}{5+2\, A \,p_{H}^{-2/5}} \; ; \\ 
\beta''(\epsilon_{H})=\frac{80 \,A \,p_H^{-7/5}}{9 \left(5+2\, A\,p_{H}^{-2/5}\right)^{3}} \nonumber \; .
%\end{gathered}
\end{eqnarray}
The procedure outlined in the last subsection for solving the trace master equation, Eq.~\eqref{master-trace-a}, in terms of Fourier transformation within a set of coordinates formally equivalent to Eulerian spatial coordinates, is still applicable in this case. We can thus substitute \eqref{poly-EOS-detail} and \eqref{poly-coeffs} in the Eulerian coordinate analogue of \eqref{master-trace-a}, and solve the corresponding ordinary differential evolution equation for each Fourier mode. This has to be performed by numerical integration as the more complicated time--evolution of the coefficients prevents an explicit analytic solution. Once initial conditions are specified we can then numerically compute the inverse Fourier transform, and formally replace the (Eulerian) spatial coordinates by the Lagrangian coordinates $X^i$ (see Section \ref{sectracesolutionw}) to obtain the solution for $\Ptil(t,X^i)$.

\subsubsection{Behavior of the first--order trace for a model overdense region}

As an instructive toy model, we will now consider the evolution of an initial spherical Gaussian deformation:\goodbreak
\begin{equation}
-\init\Ptil = {} \init{\alpha_{H}}\init{\delta \epsilon}= \init c \exp\left(-\frac{R^{2}}{2\sigma^{2}}\right)
\; ,
\label{initialdistribution}
\end{equation}
where $\sigma$ and $\init c$ respectively define the characteristic scale and maximum amplitude of the initial perturbation, and $R:=\left(\delta_{i j} X^i X^j\right)^{1/2}$ is a Lagrangian coordinate `radius'.\footnote{%
We have chosen the set of Lagrangian coordinates $X^i$ such that the components of the spatial metric at initial time, $G_{ij}$, are approximately $\delta_{ij}$ (at leading order) in these coordinates. They can thus be considered as Cartesian--like coordinates, and $R$ is thus a fluid--comoving radial coordinate. It does not, however, coincide with the spatial metric distance between the fluid elements of the respective Lagrangian coordinates $(X^i)$ and $(0,0,0)$. (This is true irrespective of a possible normalization by $a(t)$ to make it a background comoving distance.)
}
We will take $\init c > 0$ and $\init c \ll 1$. The perturbation can then be seen to describe a small initial local overdensity, since the initial rest mass density contrast,
\begin{equation}
\init\delta := \frac{\init\varrho}{\init{\varrho_H}} - 1 = \frac{F(\init\epsilon)}{F(\init{\epsilon_H})} - 1 = \frac{F(\init{\epsilon_H}[1+\delta\init\epsilon]) - F(\init{\epsilon_H})}{F(\init{\epsilon_H})} \; ,
\end{equation}
is well approximated by $\init{\alpha_H} \,\delta\init\epsilon = - \init\Ptil$ for $\init c \ll 1$.
\begin{figure*}[!ht]
\subfloat[]{\includegraphics[clip,width=\columnwidth]{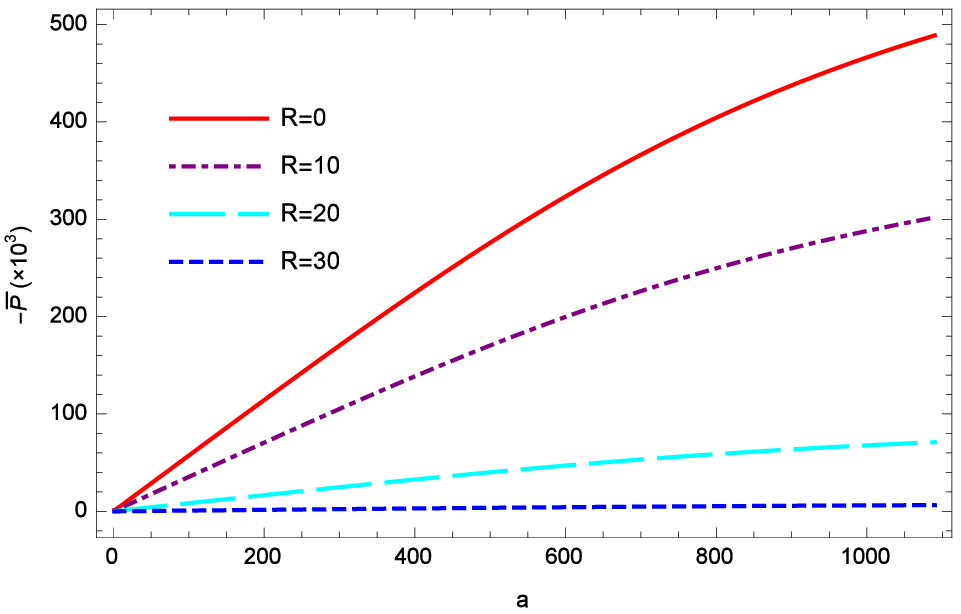}}
\subfloat[]{\includegraphics[clip,width=\columnwidth]{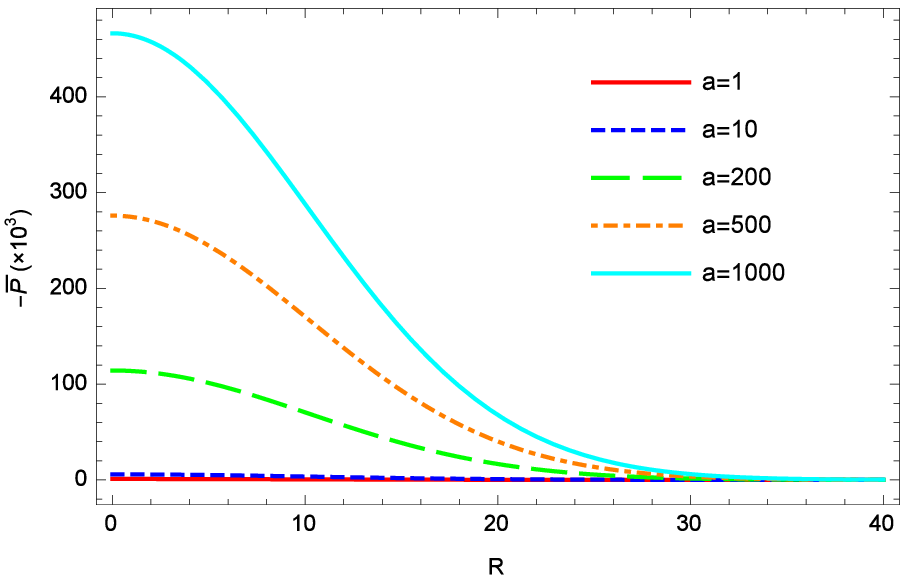}}
\caption{Numerical solution for the first--order trace $-\Ptil$ in Lagrangian space, for an initial spherical Gaussian overdensity with a peak amplitude of $10^{-3}$ at $R=0$ and a standard deviation $\sigma$ such that $k_J(\init{\epsilon_H})\,\sigma=10$. (a). Evolution of $-\Ptil$ as a function of $a$ for fixed values of the Lagrangian radius $R$. From top to bottom: $R=0$, $10$, $20$ and $30$. (b). Spatial variation of $-\Ptil$ with $R$, for several values of the background scale factor. From bottom to top: $a=1$, $10$, $200$, $500$ and $1000$. The perturbation strongly grows over time, corresponding to a collapsing structure.}
\label{sigma10}
\end{figure*}

The actual value of the amplitude $\init c$ is irrelevant for the evolution of $\Ptil$ itself, since it obeys a linear equation. However, it will matter for the nonlinear evaluation of any physical quantity such as $\varrho$ determined by the first--order solution for $\Ptil$ through the extrapolation procedure mentioned above from the Relativistic Zel'dovich Approximation. To best illustrate the effect of this procedure, we choose a rather large overdensity with the arbitrary amplitude $\init c = 10^{-3}$ at an initial time
that corresponds to the epoch of last scattering. 
As we will see, this will let the unstable perturbations enter the mildly nonlinear regime (where $|\Ptil| < 1$ but is of order $1$) around the present epoch, \textit{i.e.}, around $a=a_0 \simeq 1090$ since we set $\init a = 1$.

The other independent initial condition amounts to specifying the first time--derivative $(\partial_t \Ptil)(\init t)$. For this we simply consider an initially stationary deformation and set $(\partial_t \Ptil)(\init t) = U = 0$.

The present formalism focuses on the description of a single fluid source, as it allows for a description in terms of a single velocity field and a single EoS. We will consequently make the simplifying assumption of a model universe filled with a single--component matter fluid and a cosmological constant. The description of model universes with multicomponent fluids is beyond the scope of the present paper, and is left to future work. The background density parameters $\Omega_m$, $\Omega_\Lambda$ for the matter component and the cosmological constant respectively, satisfy $\Omega_{m} + \Omega_{\Lambda} =1$. We will take the present epoch value $\Omega_\Lambda^0=0.692$ in agreement with the best--fit $\Lambda$CDM parameters from the Planck Collaboration \cite{Planck2015}.

The background is also affected by the polytropic EoS \eqref{polytropiceos} of the source fluid. As noted above, our polytrope is exact for the background and is parametrized by the arbitrary constant $\kappa$, or equivalently $A$ as we set $A \kappa^{1/\gamma} = 1$. Specifying its value amounts to choosing the initial instability scale as determined by $k_J(\init{\epsilon_H})$. It also controls the ratio between pressure and rest mass density at a given time, and hence the deviation of the background from a dust--fluid $\Lambda$CDM model. The value we adopt for our examples below, $A \,\init{p_H}^{-2/5}=3/2$, requires the background fluid pressure to be relativistic (and radiation--like) at the initial time, $\init{p_H} = \init{\epsilon_H}/3$, with $\init{p_H}/\init{\varrho_H} = 2/3$. However, it subsequently quickly becomes negligible as $p_H/\varrho_H \propto a^{-2}$, keeping the late--time dynamics of the background very close to that of the $\Lambda$CDM model.
\begin{figure*}[!ht]
\subfloat[]{\includegraphics[clip,width=\columnwidth]{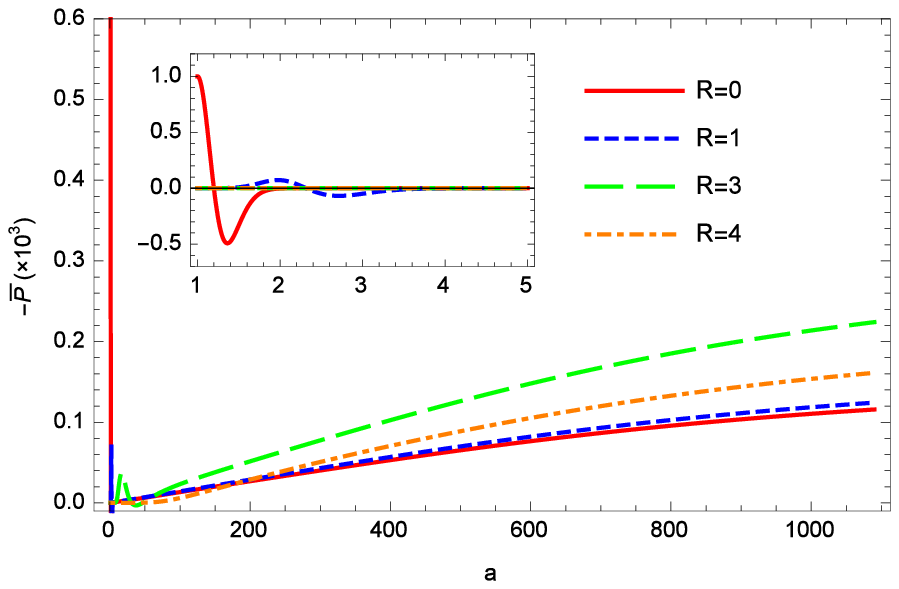}}
\subfloat[]{\includegraphics[clip,width=\columnwidth]{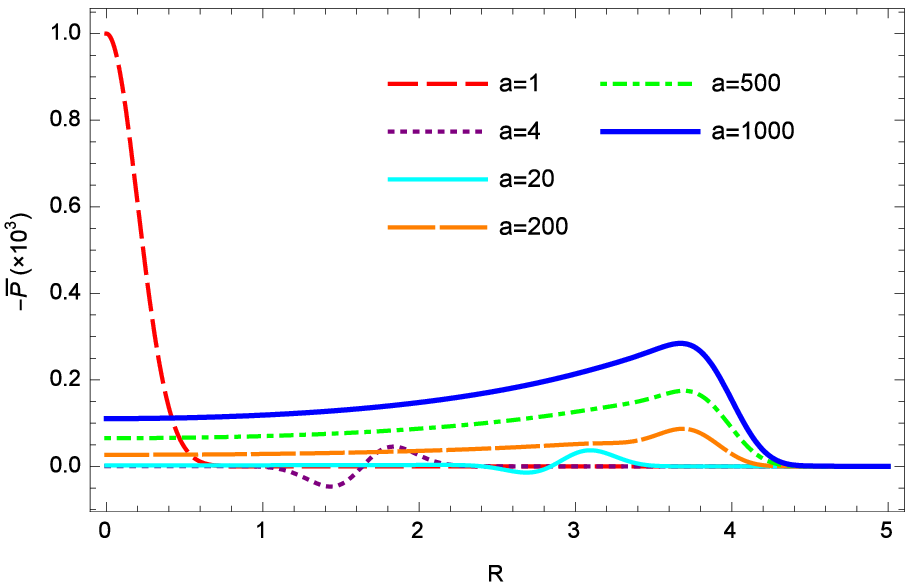}}
\caption{Numerical solution for the first--order trace $-\Ptil$ in Lagrangian space, for an initial spherical Gaussian overdensity with a peak amplitude of $10^{-3}$ at $R=0$ and a standard deviation $\sigma$ such that $k_J(\init{\epsilon_H})\,\sigma=0.2$. (a). Evolution of $-\Ptil$ as a function of $a$ at fixed distance $R$. From top to bottom at $a=1000$: $R=3$, $R=4$, $R=1$ and $R=0$. The inset panel shows a detail of the early evolution (small values of $a$), where only the $R=0$ (solid line) and $R=1$ (dashed line) are visibly nonzero. (b). Spatial variation of $-\Ptil$ with the Lagrangian radius, for several values of the background scale factor.
The structure is first damped and spread out by the Lagrangian pressure gradient, before starting to grow back after the critical wave number $a \, k_J(\epsilon_H)$ has increased, as the perturbation enters the unstable regime.}
\label{sigma0p2}
\end{figure*}
\begin{figure*}[!htp]
\subfloat[]{\includegraphics[clip,width=\columnwidth]{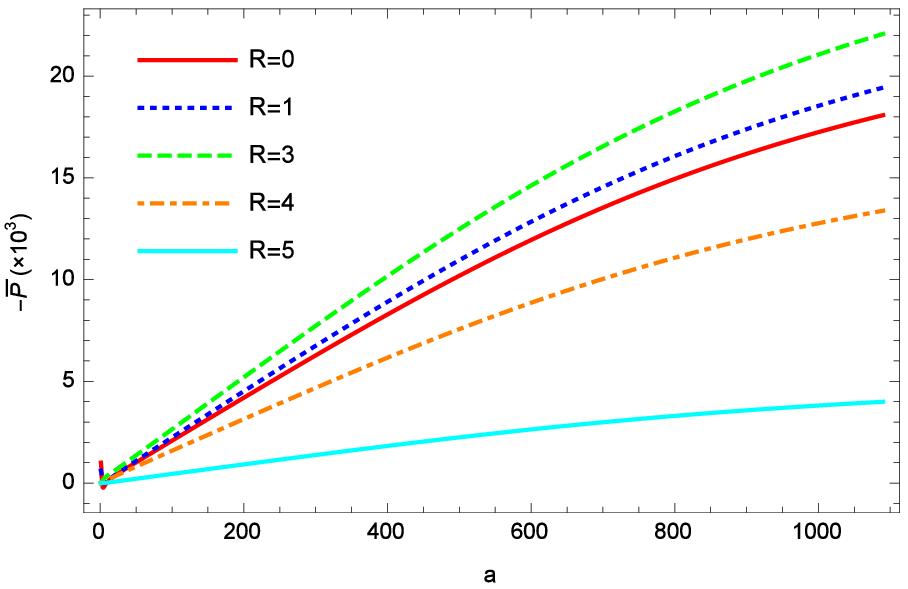}}
\subfloat[]{\includegraphics[clip,width=\columnwidth]{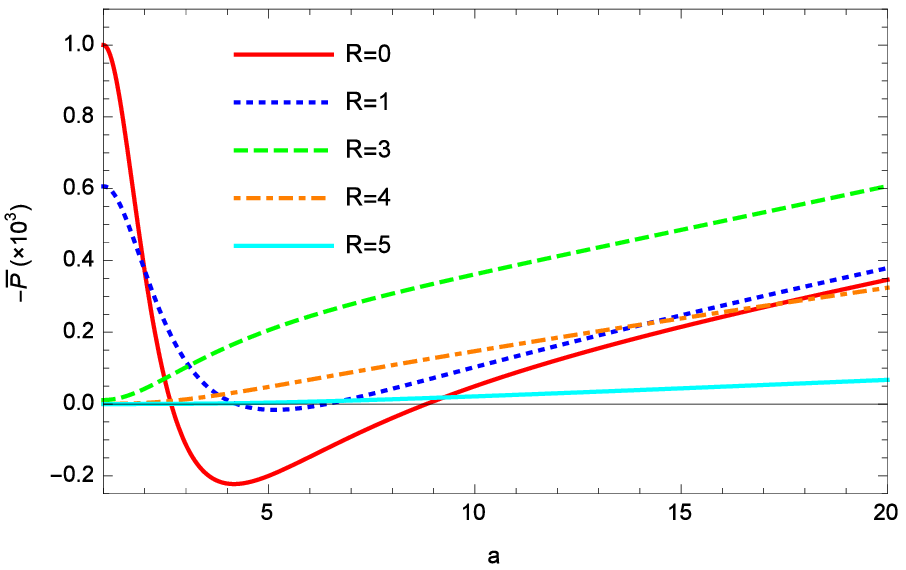}}
\\
\subfloat[]{\includegraphics[clip,width=\columnwidth]{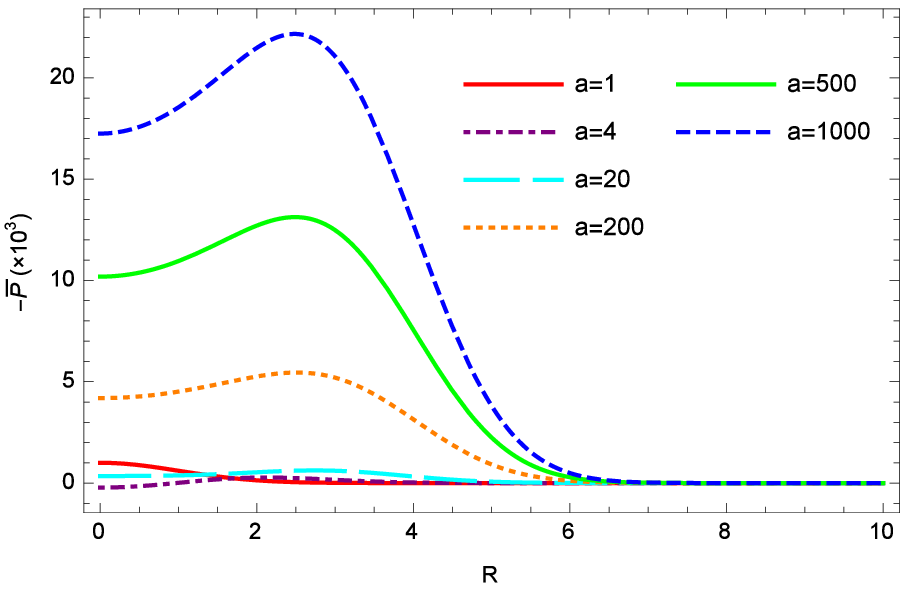}}
\caption{Numerical solution for the first--order trace $-\Ptil$ in Lagrangian space, for an initial spherical Gaussian overdensity with a peak amplitude of $10^{-3}$ at $R=0$ and a standard deviation $\sigma$ such that $k_J(\init{\epsilon_H})\,\sigma=1$. (a) and (b). Evolution of $-\Ptil$ as a function of $a$ at a given distance $R$, for late and early times, respectively. From top to bottom at $a=1000$ for (a): $R=3$, $R=1$, $R=0$, $R=4$, $R=5$; same order for (b) at $a=20$. (c). Spatial variation of $-\Ptil$ with $R$, for fixed values of the background scale factor. From top to bottom at $R=0$: $a=1000$, $a=500$, $a=200$, $a=1$, $a=20$, $a=4$. The behavior is rather similar to the previous case of $k_J(\init{\epsilon_H}) \, \sigma = 0.2$; as expected, the unstable regime is, however, reached sooner, and the perturbation then grows similarly to the case of $k_J(\init{\epsilon_H}) \, \sigma = 10$, up to much above its initial amplitude.}
\label{sigma1}
\end{figure*}
\begin{figure}[htb]
\includegraphics[clip,width=\columnwidth]{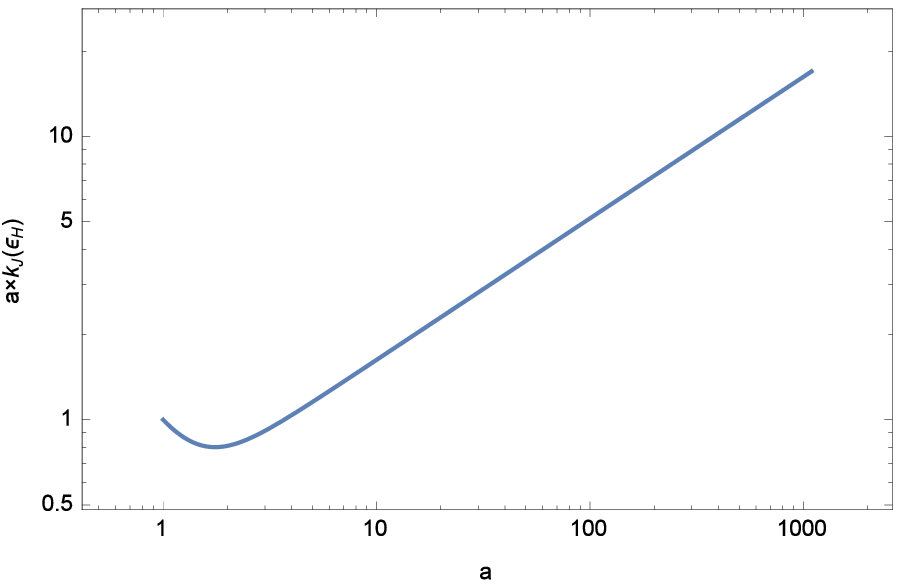}
\caption{Evolution of the instability wave number $a \, k_J(\epsilon_H)$ with the scale factor $a$ for the polytropic EoS considered here, with the unit of length convention $k_J({\epsilon_H}_{\mathbf i}) = 1$. As this wave number only depends on the background by construction, this result applies to all examples considered in this Subsection \ref{secpolytrope}. After a small initial dip, $a\, k_J(\epsilon_H)$ exceeds its initial value around $a \simeq 4$ and enters the increasing power law regime $a\, k_J(\epsilon_H) \propto \sqrt{a}$ (valid as long as $(\Omega_\Lambda / \Omega_m) (a / \init a)^{-2} \ll 1$, which is satisfied up to the present epoch) as expected from the large $a$ expansion of its expression for the present polytropic EoS.}
\label{Jwave}
\end{figure}
We choose to make the lengths $R$, $\sigma$ dimensionless by setting the initial instability scale $k_J(\init{\epsilon_H})^{-1}$ (as derived from substituting \eqref{poly-EOS-detail} into \eqref{jeans} at the initial time) to be our length unit. Thus $\sigma<1$ means that the scale of the initial perturbation is below the Jeans scale $k_J(\init{\epsilon_H})^{-1}$, and above it for $\sigma>1$.
For the value of $A$ adopted in the present example and
estimating $\init{\varrho_H}$ from $\Lambda$CDM background parameters \cite{Planck2015}, this length unit is approximately $98$~kpc. This would correspond to a large background comoving initial overdensity size of $a_0 \,k_J(\init{\epsilon_H})^{-1} \simeq 107$~Mpc.\footnote{%
Note that $k_J(\init{\epsilon_H})^{-1}$ defines an initial instability `scale' only in terms of Lagrangian coordinates, \textit{e.g.}, in terms of $R$. This means that the corresponding `background comoving' distance, $a(t) k_J(\init{\epsilon_H})^{-1}$ evaluated at present time, does not coincide with the present--day physical size of an object that would initially have been of this scale, as such a size must be evaluated using the actual, deformed, spatial metric. (See previous footnote.) $a_0 k_J({\epsilon_H}_{\mathbf i})^{-1}$ may be seen as a rough estimate of this physical size, as obtained by fully neglecting the deformations $G^{(1)}_{ab}$, $\tensor{P}{^a_i}$, in the evaluation of the integrated spatial line element.
}

Figs.~\ref{sigma10}--\ref{sigma1} show the numerical results for $\Ptil$ with the procedure, initial conditions and parameters given above, for three different values of $\sigma$.

The first case, $\sigma=10$ (Fig.~\ref{sigma10}), corresponds to a super--Jeans length, hence unstable, initial perturbation. Figs.~\ref{sigma10}(a),(b) show the numerical results for the evolution of the perturbation $-\Ptil$ as a function of the scale factor at several values of $R$, and over the whole range of radii $R$ for increasing values of $a$, respectively. As expected, this perturbation is unstable and remains so by growing at all times, the pressure gradient being insufficient to prevent the collapse of the structure. The evolution is similar to the dust case with the fast onset of a linear growth of the perturbation with $a$ before a late--time slow down due to the presence of $\Lambda$.

The second case, $\sigma=0.2$ (Fig.~\ref{sigma0p2}), illustrates the opposite situation of an initially sub--Jeans length perturbation. Figs.~\ref{sigma0p2}(a),(b) show the numerical solution for $-\Ptil$ in this situation along the same reasoning as for Figs.~\ref{sigma10}(a),(b). At the early stage, the pressure gradient dominates and opposes the gravitational collapse. The perturbation behaves as an acoustic wave and is damped as it propagates away from the initial peak at $R=0$. However, the instability wave number $a \,k_J(\epsilon_H)$ quickly starts increasing over time (cf., Fig.~\ref{Jwave}). That is why around $a=50$ to $100$ the perturbation starts to grow as its typical wave number (estimated by $\sigma^{-1} = 5$) ends up below the critical value, with $a\,k_J(\epsilon_H) = 5$ for $a\simeq 94$, and it enters the unstable regime. The peak of this growing structure remains at a mostly stationary Lagrangian position, at $R \simeq 3.7$, while its increasing amplitude still remains small and below the initial value $-\Ptil(a=1,R=0) = 10^{-3}$ up to present time ($a \simeq 1090$).

For comparison we also consider the special case where the initial scale lies at the stability threshold, $\sigma=1$. The evolution of the corresponding solution for $-\Ptil$ with $a$ at several radii is shown in Figs.~\ref{sigma1}(a),(b), with the latter highlighting the early evolution ($1\leq a \leq 20$). Fig.~\ref{sigma1}(c) shows the spatial dependence of $-\Ptil$ with $R$ at some values of the scale factor. The behavior of the perturbation in this case is as expected intermediate, with an initial acoustic damping and propagation away from $R=0$ similarly to the $\sigma=0.2$ case, but more rapidly entering an unstable regime, after $a \simeq 5$. The amplitude of the perturbation then starts growing with a dust--like behavior up to beyond $20$ times its initial value at present time, with a shifted peak as in the $\sigma=0.2$ case, that stays around $R \simeq 2.5$.

\begin{figure*}[!ht]
\subfloat[]{\includegraphics[clip,width=0.9\columnwidth]{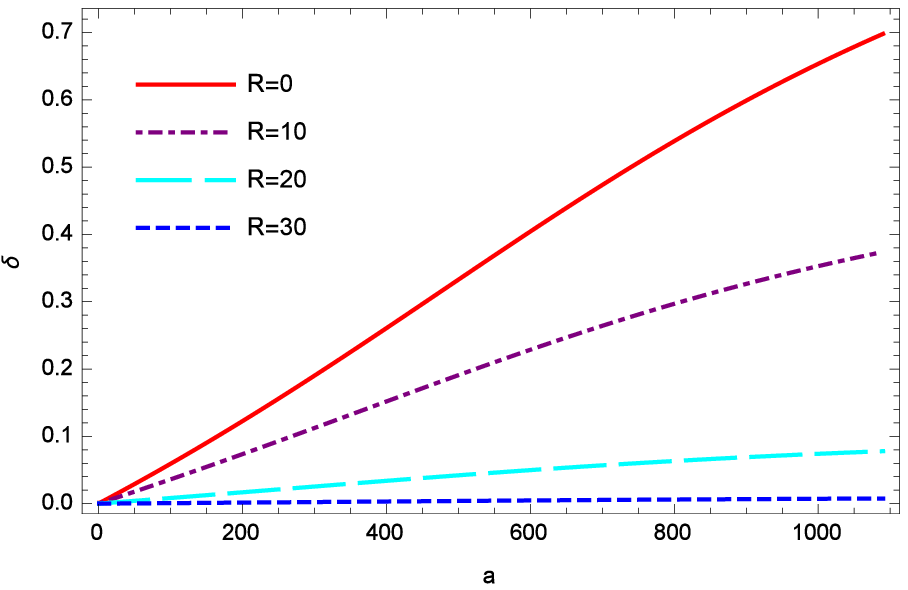}}
\subfloat[]{\includegraphics[clip,width=0.9\columnwidth]{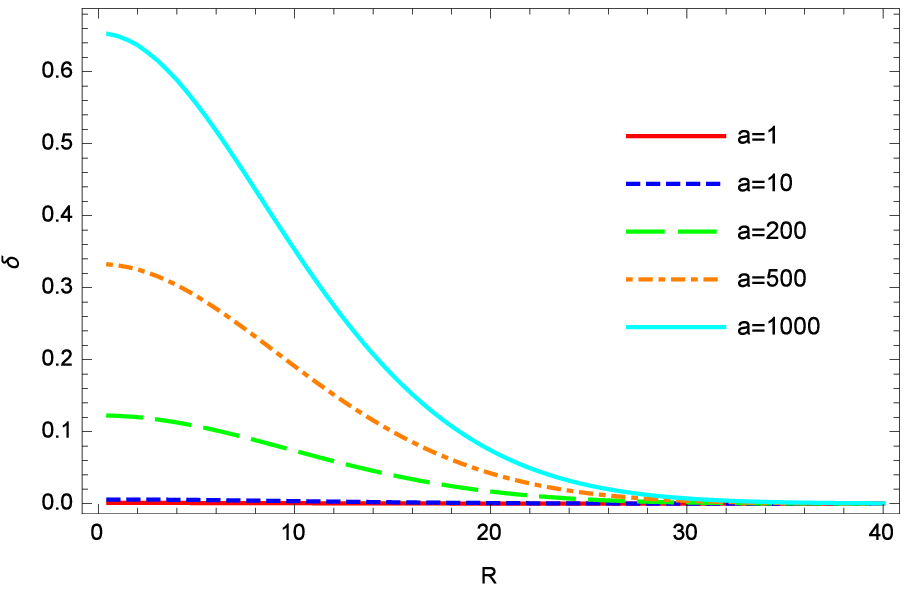}}
\caption{Numerical evaluation of the nonlinear density contrast $\delta$ as extrapolated from the first--order Lagrangian perturbation, where the initial $-\Ptil$ is the same spherical Gaussian field as for Fig.~\ref{sigma10}, with peak value of $10^{-3}$ and $k_J(\init{\epsilon_H})\,\sigma = 10$. (a). Evolution of $\delta$ with the background scale factor at fixed distances $R$. From top to bottom: $R=0$, $10$, $20$ and $30$. (b). Spatial variation of $\delta$ with the Lagrangian radius, for given values of $a$. From bottom to top: $a=1$, $10$, $200$, $500$ and $1000$. The overall behavior of $\delta$ is similar to the results of Fig.~\ref{sigma10} for the first--order $-\Ptil$ in the same situation, but the extrapolated density contrast grows faster at late times near the $R=0$ maximal overdensity. Additional nonlinear effects concerning the comparison with a standard perturbation approach, not studied here, could also be revealed by using instead as the 
$x$--axis for (b) the actual spatial metric distance to the $R=0$ fluid element (as an `Eulerian radius'), altering the spatial dependence. (See the discussion in Section \ref{discussion}.)}
\label{extrapolsigma10}
\end{figure*}
\begin{figure*}[!ht]
\subfloat[]{\includegraphics[clip,width=0.9\columnwidth]{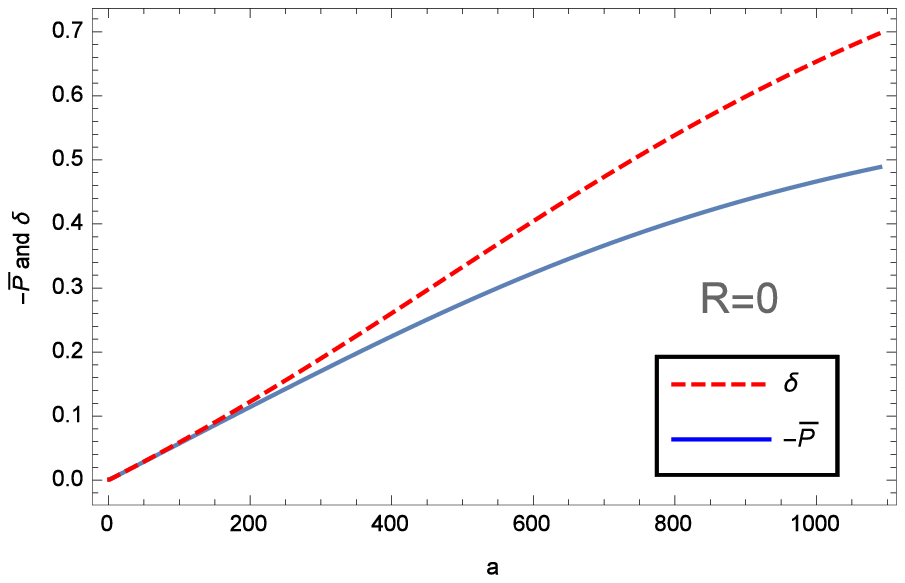}}
\subfloat[]{\includegraphics[clip,width=0.9\columnwidth]{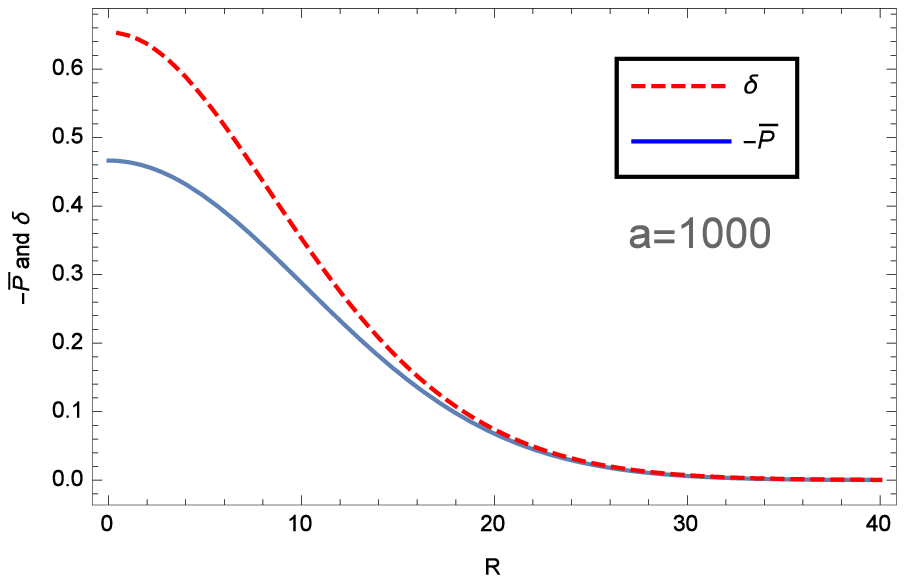}}
\caption{Comparison of the extrapolated nonlinear density contrast $\delta$ (dashed line) with the first--order solution for the sign--inverted deformation trace $-\Ptil$ (solid line) within the same setting as Figs.~\ref{sigma10} and~\ref{extrapolsigma10}. (a). Comparison of the evolution of both quantities as a function of $a$ at the centre of the overdensity ($R=0$). (b). Comparison of the spatial variation of both quantities with $R$ at a late time ($a=1000$).
In this situation, the perturbation grows large enough to enter the nonlinear regime and to render the time evolution and spatial behavior of the extrapolated $\delta$ clearly deviating from those of $-\Ptil$.}
\label{extrapolcomparsigma10}
\end{figure*}

\subsubsection{Evaluating the nonlinear density contrast}

As we recalled above, even the first--order Lagrangian perturbation scheme allows one to probe part of the nonlinear regime in the evaluation of observable quantities. This involves extrapolating these observables as exact, nonlinear functionals of the deformation field, the latter being evaluated as a solution to its first--order evolution equations and constraints.

Adopting this procedure for the rest mass density we evaluate it as the exact integral to the rest mass conservation equation \eqref{restmass_conservation}:
\begin{equation}
 \varrho = \frac{\init\varrho}{J} \quad ; \quad J = \det(\tensor{\eta}{^a_i}) = a^3 \det(\tensor{\delta}{^a_i}+\tensor{P}{^a_i}) \; ,
\end{equation}
where $\tensor{P}{^a_i}$ are the components of the deformation field. The density contrast $\delta$ is then deduced from the above:
\begin{equation}
 \delta := \frac{\varrho - \varrho_H}{\varrho_H} = \frac{\init\varrho}{\init{\varrho_H} a^{-3} J} -1 \quad ; \quad a^{-3} J = \det(\tensor{\delta}{^a_i}+\tensor{P}{^a_i}) \; ,
 \label{densitycontrast}
\end{equation}
and it is evaluated by replacing $\tensor{P}{^a_i}$ by the first--order solution.

Using the polytropic EoS and the parameters adopted here, the lapse may then be computed from
\begin{equation}
N=\frac{\varrho}{\epsilon+p} = \frac{\varrho}{\varrho + \frac{\gamma}{\gamma-1} \kappa \varrho^\gamma} = \frac{1}{1+\frac{5}{3} (1+\delta)^{2/3} \,a^{-2}} \; ,
\label{lapsepoly}
\end{equation}
with $\delta$ expressed from the deformation field as above. This formula shows that the lapse is $1$ in pressure--free (here empty) regions ($\delta=-1$) and decreases with increasing density contrast at a given time. The deviation $(1-N)$ rapidly decreases over time as $\propto a^{-2}$, with late time values of order $10^{-6}$ (when $a \simeq 1000$), as long as $\delta$ remains at most of order unity.

We will now illustrate this process for the density contrast with two examples using the same polytropic EoS as above. Note that this evaluation requires the knowledge of all components of the deformation field, including the traceless part. We specify procedures in Appendix~\ref{solutiontraceless} to obtain a particular (gravitoelectric) solution for the first--order traceless part from the initial conditions for the trace in specific cases. These procedures have been used to determine a consistent solution for the full deformation field in the examples below. We have also made use of the fact that the initial density $\init{\varrho} = F(\init{\epsilon_H}\, [1+\delta\init\epsilon])$ is well approximated by $F(\init{\epsilon_H}) (1+\init{\alpha_H} \, \delta\init\epsilon) = \init{\varrho_H} (1- \init\Ptil)$ for a small, still linear, initial density perturbation (with $\init{\alpha_H} = 3/4$ for the chosen EoS parameters) for the evaluation of $\delta$.

\paragraph{Localized overdensity:}~\\
Let us first retain the `spherical' initial overdensity example studied thus far in this section, with the initial conditions for the trace given by \eqref{initialdistribution}, with $\init c = 10^{-3}$, and $U=0$. The first--order solution for the trace in this situation has been determined above, and is complemented by a gravitoelectric solution for the first--order traceless part through the use of the procedure given in Appendix \ref{appendixFouriersec} that directly applies to this case. The determinant $J$ is then computed from this solution as in Appendix \ref{appendixfunctionals}, giving $\delta$ from Eq.~\eqref{densitycontrast}.

Note that when all components of the deformation field are very small, \textit{i.e.}, when it lies fully in the linear regime, then the extrapolated $\delta$ remains quantitatively close to $-\Ptil$, which corresponds to its expansion at first order in the deformation field. This is the case in the initially stable or marginally stable cases $\sigma=0.2$ and $\sigma=1$, where the initial acoustic damping of the perturbation keeps its amplitude small up to the present time despite the late--time growth. In both of these cases, the resulting density contrast indeed remains indistinguishable from the value of $-\Ptil$ already depicted above (Figs.~\ref{sigma0p2}--\ref{sigma1}).

We will consequently focus from now on on the case $\sigma=10$, where the unstable deformation reaches into the mildly nonlinear regime before the present time, as can be seen for the trace (whose amplitude reaches about $0.5$ at the present epoch).

Figs.~\ref{extrapolsigma10}(a),(b) show the result of the nonlinear evaluation of the density contrast in this situation, as a function of $a$ at given radii $R$, and as a function of the radius at several moments in its evolution, respectively. Although the general behavior is roughly similar to that of $-\Ptil$ (\textit{cf.} Fig.~\ref{sigma10}), nonlinear effects are visible in the amplified growth of $\delta$ at late times near $R=0$, with a maximal overdensity reaching about $0.7$ at present.

This nonlinear deviation of the density contrast functional with respect to its first--order estimate $-\Ptil$ is made explicit by the direct comparison of the peak ($R=0$) amplitude evolution of $\delta$ and $-\Ptil$ as a function of the background scale factor in Fig.~\ref{extrapolcomparsigma10}(a). The spatial dependence on $R$ of both quantities at late times, compared in Fig.~\ref{extrapolcomparsigma10}(b) at $a=1000$, is also visibly affected by the amplified growth of the density contrast where $\Ptil$ is no longer small, \textit{i.e.}, around $R=0$.

\begin{figure*}[!htp]
\subfloat[]{\includegraphics[clip,width=0.92\columnwidth]{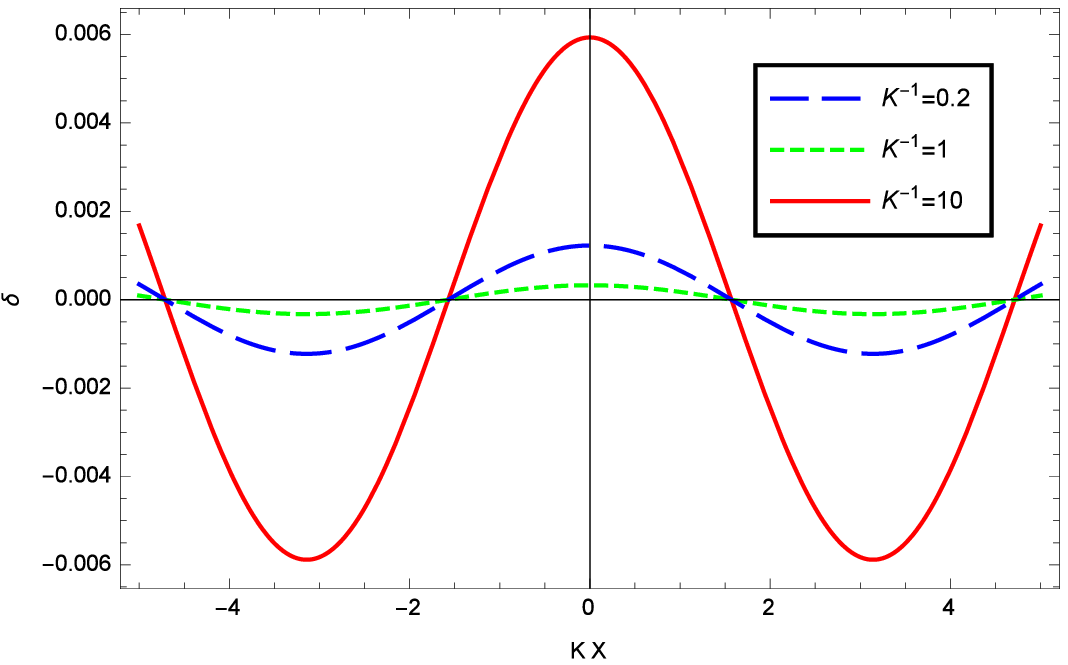}}
\subfloat[]{\includegraphics[clip,width=0.92\columnwidth]{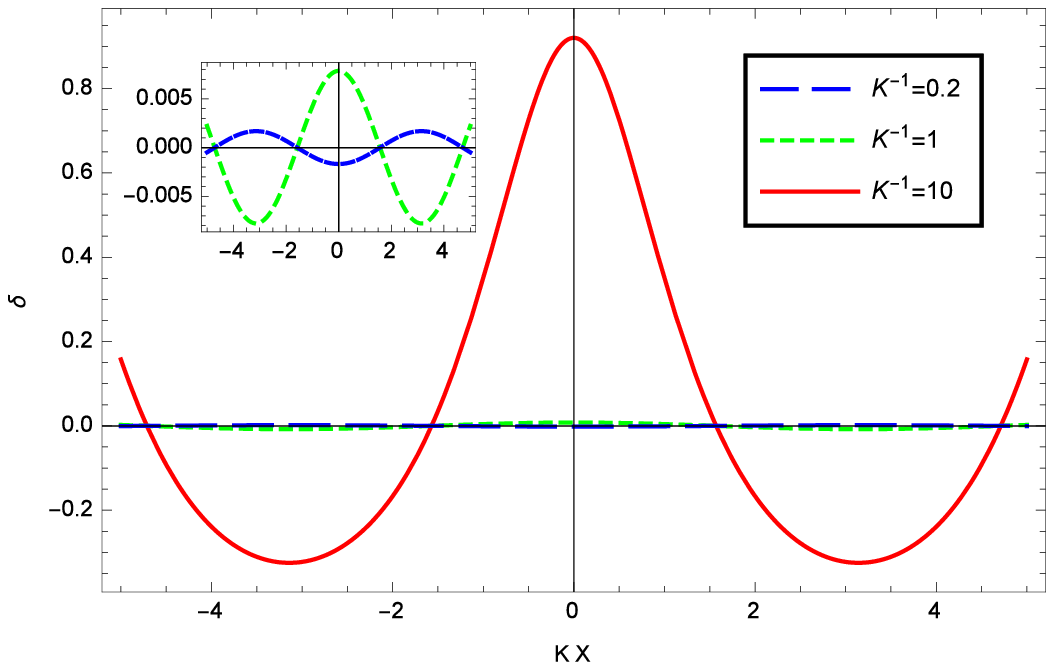}}
\caption{Numerical evaluation of the nonlinear density contrast $\delta$ as extrapolated from the first--order Lagrangian perturbation. The first--order deformation trace is taken as a plane--wave in Lagrangian coordinates of wave vector $\mathbf{K}$ (of norm $K$) along the $X$ coordinate, $-\Ptil \propto \cos(K X)$ , of initial amplitude $10^{-3}$. The result is shown at a given time as a function of $K X$ for three possible values of $K$, which is expressed in units $k_J(\init{\epsilon_H}) = 1$. (a). At $a=10$, for $K=0.1$ ($K^{-1}=10$), $K=5$ ($K^{-1}=0.2$) and $K=1$ by order of decreasing amplitude. (b). At $a=1000$, for $K=0.1$ ($K^{-1}=10$), $K=1$ and $K=5$ ($K^{-1}=0.2$) by order of decreasing amplitude. The side panel displays the (otherwise barely visible) latter two curves on a different vertical scale. The most unstable perturbation, for $K^{-1}=10$, displays a non--sinusoidal asymmetric shape at late times as it reaches the mildly nonlinear regime. This shape would be further nonlinearly modified, \textit{via} a different $x$--axis dependence, if this axis were expressed alternatively in terms of an Eulerian--type, regularly spaced (in terms of spatial metric distances), $x$ coordinate.}
\label{planewavedelta}
\end{figure*}
\paragraph{Lagrangian monochromatic wave:}~\\
The second toy model we consider is that of a single Lagrangian monochromatic wave deformation. The choices of background parameters and the length unit ($k_J(\init{\epsilon_H}) = 1$) are unchanged.
The initial perturbation is now chosen to be
\begin{equation}
 -\init\Ptil = \init c \cos(KX) \quad ; \quad U = 0 \; ,
\end{equation}
where we will again take $\init c = 10^{-3}$ as an initial amplitude. This situation corresponds to an initially stationary 
monochromatic wave in the given Lagrangian coordinate set,\footnote{%
Similarly to the interpretation of $R$ for the previous example, it is important to keep in mind that the perturbation we are considering here only has a sinusoidal dependence in the chosen Lagrangian coordinates $X^i$. It would have a different functional dependence in terms of actual physical (metric) spatial distance between two points on a given hypersurface $t=const$. One expects for instance, at a given late time $t$ and along a given spatial geodesic line, the distance between the successive perturbation nodes at $KX = -\pi/2$ and $KX = \pi/2$ (surrounding a collapsing overdensity) to be shorter than the distance between the nodes at $KX = \pi/2$ and $KX = 3 \pi/2$ (surrounding an expanding underdensity), despite all nodes being equally separated in terms of the Lagrangian coordinate $X$.
} $-\init\Ptil = \init c \cos(\delta_{ij} K^i X^j + \phi_0)$ with $\phi_0 = 0$ and a Lagrangian wave vector $\mathbf{K}$ along the first coordinate $X$, with components $K^i = (K,0,0)$.

The first--order trace solution then remains in this monochromatic mode form in the Lagrangian coordinates at all times, $\Ptil = \hat P_{\mathbf{K}}(t) \cos(KX)$. The amplitude $\hat P_{\mathbf{K}}(t)$ evolves according to the ordinary differential equation \eqref{planewavetracemastereq} which is solved by numerical integration for a given wave number $K$. A gravitoelectric solution for the traceless part is then determined along the lines of Appendix \ref{appendixplanewave}, where the relevant amplitude $\hat Q_{\mathbf{K}}(t)$ is again numerically evaluated, knowing $\hat P_{\mathbf{K}}(t)$, through its defining time integral formula \eqref{planewaveQKdef}. From these, one can calculate the density contrast in the same way as in the previous example, with the determinant $J$ evaluated as detailed in Appendix \ref{appendixfunctionals}.

Here we again study three cases distinguished by their wave number in direct analogy to the previous example, with $K^{-1}$ playing the role of the characteristic length $\sigma$. We accordingly choose $K^{-1} = 0.2$, $K^{-1} = 1$ and $K^{-1} = 10$, which at the initial time are stable, marginally stable and unstable, respectively. The corresponding spatial dependence of $\delta$ as a function of $KX$ for the three wave number choices is shown at an early time ($a=10$) in Fig.~\ref{planewavedelta}(a), and at a late time ($a=1000$) in Fig.~\ref{planewavedelta}(b).

In this situation, in the first two cases the components of the deformation field again remain small at all times, due to initial acoustic oscillations, and the density contrast thus follows the sinusoidal shape of $-\Ptil$ at all times. This is also the case for the unstable mode $K^{-1}=10$ at $a=10$ when it is still in the linear regime. At $a=1000$, however, this mode clearly deviates from this behavior as its amplitude is no longer linear. In particular, an asymmetry develops between the under-- and overdensity magnitudes as the latter is sharply amplified by the nonlinear evolution of $\delta$.

\subsubsection{Discussion}
\label{discussion}

In both examples above, the Lagrangian scheme and the proposed extrapolation procedure exhibit nonlinear effects on the overdensity for unstable perturbations when they become large enough. The amplitude of large overdensities in these examples is clearly underestimated when they are approximated by the first--order expression $-\Ptil$ instead of using the nonlinear extrapolation for~$\delta$.

An even higher initial overdensity amplitude could actually lead to a vanishing determinant $a^{-3} J$ at the maximum overdensity at a late enough time, implying $\varrho \rightarrow \infty$ with deformation coefficients still of order $1$. This situation corresponds to a shell--crossing, beyond which the first--order Lagrangian scheme in no longer valid.

The presence of pressure can delay its occurrence by damping the perturbation.
An improvement of the perturbative scheme to account for further local nonlinear effects in the dynamical evolution, e.g., allowing for a nonlinear coefficient to define the Jeans length is needed, however, to fully circumvent this problem. Velocity dispersion effects may in principle allow us to model the multistream regime, and the stabilization of structure formation in the form of virialization, which may help to avoid shell--crossings \cite{buchert:shanghai,RN50}.

We emphasize that the current Lagrangian perturbation scheme already contains another effect of nonlinear structure evolution, which lies in the exact propagation of the spatial coordinates used along the fluid flow lines. This is analogous to the inclusion of quadratic convection terms within linear Lagrangian time derivatives in the Newtonian framework.\footnote{%
In addition to the time derivatives being taken at different fixed spatial coordinates, a difference also comes from the spatial derivative operators, such as the Laplacian $\Delta_0$ appearing in the trace master equation \eqref{mastereqtrace}, being expressed in terms of Lagrangian coordinates and thus differing from the corresponding Eulerian operators. (See \cite{adlerbuchert} for the explicit transformation in the Newtonian case.)
}

Let us suggest a procedure that would be required to make these effects explicit also in the relativistic context; its concrete application is beyond the scope of this paper.

Eulerian--like coordinates could first be recovered, at least along a given spatial geodesic direction, by labeling points at equal intervals of spatial metric distances. This would involve solving for the initial metric components $G_{ab}$ such that their Ricci tensor is consistent with the initial conditions \eqref{initialconditiontl}--\eqref{initialconditionenergy} for given initial deformation field data, and then functionally evaluating and integrating the line element as given by \eqref{functionalds2} from the first--order solution for $\tensor{P}{^a_i}$. The resulting length, as a function of a Lagrangian coordinate, could then be used as an estimate of the Eulerian coordinate distance. Finally, this relation would have to be numerically inverted so that a given Lagrangian function obtained through the Relativistic Zel'dovich Approximation, such as $\varrho(X^i)$, could be expressed as a function of the Eulerian coordinate $x$ along the chosen line.

A different functional dependence on this spatial distance (which may be normalized by $a(t)$ to become a background comoving distance), as compared to the fluid--comoving coordinates $X^i$, would thus include nonlinear effects of the fluid--propagation--dependent coordinate transformation.

Recall, however, that a three--dimensional family of Eulerian observers generally does not exist in a relativistic (intrinsic) description. Strictly, a coordinate transformation to Eulerian space can only be conducted after the Minkowski Restriction of the relativistic solution has been executed.

\section{Conclusion}
\label{SecV}

In this paper we have generalized the Lagrangian perturbation approach to the
nonlinear evolution of inhomogeneous general relativistic model universes
containing a single irrotational fluid obeying a general barotropic relation.

By choosing a suitable set of coframes, we obtained the master partial differential equations for the evolution of the trace and traceless parts of the first--order deformation field that reduce to the corresponding equations in the dust case. The trace part also matches the Newtonian limit of the corresponding Lagrangian perturbation problem.

We discussed the procedure proposed in previous papers of how to find the solution for perturbations that propagate in the perturbed space, and applied this procedure to specific toy models,
illustrating the mildly nonlinear evolution of the density contrast.
We also discussed the limits of a first--order Lagrangian scheme, and we proposed ideas for a nonperturbative
generalization, which is needed especially in application to cases where the pressure term is taken to model multistreaming beyond the mildly nonlinear regime.\\

\smallskip\noindent
{\small {\bf Acknowledgements:}
This work is part of a project that has received funding from the European Research Council (ERC)
under the European Union's Horizon 2020 research and innovation program (grant agreement ERC advanced grant 740021--ARTHUS, PI: TB).
It is also supported by Catalyst grant CSG--UOC1603 administered by the Royal Society of New Zealand.
YL is supported by the China Scholarship Council. PM is supported by a `sp\'ecifique Normalien' Ph.D. grant, and acknowledges support and hospitality for a visit to the University of Canterbury. PM also acknowledges support by the National Science Centre, Poland, under grant 2014/13/B/ST9/00845. YL and DLW acknowledge support and hospitality for visits to CRAL--ENS, Lyon. We wish to thank Asta Heinesen for helpful discussions.}

\appendix
\section{Examples of solutions for the gravitoelectric traceless part}
\label{solutiontraceless}

In this paper we will not attempt to find the general solution of Equations \eqref{firstordertraceless}--\eqref{momentumconstraintEOS2} for the traceless part. We will, however, discuss a procedure for finding one possible solution for suitably chosen traceless--part initial conditions. For any barotropic EoS, this yields one example of a full gravitoelectric solution for all components of the deformation field $\tensor{P}{^a_i}$. It can then be substituted into exact nonlinear formulae to extrapolate functionals of the coframes such as metric distances or the rest mass density. 

To find such an example solution, we will focus on the gravitoelectric part which is directly coupled to the trace, and accordingly we set the gravitomagnetic part to zero.

\subsection{Case of a Lagrangian monochromatic wave}
\label{appendixplanewave}
Let us first assume that the first--order trace solution can be written as a single monochromatic wave mode in the given set of Lagrangian spatial coordinates $X^i$:
\begin{equation}
\Ptil(t,X^i) = \varphi(\mathbf{K} \cdot \mathbf{X}) \,{\hat P}_{\mathbf{K}}(t) \; ,
\end{equation}
for some constant Lagrangian wave vector $\mathbf{K}$, where\break $\mathbf{K} \cdot \mathbf{X} := \delta_{ij} K^i X^j$, and $\varphi(\mathbf{K} \cdot \mathbf{X}) = \cos(\mathbf{K} \cdot \mathbf{X} + \phi_0)$, with constant phase $\phi_0$. This form is a solution of the first--order trace master equation, if and only if $\hat P_{\mathbf{K}}(t)$ is a solution of the ordinary differential equation
\begin{align}
 & \frac{{\rm d}^2}{{\rm d}t^2} \hat P_{\mathbf{K}}+2H (1-3\beta' \!\left(\epsilon_H\right)) \, \ddt{}\hat P_{\mathbf{K}}
- \mathcal{W}(t) N_H^{2} \,\hat P_{\mathbf{K}} \nonumber \\
& \qquad = - a^{-2}N_H^{2}\,\beta'\left(\epsilon_{H}\right) K^2 \,\hat P_{\mathbf{K}} \; ,
\label{planewavetracemastereq}
\end{align}
with $K := \left(\delta_{i j} K^i K^j \right)^{1/2}$. Then $\Pold = \Ptil - \init\Ptil ={}$\break $\varphi(\mathbf{K} \cdot \mathbf{X})\,(\hat P_{\mathbf{K}}(t) - \hat P_{\mathbf{K}}(\init t))$.

\noindent
Setting
\begin{align}
 &\hat Q_{\mathbf{K}}(t) := \int_{\init t}^t N_H(t')\,\,\partial_t \!\left(\frac{\hat P_{\mathbf{K}}}{N_H} \right) \!(t')\,\,\rd t' \nonumber \\
 				&{\quad}= \hat P_{\mathbf{K}}(t) - \hat P_{\mathbf{K}}(\init t) - 3 \int_{\init t}^{t} \! H(t')\, \beta'(\epsilon_H)(t')\, \hat P_{\mathbf{K}}(t') \;\rd t' \; ,
\label{planewaveQKdef}
\end{align}
the time integral of the momentum constraints \eqref{momentumconstraintEOS2} is
\begin{equation}
 \tensor{\Pi}{^i_{j|i}} = \frac{2}{3} \hat Q_{\mathbf{K}}(t) K_j \, \varphi'(\mathbf{K} \cdot \mathbf{X}) \; .
\end{equation}
We now take $\tensor{\Pi}{^i_j}$ to be a purely longitudinal mode and get the following solution to the momentum constraints
(with $K_j := \delta_{jl}\,K^l$):
\begin{align}
 \tensor{\Pi}{^i_j} &{}= \left(\frac{K^i K_j}{K^2} - \frac{1}{3} \tensor{\delta}{^i_j} \right) \hat Q_{\mathbf{K}}(t) \,\varphi(\mathbf{K} \cdot \mathbf{X})
 \label{tracelesssolution1} \\
				&{}= \left( \frac{K^i K_j}{K^2} - \frac{1}{3} \tensor{\delta}{^i_j} \right) \left(\frac{\hat Q_{\mathbf{K}}(t)}{\hat P_{\mathbf{K}}(t) - \hat P_{\mathbf{K}}(\init t)} \right) P(t,X^i) \; .
\end{align}
Substituting this form into the master equation \eqref{firstordertraceless} shows that it is consistently a solution of both equations for the traceless part.
It is straightforward to show from the above formula that $2\, \Delta_0 \tensor{\Pi}{^i_j} + \tensor{\Pi}{^k_l_|_k^|^l}\,\tensor{\delta}{^i_j} - 3 \, \tensor{\Pi}{^i_k_|_j^|^k} = 0$, \textit{i.e.}, this $\tensor{\Pi}{^i_j}$ obeys the defining relation \eqref{laplacianEPi} for the gravitoelectric part and evolves according to \eqref{firstordertracelessE}.
This solution is thus a pure gravitoelectric one, amounting to setting the gravitomagnetic part to zero by the choice of vanishing gravitomagnetic traceless part of the initial deformation: $\tensor{\Pi}{^i_j} = \tensor[^{E}]{\Pi}{^i_j}$. 

Choosing this solution amounts to specifying the following (gravitoelectric) initial conditions:
\begin{align}
 \tensor[^{\mathrm{tl}}]{U}{^i_j} &{}= \left(\frac{K^i K_j}{K^2} - \frac{1}{3} \tensor{\delta}{^i_j} \right) \Big(U + 3 \init H \,\beta'(\init{\epsilon_H}) \,\init{\alpha_H}\,\delta\init\epsilon\Big) \; ; \\
 \tensor[^{\mathrm{tl}}]{W}{^i_j} &{}= \left(\frac{K^i K_j}{K^2} - \frac{1}{3} \tensor{\delta}{^i_j} \right) \Big(W + 3 \init H \,\beta'(\init{\epsilon_H}) \,U \nonumber \\
 &{\quad}+3 \,\big[ \partial_t (H \beta'(\epsilon_H))(\init t) + 2 \,H^2_{\mathrm i} \, \beta'(\init{\epsilon_H}) \big] \init{\alpha_H} \,\delta\init\epsilon \Big) \; .
\end{align}
This is compatible with the set of constraints on the initial conditions given in Section \ref{SubsecInitCond}, in particular the initial momentum constraints \eqref{initialconditionmomentum} and Eq.~\eqref{initialconditiontl}, provided that the latter is used to specify the traceless part of the initial first--order Ricci tensor $\tensor{\mathscr{T}}{^i_j}$.

The corresponding full perturbation field $\tensor{P}{^i_j} = \tensor{\Pi}{^i_j} + \frac{1}{3} \tensor{\delta}{^i_j} P$ then reads:
\begin{align}
 \tensor{P}{^i_j} ={}& \frac{K^i K_j}{K^2} \left(\frac{\hat Q_{\mathbf{K}}(t)}{\hat P_{\mathbf{K}}(t) - \hat P_{\mathbf{K}}(\init t)} \right) P \nonumber \\
 				&{} + \frac{1}{3} \tensor{\delta}{^i_j} \left(1-\frac{\hat Q_{\mathbf{K}}(t)}{\hat P_{\mathbf{K}}(t) - \hat P_{\mathbf{K}}(\init t)} \right) P \, .
\label{planewavePij}
\end{align}
Note that the corresponding deformation $1-$forms $\mathbf{P}^a = \tensor{\delta}{^a_k} \tensor{P}{^k_i} {\mathrm d}X^i$ are not exact due to the different time evolution of the trace and gravitoelectric traceless parts. This contrasts with the dust case where a purely gravitoelectric perturbation would lead to integrable coframes \cite{rza4}, so that only the non--flat initial metric would prevent one obtaining an Euclidean spatial metric at all times in that situation.

By linearity of the equations, a solution for $\tensor{\Pi}{^i_j}$ can also be obtained when the trace is a finite sum of such monochromatic waves, or the sum of the two time--evolution modes solutions of the evolution equation \eqref{planewavetracemastereq} for a given wave vector $\mathbf{K}$, simply by summing the corresponding solutions as given by \eqref{tracelesssolution1}.

\subsection{Case of a spatially localized solution}
\label{appendixFouriersec}
We assume here either that the spatial slices are globally diffeomorphic to the Euclidean space $\mathbb{R}^3$, \textit{i.e.}, that they can be covered by a single chart, or that the deformation field can be assumed to vanish outside a given chart. In either case it suffices to work within the Euclidean space spanned by the spatial coordinates in a given chart.

Let us now consider a spatially localized solution for the trace, \textit{e.g.}, a local overdensity evolving from an initial Gaussian perturbation in terms of the given set of spatial Lagrangian coordinates, as studied in the numerical examples of Section \ref{SecIV}. More specifically, we require the solution for the trace to always be a square--integrable function of the spatial coordinates in the chart, so that its Fourier transform in these coordinates can be performed and inverted. We can thus write:
\begin{equation}
 \Ptil(t,X^i) = \iiint e^{- i \mathbf{K} \cdot \mathbf{X}} \hat P (t,\mathbf{K}) \,\rd ^3 \mathbf{K} \; ,
\end{equation}
where $\hat P(t,\mathbf{K})$ is a solution of the evolution equation \eqref{planewavetracemastereq} at fixed $\mathbf{K}$, with the initial conditions set by the forward Fourier transform in the chart coordinates:
\begin{align}
 & \hat P(\init t, \mathbf{K}) = - \frac{1}{(2\pi)^3} \,\init{\alpha_H} \iiint e^{i \mathbf{K} \cdot \mathbf{X}}\, \delta\init\epsilon(\mathbf{X}) \, \rd ^3 X \; ; \\
 & (\partial_t \hat P)(\init t, \mathbf{K}) = \frac{1}{(2\pi)^3} \iiint e^{i \mathbf{K} \cdot \mathbf{X}}\, U(\mathbf{X}) \, \rd ^3 X \; .
\end{align}

Note that the above approach represents an alternative and complementary formulation of the method of solution presented in \cite{rza4} which formally replaces the Lagrangian coordinates by `Eulerian' ones. In the present paper it is applied in Sections \ref{sectracesolutionw} and \ref{secpolytrope}. The reformulation suggested here allows us to be more explicit about the required assumptions, as well as expressing the coordinate components of tensors such as $\tensor{\Pi}{^i_j}$ in a more convenient form. In both formulations, the use of plane--wave modes and flat--space Fourier transformations is sufficient since the Lagrangian first--order master equations to be solved only involve the metric--independent coordinate spatial derivatives ${}_{|i}$ and Laplacian $\Delta_0 = {}_{|i|j} \, \delta^{i j}$ as spatial derivative operators.

By linearity of the equations, a solution for the (gravitoelectric) traceless part is obtained by summation of the plane wave solutions for all Fourier modes:
\begin{align}
 \tensor{\Pi}{^i_j} = \tensor[^{E}]{\Pi}{^i_j} = \iiint e^{- i \mathbf{K} \cdot \mathbf{X}} \frac{K^i K_j}{K^2} \hat Q(t,\mathbf{K})\,\rd ^3 \mathbf{K} \nonumber \\
 -\frac{1}{3} \,\tensor{\delta}{^i_j} \iiint e^{- i \mathbf{K} \cdot \mathbf{X}} \,\hat Q(t,\mathbf{K}) \,\rd ^3 \mathbf{K}\; ,
\end{align}
with
\begin{equation}
\hat Q(t,{\mathbf{K}}) := \int_{\init t}^t N_H(t')\,\,\partial_t \!\left(\frac{\hat P(t,{\mathbf{K}})}{N_H(t)} \right) \!(t')\,\,\rd t' \; .
\end{equation}
Using this solution again implies a specific choice of initial conditions for the traceless deformation field (in particular taking it to be gravitoelectric) and for the traceless part of the spatial Ricci tensor.

In the case of spherically symmetric initial conditions in the chart coordinates, \textit{i.e.}, when $\delta\init\epsilon(X^i)$ and $U(X^i)$ only depend on $R := \left(\delta_{i j} X^i X^j \right)^{1/2}$, their Fourier transform will also depend only on $K$. From the evolution equation \eqref{planewavetracemastereq}, this feature is preserved over time, so that one can write $\hat P(t,\mathbf{K})$ as $ \hat P(t,K)$ and consequently $\hat Q(t,\mathbf{K})$ as $\hat Q(t,K)$ and $\Ptil(t,X^i)$ as $\Ptil(t,R)$. The above solution for $\tensor{\Pi}{^i_j}$ can then be computed as
\begin{equation}
 \tensor{\Pi}{^i_j} = \left( \frac{X^i X_j}{R^2} - \frac{1}{3} \, \tensor{\delta}{^i_j} \right) q(t,R) \; ,
\end{equation}
with $X_j := \delta_{jk} X^k$ and
\begin{align}
 & q(t,R) := \frac{4\pi}{R} \int_{0}^{\infty} K \sin(R K) \,\hat Q(t,K) \,\rd K \nonumber \\
 & {}- \frac{4\pi}{R^3} \! \int_{0}^{\infty} \!\!\left(\!\frac{\sin(R K)}{K} - R \cos(R K)\!\right) \hat Q(t,K) \,\rd K \, .
\end{align}

\subsection{Time integral of the gravitoelectric evolution equation}
The above procedure gives a way of obtaining a traceless part consistent with the momentum constraints and evolution equations in particular situations, and when only initial conditions on the trace part (or on the energy density) are explicitly specified. Alternatively, and still focusing on a purely gravitoelectric traceless part, a solution can be derived from the gravitoelectric traceless evolution equation \eqref{firstordertracelessE}, if the trace part and the (gravitoelectric) traceless initial conditions are known. It can be achieved by rewriting this evolution equation as follows: 
\begin{align}
 & \partial_t\left(\frac{a^3}{N_H} \,\partial_t \!\tensor[^E]{\Pi}{^i_j} \right) =- \frac{a N_H}{3} \tensor{\mathcal{D}}{^i_j} \left( \int_{\init t}^t \! N_H \, \partial_t \! \left( \frac{\Ptil}{N_H} \right) {\rm d}t' \right) \nonumber \\
 &{} + \frac{a N_H}{3} \Big( \big[1 + 3 \beta'(\epsilon_H)\big] \, \tensor{\CD}{^i_j} \Ptil
-\big[1 + 3 \beta'(\init{\epsilon_H})\big] \, \tensor{\CD}{^i_j} \init\Ptil\Big) \nonumber \\
&{}+\frac{a N_H}{N^{2}_{H_\mathbf{i}}} \, \Big( \tensor[^{\rm{tl},E}]{W}{^i_j} \!+ \init H \, \big[1- 3 \beta'(\init{\epsilon_H})\big] \tensor[^{\rm{tl},E}]{U}{^i_j} \Big)\; ,
\end{align}
after replacing $\Delta_0 \tensor[^{E}]{\Pi}{^i_j}$ by its integral expression \eqref{laplacianEPi2} in terms of $\Ptil$. It can be readily time--integrated twice to give $\tensor[^{E}]{\Pi}{^i_j}$. This yields the full $\tensor{\Pi}{^i_j}$ if the initial conditions are chosen such that the gravitomagnetic part vanishes.

In contrast to the previous subsections, this procedure can be applied in general, allowing the gravitoelectric initial conditions for the traceless part to be freely set. However, this requires the initial conditions $\tensor[^{\rm{tl}}]{U}{^i_j} = \tensor[^{\rm{tl},E}]{U}{^i_j}$ and $\tensor[^{\rm{tl}}]{W}{^i_j} = \tensor[^{\rm{tl},E}]{W}{^i_j}$ to be explicitly specified. While the trace parts relate to the energy density and spatial scalar curvature, the tracefree parts are related to properties of the gravitational wave components at the initial time. The latter have to be set in such a way as to fulfill the momentum constraints and their time derivative at the initial time, as well as the geometric constraints \eqref{electricconstraintU}--\eqref{electricconstraintW} for the gravitoelectric parts.

\subsection{On the evaluation of physical quantities}
\label{appendixfunctionals}
From given solutions for the trace and traceless parts, the full deformation field is straightforwardly obtained as $\tensor{P}{^i_j} = \tensor{\Pi}{^i_j} + (1/3) P \,\tensor{\delta}{^i_j}$, with $P = \Ptil - \init \Ptil$. This expression can then be inserted into the Lagrangian functional expressions for various physical quantities in terms of the deformation field. They can then be directly evaluated without any further linearization. This extrapolation is a crucial part of the Relativistic Zel'dovich Approximation as defined in \cite{rza1}, and it generally requires the knowledge of all components of the deformation field.

One would for instance directly compute a spatial distance from the line element
\begin{equation}
 \rd s^2 = a(t)^2 G_{a b} \left( \tensor{\delta}{^a_i} + \tensor{P}{^a_i} \right) \left( \tensor{\delta}{^b_j} + \tensor{P}{^b_j} \right) \rd X^i \rd X^j \; ,
\label{functionalds2}
\end{equation}
where knowledge of $G_{ab}(X^k)$ is also required. In turn, the rest mass density (with initial conditions set in such a way that it does coincide with $\varrho = F(\epsilon)$) would be computed as
\begin{equation}
 \varrho = \frac{\init{\varrho}}{J} = \frac{\init{\varrho_H} \left(1 + \init{\alpha_H} \, \delta\init\epsilon \right)}{a^3 \det \left( \tensor{\delta}{^a_i} + \tensor{P}{^a_i} \right)} \; .
\end{equation}
For the evaluation of the latter, note that in the case of a monochromatic wave (with one or both time--evolution modes), the deformation field components can be written as follows:
\begin{equation}
 \tensor{P}{^i_j} = \lambda_1 \frac{K^i K_j}{K^2} + \lambda_2 \, \tensor{\delta}{^i_j} \;,
\end{equation}
and similarly in the case of a localized spherically symmetric perturbation,
\begin{equation}
\tensor{P}{^i_j} = \lambda_1 \frac{X^i X_j}{K^2} +\lambda_2 \, \tensor{\delta}{^i_j} \; .
\end{equation}
The coefficients $\lambda_1(t,X^k)$, $\lambda_2(t,X^k)$ for the monochromatic case are directly deduced from \eqref{planewavePij} or from a sum of two such solutions, while in the localized spherically symmetric case, $\lambda_1(t,X^k) = q(t,R)$ and $\lambda_2(t,X^k) = \big(P(t,R) - q(t,R)\big) / 3$.
The determinant of the spatial coframe coefficients, from which $\varrho$ is evaluated, is then expressed in both cases by
\begin{equation}
 J = a^3 (1+\lambda_2)^2 \,(1+\lambda_1+\lambda_2) \; ,
 \label{sphericalJ}
\end{equation}
leading to an infinite rest mass density (from shell--crossing) whenever $\lambda_2 \rightarrow -1$ or $\lambda_1 + \lambda_2 \rightarrow -1$.

Such an extrapolation procedure provides the exact metrical distances, density and other physical properties as produced by the deformation field at a given order. In particular, this gives powerful approximations for the Ricci and Weyl curvatures that are not available in standard perturbation theory. It is, however, clear that the resulting expressions are approximations that must be controlled.

We can further combine the exact functionals for a given deformation with exact averages of Einstein's equations. An example was given in \cite{rza2} that also showed that the resulting prescription can even lead to exact results. For example, the combination of the first--order Lagrangian dust model with exact averages led to an exact formula for the kinematical backreaction within a class of averaged Lema\^\i tre--Tolman--Bondi solutions \cite{rza2}.

%--------------Reference macros
\def\art#1{{\em``#1'',}}
\def\PR#1#2{Phys.\ Rev.\ #1 {\bf#2}}
\def\PRL#1{Phys.\ Rev.\ Lett.\ {\bf#1}}
\def\CQG#1{Classical Quantum Gravity {\bf#1}}
\def\JMP#1{J.\ Math.\ Phys.\ {\bf#1}}
\def\GRG#1{Gen.\ Relativ.\ Gravit.\ {\bf#1}}
\def\AaA#1{Astron.\ Astrophys.\ {\bf#1}}
\def\ApJ#1{Astrophys.\ J.\ {\bf#1}}
\def\MNRAS#1{Mon.\ Not.\ R.\ Astron.\ Soc.\ {\bf#1}}
\def\JCAP#1{J.\ Cosmol.\ Astropart.\ Phys.\ {#1}}
%--------------

\end{document}